\documentclass[%
 reprint,
superscriptaddress,
 amsmath,amssymb,
 aps,
floatfix,
]{revtex4-2}

\usepackage[english]{babel}
\usepackage{mathtools}
\usepackage{graphicx}
\usepackage{dcolumn}
\usepackage{bm}
\usepackage{physics}
\usepackage{comment}
\usepackage{color}
\usepackage{amsthm}
\usepackage{enumerate}
\usepackage{stmaryrd}
\usepackage{zx-calculus}
\usepackage{tikz}

\definecolor{myyellow}{RGB}{255,230,153}
\usetikzlibrary{external}
\usepackage{hyperref}
\allowdisplaybreaks[4]
\hypersetup{
colorlinks=true, 
bookmarks=true, 
bookmarksnumbered=true,
pdfborder={0 0 0},
bookmarkstype=toc
}

\usetikzlibrary{decorations.markings, arrows.meta}
\usetikzlibrary{fit}
\tikzset{
outarrow/.style={decoration={markings,mark=at position 0.4 with {\arrow{#1}}}, postaction=decorate},
inarrow/.style={decoration={markings,mark=at position 0.7 with {\arrow{#1}}}, postaction=decorate},
marrow/.style={decoration={markings,mark=at position 0.5 with {\arrow{#1}}}, postaction=decorate},
numarrow/.style 2 args={decoration={markings,mark=at position #2 with {\arrow{#1}}}, postaction=decorate}
}
\zxNewNodeFromPic{F}[
every node/.append style={transform shape},
scale=0.8
]{
\node[regular polygon, regular polygon sides=3, shape border rotate=90, draw=black,fill=white!50, inner sep=1pt, font=\footnotesize, zx main node] {$F$};
}
\zxNewNodeFromPic{Fdag}[
every node/.append style={transform shape},
scale=0.8
]{
\node[regular polygon, regular polygon sides=3, shape border rotate=270, draw=black,fill=white!50, inner sep=-1pt, font=\footnotesize, zx main node] {$F^\dagger$};
}
\zxNewNodeFromPic{Fsq}[
every node/.append style={transform shape},
scale=1
]{
\node[regular polygon, regular polygon sides=4, shape border rotate=45, draw=black,fill=white!50, inner sep=-1pt, font=\footnotesize, zx main node] {$F^2$};
}
\zxNewNodeFromPic{GKP}[
every node/.append style={transform shape},
scale=1
]{
\node[regular polygon, regular polygon sides=4, shape border rotate=45, draw=black,fill=yellow, inner sep=2pt,zx main node] {};
}

\theoremstyle{plain}
\newtheorem{thm}{Theorem}
\newtheorem*{thm*}{Theorem}
\newtheorem{lem}{Lemma}
\newtheorem*{lem*}{Lemma}

\newtheorem*{prf*}{Proof}

\theoremstyle{definition}
\newtheorem{dfn}{Definition}

\newcommand{\mat}[1]{\left( \begin{matrix} #1 \end{matrix} \right)}
\makeatletter
\newcommand{\subscripts}[3]{%
  \@mathmeasure\z@\displaystyle{#2}%
  \global\setbox\@ne\vbox to\ht\z@{}\dp\@ne\dp\z@
  \setbox\tw@\box\@ne
  \@mathmeasure4\displaystyle{\copy\tw@_{#1}}%
  \@mathmeasure6\displaystyle{{#2}_{#3}}%
  \dimen@-\wd6 \advance\dimen@\wd4 \advance\dimen@\wd\z@
  \hbox to\dimen@{}\mathop{\kern-\dimen@\box4\box6}%
}
\makeatother

\begin{document}

\preprint{APS/123-QED}

\title{ZX~Graphical Calculus for Continuous-Variable Quantum Processes}

\author{Hironari Nagayoshi}
\email{nagayoshi@alice.t.u-tokyo.ac.jp}
\affiliation{
 Department of Applied Physics, School of Engineering, The University of Tokyo, 7-3-1 Hongo, Bunkyo-ku, Tokyo 113-8656, Japan
}
\author{Warit Asavanant}
\email{warit@alice.t.u-tokyo.ac.jp}
\affiliation{
 Department of Applied Physics, School of Engineering, The University of Tokyo, 7-3-1 Hongo, Bunkyo-ku, Tokyo 113-8656, Japan
}
\affiliation{Optical Quantum Computing Research Team, RIKEN Center for Quantum Computing, 2-1 Hirosawa, Wako, Saitama 351-0198, Japan
}
\author{Ryuhoh Ide}
\affiliation{
 Department of Applied Physics, School of Engineering, The University of Tokyo, 7-3-1 Hongo, Bunkyo-ku, Tokyo 113-8656, Japan
}
\author{Kosuke Fukui}
\affiliation{
 Department of Applied Physics, School of Engineering, The University of Tokyo, 7-3-1 Hongo, Bunkyo-ku, Tokyo 113-8656, Japan
}
\author{Atsushi Sakaguchi}
\affiliation{Optical Quantum Computing Research Team, RIKEN Center for Quantum Computing, 2-1 Hirosawa, Wako, Saitama 351-0198, Japan
}
\author{Jun-ichi Yoshikawa}
\affiliation{Optical Quantum Computing Research Team, RIKEN Center for Quantum Computing, 2-1 Hirosawa, Wako, Saitama 351-0198, Japan
}
\author{Nicolas C. Menicucci}
\affiliation{
 Centre for Quantum Computation \& Communication Technology, School of Science, RMIT University, Melbourne, VIC 3000, Australia
}
\author{Akira Furusawa}
 \email{akiraf@ap.t.u-tokyo.ac.jp}
\affiliation{
 Department of Applied Physics, School of Engineering, The University of Tokyo, 7-3-1 Hongo, Bunkyo-ku, Tokyo 113-8656, Japan
}
\affiliation{Optical Quantum Computing Research Team, RIKEN Center for Quantum Computing, 2-1 Hirosawa, Wako, Saitama 351-0198, Japan
}

\date{\today}

\begin{abstract}


Continuous-variable (CV) quantum information processing is a promising candidate for large-scale fault-tolerant quantum computation. However, analysis of CV quantum process relies mostly on direct computation of the evolution of operators in the Heisenberg picture, and the features of CV space has yet to be thoroughly investigated in an intuitive manner. One key ingredient for further exploration of CV quantum computing is the construction of a computational model that brings visual intuition and new tools for analysis. In this paper, we delve into a graphical computational model, inspired by a similar model for qubit-based systems called the ZX~calculus, that enables the representation of arbitrary CV quantum process as a simple directed graph. We demonstrate the utility of our model as a graphical tool to comprehend CV processes intuitively by showing how equivalences between two distinct quantum processes can be proven as a sequence of diagrammatic transformations in certain cases. We also examine possible applications of our model, such as measurement-based quantum computing, characterization of Gaussian and non-Gaussian processes, and circuit optimization.

\end{abstract}

\maketitle


\section{\label{sec:Intro}Introduction}
Quantum computing is expected to become the next-generation technology by surpassing the performance of ordinary computers for certain tasks. Recent experiments have already demonstrated this in several platforms~%
\cite{arute_quantum_2019,zhong_quantum_2020}, though practical applications remain yet to be demonstrated. Toward the ultimate goal of universal quantum computing, further investigation into the ability of quantum computation is needed, as well as construction of feasible computational architectures.


In the field of classical computational theory, various computational models have been proposed~\cite{fernandez_models_2009}, such as the Turing machine, lambda calculus, and process algebra, to name a few. These models not only provide new perspectives from which to comprehend the theoretical nature of computation but also contribute to the designs of real computer architectures and programming languages. As the development of quantum devices is progressing rapidly, the necessity for a high-level quantum computational model has increased~\cite{miszczak_high-level_2012}.

Most of the computational models for quantum computation proposed so far are natural generalization of conventional computational models, including the quantum Turing machine~\cite{deutsch_quantum_1985} and the quantum lambda calculus~\cite{selinger_lambda_2005}. These models are based on a discrete-variable (DV) approach based on qubits.
In the quantum circuit model---the most widely used quantum computational model---computing processes are represented as sequences of quantum gates~\cite{nielsen_quantum_2000}, whose concept originates from the logic circuit of classical bit registers and has a natural interpretation as an array of physical unitary transformations evolving under specific Hamiltonians.

Another interesting DV quantum model that emerged from a different point of view is what is called a \emph{graphical calculus} model. This idea derives from categorical quantum mechanics, a research field that investigates the general mathematical structure of quantum information theory, providing insights into the unique features of
quantum theory such as entanglement and
superposition
\cite{heunen_categories_2019}. One of the most useful variations is called the \emph{ZX~calculus}~\cite{coecke_graphical_2007,coecke_interacting_2011}, proposed by Coecke and Duncan. It provides a graphical language with which to visualize qubit processes and manipulate them in an intuitive manner. In the framework of the ZX~calculus, any quantum process is represented as a simple diagram consisting of and undirected graph with parameters on the vertices. By following clearly defined procedures, one can easily verify the equivalence between two quantum processes via graphical transformations of corresponding diagrams. Since the ZX~calculus cleverly exploits the algebraic properties of the qubit Hilbert space, it has various applications such as quantum circuit optimization~\cite{kissinger_reducing_2020,de_beaudrap_techniques_2020}, analysis of measurement-based quantum computing~\cite{backens_there_2021,kissinger_universal_2019}, graphical reasoning of the surface code~\cite{horsman_quantum_2011}, quantum machine learning~\cite{toumi_diagrammatic_2021}, and others.

Aside from the DV approach, a continuous-variable (CV) approach~\cite{lloyd_quantum_1999} is attracting attentions in recent years.
Unlike DV computing with qubits, a CV approach exploits the infinite-dimensional nature of bosonic systems. As the bosonic Hilbert space is much larger than two-dimensional qubit space, it is possible to build error resilience directly into the states used as the fundamental quantum information carriers.
For instance, the CV approach is considered to be compatible with fault-tolerant quantum computation for various types
of logical qubit encoding, such as the Gottesman-Kitaev-Preskill qubit~\cite{gottesman_encoding_2001} and the binomial code~\cite{michael_new_2016}. Experimentally, the CV approach has been intensively investigated in recent years in multiple platforms, including optics~\cite{furusawa_unconditional_1998,asavanant_generation_2019,larsen_deterministic_2019,konno_logical_2024}, trapped ions~\cite{fluhmann_encoding_2019,podhora_quantum_2022,de_neeve_error_2022}, and superconducting resonators~\cite{wang_schrodinger_2016,gao_programmable_2018,campagne-ibarcq_quantum_2020,sivak_real-time_2023}. 

However, the major difficulty with the CV approach is the same as its main feature: the infinite-dimensional Hilbert space at its core and the continuous nature of the quantum variables used for computation.
As DV computing can be seen as natural expansion of classical computing, where bits are expanded to superposition of bits (qubits), the operational semantics of qubit computing is largely comprehensible. For example, some of the DV quantum gates have a natural interpretation as Boolean logic gates, such as the correspondences between the quantum $X$ gate and classical NOT gate, the quantum CNOT gate and classical XOR gate, etc. The notion of quantum circuits originally arose from classical logic circuits~\cite{feynman_quantum_1986}, and nowadays it is used as the standard tool to describe DV quantum computing. In contrast, CV processes are often treated as purely physical operations, and their computational meanings are less obvious to translate to computational language. This is part of the reason that DV computing has been actively investigated at the level of
abstract algorithms and multi-qubit quantum error correction, while most theoretical research on CV computing either remains at the operational level or includes a reduction into DV computing through logical qubit encoding.

One of the greatest difficulties in studying the CV approach is the absence of useful tools for physicists to describe CV quantum processes. There is an established way to represent Gaussian quantum states as graphs, and Gaussian operations appear as well-defined graph transformations~\cite{menicucci_graphical_2011}. We would like a graphical representation, though, that goes beyond Gaussian to include all CV states and operations.
Previous research has also shown that the methodology of the ZX~calculus can be extended into three and higher Hilbert-space dimensions~\cite{wang_qufinite_2021}. However, these attempts are confined only to finite-dimensional spaces, partially due to complicated properties of the CV space involving infinities.

Despite these obstacles, here we present the key algebraic structure for a ZX~calculus in the case of CV systems.
We give definitions for CV diagrams and rewrite rules and demonstrate how various quantum protocols can be expressed intuitively through diagram transformations.
Moreover, we discuss the possibility of applications of our graphical model, such as CV quantum circuit optimization, characterization of Gaussian processes, and graphical representation of measurement-based quantum computing. We expect that our model may serve as education material for those not familiar with CV approaches, as well as a convenient calculation tool for experienced physicists and a high-level language for computer scientists.

\section{\label{sec:Pre}Preliminaries}
In this chapter we review the basics of CV quantum computing~\cite{asavanant_optical_2022} and the original ZX~calculus~\cite{coecke_interacting_2011}. Throughout this paper, we assume a natural unit system where $\hbar = 1$.

\subsection{Continuous-variable system and quadrature eigenstates \label{ssec:CV_quad}}
In the CV (bosonic) quantum system, the quadrature operators $\hat{q}$ and $\hat{p}$ are defined as
\begin{align}
\hat{q} = \frac{1}{\sqrt{2}}(\hat{a}^\dagger+\hat{a}), \hat{p} = \frac{i}{\sqrt{2}}(\hat{a}^\dagger-\hat{a})
\label{eq:quad}
\end{align}
where $\hat{a}$ and $\hat{a}^\dagger$ represent the annihilation and creation operators of a bosonic mode, respectively.
By definition, these quadrature operators are canonical variables that satisfy the canonical commutation relation $[\hat{q},\hat{p}]=i$. Quadrature operators are Hermitian operators, and their eigenvalues ranges over all real numbers $\mathbb{R}$. We shall denote corresponding eigenstates as $\ket{s}_q$ and $\ket{t}_p$, respectively, satisfying
\begin{align}
\hat{q} \ket{s}_q=&s\ket{s}_q, & \hat{p} \ket{t}_p=&t\ket{t}_p.
\end{align}

Using the Fock basis, $\ket{s}_q$ and $\ket{t}_p$ can be expanded as
\begin{align}
\ket{s}_q=\sum_{n=0}^{\infty} \frac{1}{\sqrt{n!2^n\sqrt{\pi}}}H_n(s)\exp(-\frac{1}{2}s^2)\ket{n}\\
\ket{t}_p=\sum_{n=0}^{\infty} \frac{i^n}{\sqrt{n!2^n\sqrt{\pi}}}H_n(t)\exp(-\frac{1}{2}t^2)\ket{n}
\end{align}
where $H_n(x)$ denotes the $n$-th Hermite polynomial. These Fock expansions are associated with the definitions of quadrature operators \eqref{eq:quad} through the following recurrence relation:
\begin{equation}
    2xH_n(x)=2nH_{n-1}(x)+H_{n+1}(x).
\end{equation}
Note that quadrature eigenstates are unphysical because their norms are infinite. A Dirac~$\delta$ normalization is thus chosen for these states:
\begin{align}
\subscripts{q}{\braket{s}{s'}}{q}=&\delta(s-s'),& \subscripts{p}{\braket{t}{t'}}{p}=&\delta(t-t').
\end{align}
Also, each family of quadrature eigenstates forms a complete orthonormal basis, satisfying
\begin{align}
\int_\mathbb{R}\dd{s}\subscripts{}{\ket{s}}{q}\subscripts{q}{\bra{s}}{}=\int_\mathbb{R}\dd{t}\subscripts{}{\ket{t}}{p}\subscripts{p}{\bra{t}}{}=\hat{I}. \label{eq:qp_comp}
\end{align}
For an arbitrary pure state $\ket{\psi}$, its wave functions $\psi(s)$ and $\Tilde{\psi}(t)$ are defined as follows:
\begin{align}
\psi(s)=&\subscripts{q}{\braket{s}{\psi}}{},& \Tilde{\psi}(t)=&\subscripts{p}{\braket{t}{\psi}}{}
\end{align}
These definitions, along with \eqref{eq:qp_comp}, offer two representations of CV quantum states:
\begin{align}
\ket{\psi}=\int_\mathbb{R}\psi(s)\ket{s}_q=\int_\mathbb{R}\Tilde{\psi}(t)\ket{t}_p
\end{align}

The eigenbases $\{\ket{s}_q\}$ and $\{\ket{t}_p\}$ are mutually associated via Fourier transform. In fact, the inner product of the quadrature eigenstates satisfies
\begin{align}\label{eq:qp_inner}
\subscripts{q}{\braket{s}{t}}{p}=\frac{1}{\sqrt{2\pi}}\exp(ist)
\end{align}
and thus, using \eqref{eq:qp_comp}, we obtain
\begin{align}
\Tilde{\psi}(t) =& \frac{1}{\sqrt{2\pi}}\int_\mathbb{R}\dd{s}\exp(-ist)\psi(s)\\
\psi(s)=& \frac{1}{\sqrt{2\pi}}\int_\mathbb{R}\dd{t}\exp(ist)\Tilde{\psi}(t)
\end{align}
which exactly corresponds to Fourier transform of complex functions.

A CV quantum state has another convenient representation other than wave function and density operator: the Wigner function. 
This is a quasi-probability function defined on the phase space that depicts quantum state in quadrature picture~\cite{wigner_quantum_1932}. Importantly, a quantum state is called a Gaussian state when its Wigner function is a (single- or multi-variable) Gaussian function. For example, the vacuum state, coherent states and squeezed states are all Gaussian states.

\subsection{Quantum gates in continuous-variable system}

In a CV quantum system, any unitary transformation can be uniquely specified by the transformation of quadrature operators it induces in the Heisenberg picture. That is, unitary operation on an $n$-mode CV systems uniquely corresponds to a $2n$-variable function $f$ that transforms the quadrature operators as
\begin{equation}
\hat{\vb{x}}^\mathrm{out}=f(\hat{\vb{x}}^\mathrm{in}) \label{eq:trans}
\end{equation}
where $\hat{\vb{x}}=(\hat{q}_1,\hat{p}_1,\ldots,\hat{q}_n,\hat{p}_n)^{\top}$. Note that $f$ must preserve the canonical commutation relations, namely $[\hat{q}_j,\hat{p}_k]=i\delta_{jk}$. When $f$ is
a linear affine map, then the corresponding quantum operation is called a Gaussian operation since it converts any Gaussian state only into another Gaussian state.

As is shown in Table~\ref{tab:table_gate}, Gaussian operations correspond to time evolution processes under (at most) quadratic Hamiltonians. In contrast, non-Gaussian operations require Hamiltonians of cubic or higher order in the quadrature operators. A non-Gaussian operation is also called a non-linear operation because it adds non-linear terms to the quadrature operators in the Heisenberg picture.

The relation between Gaussian and non-Gaussian operations is analogous to that of Clifford and non-Clifford operations in qubit systems. A CV quantum process limited to starting with Gaussian states and using only Gaussian operations and measurements
can be efficiently simulated by matrix multiplication~\cite{bartlett_efficient_2002}. Meanwhile, it is also known that one can approximate an arbitrary CV quantum operation if at least one single-mode non-Gaussian operation is available, in addition to a universal multi-mode Gaussian gate set~\cite{lloyd_quantum_1999}. In other words, the availability of a single non-Gaussian operation is necessary and sufficient for a multi-mode CV universal gate set when arbitrary Gaussian operations are available.

\begin{table*}
\caption{\label{tab:table_gate}List of representative CV quantum gates and their properties. Each index on a quadrature operator in an entangling gate specifies the mode it belongs to. The unitary action of each gate is given by $\hat{U}=e^{-i\hat{H}}$, and the quadrature transformation is defined using the Heisenberg picture. Note that all parameters except $\alpha$ need to be real since Hamiltonians are Hermitian.}
\begin{ruledtabular}
\begin{tabular}{cccccc}
Gate   & Notation & Number of modes & Hamiltonian & Quadrature transformation& Gaussianity \\
\colrule
Displacement gate  & $\hat{D}(\alpha)$ & 1 & $i(\alpha\hat{a}^\dagger-\alpha^*\hat{a})$& $\mat{\hat{q}\\\hat{p}}\mapsto\mat{\hat{q}+\sqrt{2}\Re(\alpha)\\\hat{p}+\sqrt{2}\Im(\alpha)}$ & Gaussian\\
Phase rotation gate& $\hat{R}(\theta)$& 1 & $\theta\hat{a}^\dagger\hat{a}$  & $\mat{\hat{q}\\\hat{p}}\mapsto\mat{\cos\theta\hat{q}+\sin\theta\hat{p}\\-\sin\theta\hat{q}+\cos\theta\hat{p}}$ & Gaussian\\
1-mode squeezing gate&$\hat{S}(r)$ & 1 & $-\frac{r}{2}(\hat{q}\hat{p}+\hat{p}\hat{q})$& $\mat{\hat{q}\\\hat{p}}\mapsto\mat{e^{-r}\hat{q}\\e^r\hat{p}}$& Gaussian\\
Controlled-sum gate & $\widehat{\mathrm{CS}}_{1,2}(g)$& 2 & $g\hat{q}_1\hat{p}_2$   & $\mat{\hat{q}_1\\\hat{p}_1\\\hat{q}_2\\\hat{p}_2}\mapsto\mat{\hat{q}_1\\\hat{p}_1-g\hat{p}_2\\g\hat{q}_1+\hat{q}_2\\\hat{p}_2}$   & Gaussian\\
Controlled-Z gate  & $\widehat{\mathrm{CZ}}(g)$& 2 & $-g\hat{q}_1\hat{q}_2$  & $\mat{\hat{q}_1\\\hat{p}_1\\\hat{q}_2\\\hat{p}_2}\mapsto\mat{\hat{q}_1\\\hat{p}_1+g\hat{q}_2\\\hat{q}_2\\g\hat{q}_1+\hat{p}_2}$   & Gaussian\\
Beamsplitter gate  & $\widehat{\mathrm{BS}}(\theta)$& 2 & $\theta(\hat{q}_1\hat{p}_2-\hat{p}_1\hat{q}_2)$ & $\mat{\hat{q}_1\\\hat{p}_1\\\hat{q}_2\\\hat{p}_2}\mapsto\mat{\cos\theta\hat{q}_1-\sin\theta\hat{q}_2\\\cos\theta\hat{p}_1-\sin\theta\hat{p}_2\\\sin\theta\hat{q}_1+\cos\theta\hat{q}_2\\\sin\theta\hat{p}_1+\cos\theta\hat{p}_2}$ & Gaussian\\
Cubic phase gate   & $\widehat{\mathrm{CPG}}(\gamma)$& 1 & $-\gamma\hat{q}^3$  & $\mat{q\\p}\mapsto\mat{q\\p+3\gamma q^2}$  & Non-Gaussian  
\end{tabular}
\end{ruledtabular}
\end{table*}

\subsection{Categorical quantum mechanics}
\label{subsec:catQM}

Categorical quantum mechanics is a relatively new field that investigates the mathematical backgrounds of quantum theory in terms of monoidal category theory~\cite{heunen_categories_2019}. In more detail, it focuses on the mathematical structures and relations of quantum processes by regrading quantum processes as linear maps between Hilbert spaces. One of the most significant feature of categorical quantum theory is that it comes already equipped with many structures that occupy important places in quantum theory, such as tensor product, dual space, transpose, adjoint, and superposition, to name a few. Another interesting aspect of this field is that a monoidal category is associated with a graphical language called a string diagram~\cite{selinger_survey_2011}, and thus it enables one to visualize quantum processes as simple diagrams. These diagrams provide useful tools for finding intuitive reasoning for various quantum processes, including quantum teleportation.

Complementarity and bialgebra~\cite{heunen_categories_2019} are the key mathematical features of quantum processes that must be present in any faithful representation using a graphical language.
These two properties, collectively referred as strong complementarity, play important roles in describing how different bases mutually relate to each other. This is made explicit in the ZX~calculus for finite-dimensional systems. In the following definitions, let $\mathcal{H}$ be a finite-dimensional Hilbert space with $\dim\mathcal{H}=n$.

\begin{dfn}[Swap operator]
We define the \emph{swap operator} as a linear map
$\hat{\sigma}: \mathcal{H}\otimes\mathcal{H}\to \mathcal{H}\otimes\mathcal{H}$ with $\hat{\sigma}(\ket{\psi}\otimes \ket{\phi}) = \ket{\phi}\otimes\ket{\psi}$.
\end{dfn}

\begin{dfn}[Complementary bases]
Let $\{\ket{u_i}\}_{i=1}^n$ and $\{\ket{v_j}\}_{j=1}^n$ each be an orthogonal basis in $\mathcal{H}$. This pair of bases is called \emph{complementary} (or equivalently, \emph{unbiased}) when there exists a
positive real number~$c > 0$ such that
\begin{equation}\label{eq:comp}
\abs{\braket{u_i}{v_j}}^2
= c
\end{equation}
for all $i,j$. In other words, this condition requires that the inner products of the two bases have constant absolute value---i.e.,~it is independent of the basis elements chosen.
\end{dfn}

To define a bialgebra, we first need to introduce the notion of a \emph{monoid} for a Hilbert space. 

\begin{dfn}[Monoid and commutative monoid]
    Let $\hat{\mu}: \mathcal{H}\otimes \mathcal{H}\to \mathcal{H}$ be a linear map and $\ket{\eta}\in \mathcal{H}$ be a (not necessarily normalized) state. The pair $(\hat{\mu},\ket{\eta})$ is called a \emph{monoid} if the following two conditions are satisfied:
    \begin{enumerate}
        \item $\hat{\mu}\circ(\mathrm{id}_{\mathcal{H}}\otimes\hat{\mu})=\hat{\mu} \circ (\hat{\mu} \otimes \mathrm{id}_{\mathcal{H}})$ (associativity)
        \item For arbitrary $\ket{\psi} \in \mathcal{H}$, $\hat{\mu}\ket{\psi}\otimes\ket{\eta}=\hat{\mu}\ket{\eta}\otimes\ket{\psi}=\ket{\psi}$ (unitality)
    \end{enumerate}
    where $\mathrm{id}_{\mathcal{H}}$ denotes the identity map on $\mathcal{H}$. If $\hat{\mu} = \hat{\mu}\circ \hat{\sigma}$, then the monoid is called \emph{commutative}.
\end{dfn}

The first condition implies that $\hat{\mu}$ induces an associative binary operation on $\mathcal{H}$, while the second requires $\ket{\eta}$ to be the identity element for that operation. Therefore, the definition above matches that for a monoid on $\mathcal{H}$. An important example of a monoid is one that is induced by an orthogonal basis:

\begin{dfn}[Monoid induced by a basis]\label{exa:monoid}
    Let $\{\ket{e_i}\}_{i=1}^n$ be an orthogonal basis in $\mathcal{H}$. Then, the pair $(\hat{\mu}, \ket{\eta})$, with
    \begin{align}
        \hat{\mu} =& \sum_{i=1}^n \frac{1}{\braket{e_i}{e_i}}\ket{e_i}\bra{e_i}\bra{e_i}\\*
        \ket{\eta} =&  \sum_{i=1}^n \frac{1}{\braket{e_i}{e_i}}\ket{e_i}
    \end{align}
    forms a \emph{monoid induced by the basis} $\{\ket{e_i}\}_{i=1}^n$.
\end{dfn}

We are now ready to define a bialgebra.

\begin{dfn}[Bialgebra]
    Let $(\hat{\mu}_1,\ket{\eta_1})$ and $(\hat{\mu}_2,\ket{\eta_2})$ each be monoids induced by two different bases on the same Hilbert space~$\mathcal{H}$. The pair of monoids is called a \emph{bialgebra} if the following four conditions are satisfied:
    \begin{align}
        \hat{\mu}_1^\dagger\circ\hat{\mu}_2 =& (\hat{\mu}_2 \otimes \hat{\mu}_2)\circ (\mathrm{id}\otimes \hat{\sigma} \otimes\mathrm{id}) \circ (\hat{\mu}_1^\dagger \otimes \hat{\mu}_1^\dagger)\\
        \hat{\mu}_1^\dagger \ket{\eta_2} =& \ket{\eta_2}\otimes\ket{\eta_2}\\
        \hat{\mu}_2^\dagger \ket{\eta_1} =& \ket{\eta_1}\otimes\ket{\eta_1}\\
        \braket{\eta_1}{\eta_2}=&1
    \end{align}
We will often refer to the bases that induced the monoids within a bialgebra as \emph{bases that form a bialgebra}.
\end{dfn}

Orthogonal bases that form a bialgebra can be characterized by the next theorem.

\begin{thm}\cite{coecke_strong_2012} \label{thm:group}
    Let $\{\ket{u_i}\}_{i=1}^n$ and $\{\ket{v_i}\}_{i=1}^n$ each be orthogonal bases, and $(\hat{\alpha},\ket{\phi})$ and $(\hat{\beta},\ket{\chi})$ be monoids induced
    by the bases, respectively.
    Then the pair of monoids forms a bialgebra if and only if there exists a commutative group $(G,\cdot)$ with $|G|=n$, and a bijective map~$\pi: G \to (1, \dotsc, n)$, with $\ket{u_g} = \ket{u_{\pi(g)}}$ (similarly for $\ket{v_g}$), such that
    \begin{align}
        \hat{\alpha} =& \sum_{g,h\in G} \frac{1}{\norm{\ket{v_g}}^2\norm{\ket{v_h}}^2}\ket{v_{g\cdot h}}\bra{v_g}\bra{v_h}\label{eq:group_1}\\
        \hat{\beta} =& \sum_{g,h\in G} \frac{1}{\norm{\ket{u_g}}^2\norm{\ket{u_h}}^2}\ket{u_{g\cdot h}}\bra{u_g}\bra{u_h}\label{eq:group_2}
    \end{align}
    and
    \begin{align}
        \ket{\phi} =& \ket{v_e}\label{eq:group_3}\\
        \ket{\chi} =& \ket{u_e}\label{eq:group_4}\\
        \braket{u_e}{v_e} =&1
    \end{align}
    where $e\in G$ denotes the identity element of $(G,\cdot)$.
\end{thm}

\begin{dfn}[Strongly complementary bases]
A pair of bases is called \emph{strongly complimentary} if it is both complimentary and forms a bialgebra.
\end{dfn}

Typically, orthogonal bases satisfying
\begin{align}
    \ket{u_j} =& \sum_{k=1}^{n} \exp\left(\frac{2\pi ijk}{n}\right)\ket{v_k}\\
    \ket{v_k} =& \sum_{j=1}^{n} \exp\left(-\frac{2\pi ijk}{n}\right)\ket{v_j}
\end{align}
correspond to the special case where $(G,\cdot)$ is isomorphic to $\mathbb{Z}/n\mathbb{Z}$, up to trivial scalar factor.
In this case, $\hat{\alpha}$ and $\hat{\beta}$ induced, respectively, by $\{\ket{u_i}\}_{i=1}^n$ and $\{\ket{v_i}\}_{i=1}^n$,
satisfy following relations
\begin{align}
    \hat{\alpha} &\propto \sum_{l,m=1}^{n} \ket{v_{l\oplus m}}\bra{v_l}\bra{v_m}\\
    \hat{\beta} &\propto \sum_{l,m=1}^{n} \ket{u_{l\oplus m}}\bra{u_l}\bra{u_m}
\end{align}
where $\oplus$ denotes addition modulo $n$.
These bases define the generalized Pauli $Z$-basis and generalized Pauli $X$-basis in finite-dimensional Hilbert spaces, whose properties are fully exploited in the ZX~calculus, as its name represents.



\subsection{ZX~calculus for qubits}\label{ssec:ZX_qubit}

The ZX~calculus for qubits mainly consists of two elements: (1)~ZX~diagrams and (2)~rewrite rules. A ZX~diagram comprises generators that represent certain projective linear operators individually, and connecting wires within these generators corresponds to composition of linear maps. (Note that here we restrict to quantum processes involving pure states only.)
A generator comprises two types of ``spiders'' with an arbitrary number of inputs and outputs,
\begin{align}
    \begin{ZX}
    \leftManyDots{}\zxZ[fill=white]{\alpha}\rightManyDots{}
\end{ZX}\coloneqq& \ket{0\ldots 0}\bra{0\ldots 0}+e^{i\alpha}\ket{1\ldots 1}\bra{1\ldots 1}\\
\begin{ZX}
    \leftManyDots{}\zxX[fill=lightgray]{\alpha}\rightManyDots{}
\end{ZX}\coloneqq& \ket{+\ldots +}\bra{+\ldots +}+e^{i\alpha}\ket{-\ldots -}\bra{-\ldots -}
\end{align}
and a ``box'' denoted by a blank box,
\begin{equation}
   \begin{ZX}
\zxNone{} \rar &[\zxwCol]\zxH{} \rar &[\zxwCol] \zxNone{}
\end{ZX}\coloneqq \ket{+}\bra{0}+\ket{-}\bra{1}
\end{equation}
representing the Hadamard gate. Though a ZX~diagram has no notion of direction, it should be read from left to right according to standard conventions to explicitly specify input-output relations.

Rewrite rules are formulated as equations between ZX~diagrams, identifying how a diagram can be transformed into another one while preserving the quantum process it represents. Below are examples of rewrite rules:
\begin{equation}
    \begin{ZX}
   \leftManyDots{}\zxZ[fill=white]{\alpha}\ar[dd,C=1,start anchor=south,end anchor=north]\ar[dd,C-=1,start anchor=south,end anchor=north]\rightManyDots{}\\
   &\ldots\\
   \leftManyDots{}\zxZ[fill=white]{\beta}\rightManyDots{}
\end{ZX}=\begin{ZX}
   \leftManyDots{}\zxZ[fill=white]{\alpha + \beta} \rightManyDots{}
\end{ZX}, \begin{ZX}
   \leftManyDots{}\zxZ[fill=lightgray]{\alpha}\ar[dd,C=1,start anchor=south,end anchor=north]\ar[dd,C-=1,start anchor=south,end anchor=north]\rightManyDots{}\\
   &\ldots\\
   \leftManyDots{}\zxZ[fill=lightgray]{\beta}\rightManyDots{}
\end{ZX}=\begin{ZX}
   \leftManyDots{}\zxZ[fill=lightgray]{\alpha + \beta} \rightManyDots{}
\end{ZX}
\end{equation}
\begin{equation}
    \begin{ZX}
\zxN{}\rar&\zxH{}\ar[dr,-N]&[\zxwCol]&[\zxwCol]\zxH{}\rar&\zxN{}\\
&\makebox[0pt][l]{\scalebox{0.8}{$\cvdotsCenterMathline$}}&\zxZ[fill=white]{}\ar[dr,N-]\ar[ur,N-]&\makebox[0pt][r]{\scalebox{0.8}{$\cvdotsCenterMathline$}}\\
\zxN{}\rar&\zxH{}\ar[ur,-N]&&\zxH{}\rar&\zxN{}
\end{ZX}=
\begin{ZX}
\zxN{}\ar[dr,-N]&[\zxWCol]&[\zxWCol]\zxN{}\\[\zxwRow]
\makebox[0pt][l]{\scalebox{0.8}{$\cvdotsCenterMathline$}}&\zxX[fill=lightgray]{}\ar[dr,N-]\ar[ur,N-]&\makebox[0pt][r]{\scalebox{0.8}{$\cvdotsCenterMathline$}}\\[\zxwRow]
\zxN{}\ar[ur,-N]&&\zxN{}
\end{ZX}
\end{equation}
Using rewrite rules, one may comprehend quantum dynamics by a simple graphical calculus. For instance, the quantum circuit shown below is a schematic illustration of qubit quantum teleportation using conventional quantum circuits.
\zxNewNodeFromPic{MyBox}[
main/.style={
/zx/picCustomStyleMyBoxMainNode/.append style={####1},
},
]{%
\node[draw, inner sep=1.3mm, rectangle, zx main node, minimum width=1.7em,minimum height=1.7em, execute at begin node=$, execute at
end node=$,
/zx/picCustomStyleMyBoxMainNode/.append style={}, 
/zx/picCustomStyleMyBoxMainNode,
]{\tikzpictext};
}
$$
\begin{ZX}[circuit]
\centering
\zxInput{\ket{\psi}} \ar[rr] &  & \zxNot{} \dar \ar[r]
& \zxBox[add label={below:\{$\ket{0},\ket{1}$\}}]{\zxMeter{}}\ar[ddrrr,classical,--|]\\[-3mm]
\zxInput{\ket{0}}\ar[r] & \zxNot{}\rar & \zxCtrl{} \rar & \zxMyBox{H} \ar[r]
& \zxBox[add label={below:\{$\ket{0},\ket{1}$\}}]{\zxMeter{}}\ar[dr,classical,--|]\\[-3mm]
\zxInput{\ket{+}}\ar[rrrrr] & \zxCtrl{}\ar[u] & &  
& & \zxMyBox{Z} \rar & \zxMyBox{X}\rar&\zxOutput{\ket{\psi}}
\end{ZX}
$$
It is easy to verify that this process is equivalent to the identity operation using the ZX~calculus~\cite{coecke_interacting_2011}:
\begin{align}
    \begin{ZX}
\zxN{}\rar& \zxX[fill=lightgray]{}\dar \ar[rr]&&\zxX[fill=lightgray]{a\pi}\\
\zxN{}\rar\dar[C]& \zxZ[fill=white]{}\rar&\zxH{}\rar&\zxX[fill=lightgray]{b\pi}\\
\zxN{}\ar[rrrr]&&&& \zxZ[fill=white]{b\pi}\rar&\zxX[fill=lightgray]{a\pi}\rar&\zxN{}
\end{ZX}=&
\begin{ZX}
\zxN{}\rar& \zxX[fill=lightgray]{a\pi}\dar\\
\zxN{}\rar\dar[C]& \zxZ[fill=white]{}\rar&\zxZ[fill=white]{b\pi}\\
\zxN{}\ar[rr]&& \zxZ[fill=white]{b\pi}\rar&\zxX[fill=lightgray]{a\pi}\rar&\zxN{}
\end{ZX}\\=&
\begin{ZX}
\zxN{}\rar& \zxX[fill=lightgray]{a\pi}\rar& \zxZ[fill=white]{2b\pi}\rar&\zxX[fill=lightgray]{a\pi}\rar&\zxN{}
\end{ZX}
\\=&
\begin{ZX}
\zxN{}\rar&[10mm]\zxN{}
\end{ZX}
\end{align}
In the procedure of the ZX~calculus, each diagram transformation is a simple application of a rewrite rule onto a fraction of the diagram, and thus one can straightforwardly confirm that the quantum process remains unchanged. This property is called \emph{soundness}---i.e., equivalence of diagrams implies equivalence of quantum processes. \emph{Completeness} of the ZX~calculus is the converse---i.e.,~whether two diagrams representing the same process are always graphically convertible. The rewrite rules of the ZX calculus (for qubits) are designed to represent equations of the qubit system in the standard interpretation, which means the qubit ZX calculus is sound. Although the original proposal of the ZX calculus is not complete~\cite{schroder_de_witt_zx-calculus_2014}, there are several variations of the model equipped with additional diagrams and rewrite rules to assure completeness~\cite{ng_universal_2017,jeandel_complete_2018}, making it an alternative \emph{formalism} (to Hilbert space) for representing quantum processes. 
Even without the assurance of completeness, one can still intuitively manipulate diagrams and straightforwardly simplify them by applying rewrite rules one by one to prove two diagrams are equivalent. For more detail, see~\cite{van_de_wetering_zx-calculus_2020,heunen_categories_2019,coecke_picturing_2017}.

The infinite-dimensional nature of CV systems creates challenges for proving soundness and completeness in a mathematically rigorous way. In our proposal for a CV ZX calculus below, we carefully design rewrite rules to achieve soundness within the standard framework of CV information processing and prove completeness of our model when confined to one-mode Gaussian states and operations. We discuss inherent limitations of the CV graphcal model in the following sections, including suggested avenues of investigation for overcoming them. 



The ZX~calculus has a helpful property of being topological, by which we mean that diagrams can be freely deformed as long as the connectivity between nodes is unchanged. This property lets us interpret and reshape diagrams in any way to obtain a different perspective as long as the input-output relationships are maintained. The topological feature is a special feature of the two-dimensional nature of the qubit space, which leads to the Pauli operators (and entangling gates such as CNOT) being self-inverse.
This degeneracy is a peculiarity of two-dimensional Hilbert space and generally would not hold for a higher-dimensional ZX~calculus. It also does not hold in our CV version. We will discuss this issue in the following sections.


Thanks to the flexibility of diagrams, the ZX~calculus has been actively applied to multiple fields of quantum computation. The most typical and successful application is quantum circuit optimization, which seeks a simplified quantum circuit for a given one so that the number of quantum gates and circuit depth are reduced. It has achieved a certain level of success so far~\cite{kissinger_reducing_2020,de_beaudrap_techniques_2020}, and there even exists a convenient Python library to demonstrate this optimization~\cite{kissinger_pyzx_2020}. Another interesting example is diagrammatic reasoning of measurement-based quantum computing (MBQC) using a diagramatic representation of graph states and projective measurements. This was the first use case of the ZX~calculus~\cite{duncan_rewriting_2010}. Recent research extends to the design and verification of quantum error correcting codes~\cite{horsman_quantum_2011}, graphical reasoning of lattice surgery~\cite{de_beaudrap_zx_2020}, analysis of certain computational complexity classes~\cite{de_beaudrap_tensor_2021}, an extended framework with diagrammatic differentiation for quantum machine learning~\cite{toumi_diagrammatic_2021}, among others.

Our goal in the next section is to propose the basic building blocks of a CV generalization of the ZX calculus. In later sections, we will prove a set of rewrite rules for the diagrams, discuss what properties of our construction can be proven with the tools at hand, and offer suggestions for eventually proving full soundness and completeness of a CV ZX calculus based on the starting point we propose.

\section{Graphical representation of continuous-variable processes}

Comparing \eqref{eq:qp_inner} and \eqref{eq:comp}, one might notice that position and momentum eigenbases in CV space satisfy complementarity. This is also true for bialgebra laws by taking an integral instead of finite sum. In fact, the following relations hold:
\begin{align}
\int_\mathbb{R}\dd{s}\subscripts{}{\ket{s}}{q_1}\subscripts{q_2, q_3}{\bra{s, s}}{} =& \frac{1}{\sqrt{2\pi}}\int_\mathbb{R}\dd{t}\dd{u}\subscripts{}{\ket{t+u}}{p_1}\subscripts{p_2,p_3}{\bra{t, u}}{}\\
\int_\mathbb{R}\dd{t}\subscripts{}{\ket{t}}{p_1}\subscripts{p_2, p_3}{\bra{t, t}}{} =& \frac{1}{\sqrt{2\pi}}\int_\mathbb{R}\dd{s}\dd{u}\subscripts{}{\ket{s+u}}{q_1}\subscripts{q_2,q_3}{\bra{s, u}}{}\\
\ket{0}_q =& \frac{1}{\sqrt{2\pi}}\int_\mathbb{R}\dd{t}\ket{t}_p\\
\ket{0}_p =& \frac{1}{\sqrt{2\pi}}\int_\mathbb{R}\dd{s}\ket{s}_q
\end{align}
This coincides with~\eqref{eq:group_1}--\eqref{eq:group_4}. For this reason, one may regard the position and momentum eigenbases as strongly complementary by the construction in Theorem.~\ref{thm:group} with the commutative group being $(\mathbb{R}, +)$. This perspective suggests that a CV graphical computational model theory may be constructed as well by simply replacing Pauli-$Z$ and $X$ eigenbases, respectively, with the position and momentum bases.

In this section, we first introduce two phase spiders as generators of our CV ZX calculus and their standard interpretations with the definition of diagram contraction. Consequently, we show several example of representation of quantum states and basic decompositions of each quantum gate in Table~\ref{tab:table_gate} as an array of phase spiders.

\subsection{Proper diagrams, operators, and their equivalence relations}

We begin with some important definitions.
\begin{dfn}[Diagram, inputs, outputs]
A \emph{diagram}~$D$ is an open, directed graph with labeled nodes. An \emph{input} of~$D$ is an open edge pointing inward. An \emph{output} of~$D$ is an open edge pointing outward.
\end{dfn}

We choose to draw diagrams from right to left so that input-output relations are matched with bra-ket notation (when the diagram does not contain loops).  Note this convention is only for readability and not substantial since our model employs directed graphs and is, in fact, topological. Wires in our diagrams are marked with arrows unless their direction can be freely reversed (we show examples later).

\begin{dfn}[Generator]
A diagram is called a \emph{generator} for the CV ZX calculus if it appears in Table~\ref{tab:table_gen}.
\end{dfn}

Diagrams can be combined in two important ways.

\begin{dfn}[Parallelization of diagrams]
For two diagrams $D_1$ and $D_2$, let $D_1\otimes D_2$ denote the \emph{parallelization} of $D_1$ and $D_2$. This is a new diagram where $D_1$ and $D_2$ are placed in parallel vertically as shown:
\begin{equation}
   \begin{ZX}[circuit]
\leftArrowedManyDots{}\zxGateMulti{1}{2}{D_1}&\rightArrowedManyDots{} \\[\zxZeroRow,1mm]
\leftArrowedManyDots{}\zxGateMulti{1}{2}{D_2}&\rightArrowedManyDots{}
\end{ZX}
\end{equation}
\end{dfn}

\begin{dfn}[Composition of diagrams]
Given two diagrams~$D_1$, $D_2$ with the number of outputs in $D_1$ equal to the number of inputs in $D_2$, let $D_2\circ D_1$ denote the \emph{composition} of $D_2$ and $D_1$, which is a diagram with all outputs of $D_1$ connected to inputs of $D_2$ as shown:
\begin{equation}
\begin{ZX}[circuit]
\leftArrowedManyDots{}\zxGateMulti{1}{2}{D_2}&\ar[r,3 vdots] \ar[r,o'={angle=25},marrow=<] \ar[r,o.={angle=25},marrow=<]&[\zxWCol]\zxGateMulti{1}{2}{D_1}&\rightArrowedManyDots{}
\end{ZX}
\end{equation}
\end{dfn}

This lets us recursively define the important concept of a proper diagram, which will be the main object of the CV ZX calculus.

\begin{dfn}[Proper diagram]
A diagram $D$ is called a \emph{proper diagram} if any of the following hold:
\begin{enumerate}
\item
$D$ is a generator,
\item
$D$ is a parallelization of any two proper diagrams, or
\item
$D$ is a composition of any two proper diagrams,
\end{enumerate}
\end{dfn}

The purpose of proper diagrams is to represent operators, so we need the following definition.

\begin{dfn}[Operator represented by a proper diagram]
Given a proper diagram~$D$, the \emph{operator represented by $D$} is denoted~$\llbracket {D}\rrbracket$ and defined as follows:
\begin{enumerate}
\item
When $D$ is a generator, $\llbracket {D}\rrbracket$ is defined in Table~\ref{tab:table_gen}.
\item
$\llbracket D_1 \otimes D_2 \rrbracket$ is the tensor product of the operators representing the two parallelized diagrams, i.e., $\llbracket D_1 \rrbracket \otimes \llbracket D_2 \rrbracket$.
\item
$\llbracket D_2 \circ D_1 \rrbracket$ is the composition (operator multiplication) of the operators representing the two composed diagrams, i.e., $\llbracket D_2 \rrbracket \circ \llbracket D_1 \rrbracket$ (or $\llbracket D_2 \rrbracket \llbracket D_1 \rrbracket$).
\end{enumerate}
\end{dfn}

In order to relate diagrams to each other, we need the following notions of equivalence for operators and diagrams.

\begin{dfn}[Equivalent operators]
Two operators~$\hat{A}$ and $\hat{B}$ are called \emph{equivalent}, denoted $\hat{A} \sim \hat{B}$, if there exists nonzero $c \in \mathbb{C} \backslash \{0\}$ such that $\hat{A} = c \hat{B}$.
\end{dfn}

\begin{dfn}[Equivalent diagrams]
Given two proper diagrams~$D_1$ and~$D_2$ and a set of rewrite rules~$\mathcal S$ for proper diagrams, $D_1$ and $D_2$ are called \emph{equivalent (modulo~$\mathcal S$)}, denoted $D_1 =_{\mathcal S} D_2$, if $D_1$ and $D_2$ are mutually rewritable using elements of~$\mathcal S$.  (Note: we will typically fix $\mathcal S$ to be the whole set of basic rewrite rules given in Sec.~\ref{ssec:CV_basic_rules} and drop the subscript $\mathcal S$ for simplicity.)
\end{dfn}

Finally, we can describe the important concept of soundness, which we define with respect to sets of proper diagrams.

\begin{dfn}[Soundness]
\emph{Soundness} of the calculus with respect to a set of diagrams~$\mathcal D$ and a set of rewrite rules~$\mathcal S$ is represented formally by $(D_1 =_{\mathcal S} D_2)\Rightarrow (\llbracket D_1\rrbracket\sim \llbracket D_2\rrbracket)$ for all $D_1, D_2 \in \mathcal D$.
\end{dfn}
If the set of all possible proper diagrams is sound with respect to a set of rewrite rules, then the entire calculus is said to be sound with respect to those rewrite rules.

\subsection{Phase spider and diagram contraction}

Functions written inside of a $q$-spider or a $p$-spider are what are called \emph{phase functions} that add quadrature-dependent phase factors. In this paper we impose a restriction for phase functions to be real polynomial functions. 
We denote phase functions of zero by blank spiders.
Note that spiders are symmetric under mode permutation, and thus combining swap diagrams with a single spider can be absorbed in the following way:
\begin{equation}
\begin{ZX}
   \zxN{}\ar[r,inarrow=<]&[\zxwCol]\zxN{}\ar[ddr,S]&[\zxwCol,2mm]\zxN{}\ar[dr,-N, outarrow=<]&[\zxwCol]&[\zxwCol,1mm]\\
   &&&  \zxZ{}\ar[r,inarrow=<]&\zxN{}\\
   \zxN{}\ar[r,inarrow=<]&\zxN{}\ar[uur,S]&\zxN{}\ar[ur,-N, outarrow=<]
\end{ZX}
=
\begin{ZX}
   \zxN{}\ar[dr,-N, outarrow=<]&[\zxwCol,1mm]&[\zxwCol,1mm]\\
   &  \zxZ{}\ar[r,inarrow=<]&\zxN{}\\
   \zxN{}\ar[ur,-N, outarrow=<]
\end{ZX}
\end{equation}
Using spiders, various quantum states can be drawn as simple diagrams. For example, quadrature eigenstates can be represented by linear phase functions as is shown here:
%
\begin{table*}
\caption{\label{tab:table_gen}List of generators for our graphical model and operators they represent. Note that a ket or bra subscript such as $q^m$ is a shorthand for the $q$ basis in $m$ modes.}
\begin{ruledtabular}
\begin{tabular}{cccc}
Name   & Diagram $D$  & Arity (outputs $\leftarrow$ inputs)& $\llbracket D \rrbracket$ \\
\colrule
$q$-spider  &
$$
\begin{ZX}\\[\zxHRow]
\leftArrowedManyDots{m} \zxZ{f(x)}\rightArrowedManyDots{n}\\[\zxHRow]
\end{ZX}
$$ & $m \leftarrow n$ &
$\displaystyle\int_\mathbb{R}\dd{s}e^{if(s)}\subscripts{}{\ket{s\ldots s}}{q^{m}}\subscripts{{q^{n}}}{\bra{s\ldots s}}{}$
\\
$p$-spider  &
$$
\begin{ZX}\\[\zxHRow]
\leftArrowedManyDots{m} \zxX{f(x)}\rightArrowedManyDots{n}\\[\zxHRow]
\end{ZX}
$$
& $m \leftarrow n$ &
$\displaystyle\int_\mathbb{R}\dd{t}e^{-if(t)}\subscripts{}{\ket{t\ldots t}}{p^{m}}\subscripts{{p^{n}}}{\bra{t\ldots t}}{}$\\
Swap&
$$
\begin{ZX}\\[\zxHRow]
\zxN{}& [\zxWCol] \zxN{} \ar[l,inarrow=>] \ar[dr,s] &[\zxWCol] \zxN{}\ar[r,inarrow=<]  &[\zxWCol] \zxN{} \\[\zxWRow]
\zxN{}& [\zxWCol] \zxN{} \ar[l,inarrow=>] \ar[ru,s] & \zxN{}\ar[r,inarrow=<] &[\zxWCol] \zxN{} \\[\zxWRow]
\end{ZX}
$$& $2\leftarrow 2$  &
$\displaystyle\iint_{\mathbb{R}^2}\dd{s}\dd{s'}\subscripts{}{\ket{s',s}}{q_1,q_2}\subscripts{q_1,q_2}{\bra{s, s'}}{}$\\
Fourier&
$$
\begin{ZX}\\[\zxHRow]
\zxN{}\rar[inarrow=<]&[\zxWCol]\zxF{}\rar[inarrow=<]&[\zxWCol]\zxN{}\\[\zxHRow]
\end{ZX}
$$
& $1\leftarrow 1$  &
$\displaystyle\int_\mathbb{R}\dd{u}\subscripts{}{\ket{u}}{p}\subscripts{q}{\bra{u}}{}$\\
Inverse Fourier &
$$
\begin{ZX}\\[\zxHRow]
\zxN{}\rar[inarrow=<] &[\zxWCol]\zxFdag{}\rar[inarrow=<]&[\zxwCol]\zxN{}\\[\zxHRow]
\end{ZX}
$$
&$1\leftarrow 1$&
$\displaystyle\int_\mathbb{R}\dd{u}\subscripts{}{\ket{-u}}{p}\subscripts{q}{\bra{u}}{}$\\
Squared Fourier &
$$
\begin{ZX}\\[\zxHRow]
\zxN{}\rar[inarrow=<] &[\zxWCol]\zxFsq{}\rar[inarrow=<]&[\zxwCol]\zxN{}\\[\zxHRow]
\end{ZX}
$$
&$1\leftarrow 1$&
$\displaystyle\int_\mathbb{R}\dd{s}\subscripts{}{\ket{-s}}{q}\subscripts{q}{\bra{s}}{}$
\end{tabular}
\end{ruledtabular}
\end{table*}
%
%
\begin{align}
\left\llbracket
\begin{ZX}
   \zxN{}& & [\zxwCol] \zxZ{} \ar[ll,N-, inarrow=>]
\end{ZX}
\right\rrbracket
=& \int_\mathbb{R}\dd s \ket{s}_{q}\sim\ket{0}_p \\
\left\llbracket
\begin{ZX}
   \zxN{}& & [\zxwCol] \zxX{} \ar[ll,N-, inarrow=>]
\end{ZX}
\right\rrbracket
=& \int_\mathbb{R}\dd t \ket{t}_{p}\sim\ket{0}_q
\label{eq:zeroq}
\end{align}
As for multi-mode states, an EPR state and a GHZ state can be represented as follows:
\begin{align}
\left\llbracket
\begin{ZX}
   \zxN{}&\\
   \zxN{}& & [\zxwCol] \zxZ{} \ar[ull,N-, inarrow=>] \ar[dll,N-, inarrow=>]\\
   \zxN{}&
\end{ZX}
\right\rrbracket
\sim& \int_\mathbb{R}\dd s \ket{s,s}_{q_1,q_2}=\int_\mathbb{R}\dd t \ket{t,-t}_{p_1,p_2}\label{eq:epr_diag_1}\\
\left\llbracket
\begin{ZX}
   \zxN{}&\\
   \zxN{}& & [\zxwCol] \zxX{} \ar[ull,N-, inarrow=>] \ar[dll,N-, inarrow=>]\\
   \zxN{}&
\end{ZX}
\right\rrbracket
\sim& \int_\mathbb{R}\dd t \ket{t,t}_{p_1,p_2}=\int_\mathbb{R}\dd s \ket{s,-s}_{q_1,q_2}\label{eq:epr_diag_2}\\
\left\llbracket
\begin{ZX}
   \zxN{}&\\
   \zxN{}& & [\zxwCol] \zxZ{} \ar[ll, inarrow=>] \ar[ull,N-, inarrow=>] \ar[dll,N-, inarrow=>]\\
   \zxN{}&
\end{ZX}
\right\rrbracket
\sim& \int_\mathbb{R}\dd s \ket{s,s,s}_{q_1,q_2,q_3}\\
\left\llbracket
\begin{ZX}
   \zxN{}&\\
   \zxN{}& & [\zxwCol] \zxX{} \ar[ll, inarrow=>] \ar[ull,N-, inarrow=>] \ar[dll,N-, inarrow=>]\\
   \zxN{}&
\end{ZX}
\right\rrbracket
\sim& \int_\mathbb{R}\dd t \ket{t,t,t}_{p_1,p_2,p_3}
\end{align}

Reversing the direction of each arrow gives interconversion between bra and ket, which can be understood as a generalized Choi-Jamio\l{}kowski isomorphism. For example, reversing one of the two output arrows of \eqref{eq:epr_diag_1} and \eqref{eq:epr_diag_2} yields the identity operation, and reversing both of them yields the projection process onto an EPR state that models a post-selected bell measurement. Also, the two kinds of EPR states above correspond to $q$- and $p$-correlated states respectively, both of which are anti-correlated in the other basis.


Connecting the input(s) of one diagram to the output(s) of another represents tensor contraction along those indices, a simple example of which is operator multiplication. However, as our calculus model handles directed diagrams and free connections, it is necessary to establish a generalized rule to admit loops to appear in the diagram. For this purpose, we define the following contraction rule:
\begin{equation}
\scalebox{0.9}{$
\left\llbracket
 \begin{ZX}[circuit]
\leftArrowedManyDots{}\zxGateMulti[
additional code={
\node[
inner sep=0pt,
at={($(zxMainNode.south west)+(1mm,0mm)$)},
zx subnode={d1l1}
] {};
\node[
inner sep=0pt,
at={($(zxMainNode.south west)+(2mm,0mm)$)},
zx subnode={d1l2}
] {};
\node[
inner sep=0pt,
at={($(zxMainNode.south east)+(-2mm,0mm)$)},
zx subnode={d1r1}
] {};
\node[
inner sep=0pt,
at={($(zxMainNode.south east)+(-1mm,0mm)$)},
zx subnode={d1r2}
] {};
}%
]{1}{2}{D_1} \ar[d,"\cdots","j_1"',start subnode={d1l1},end subnode={d2l1},C=0.5, marrow=<]\ar[d,"j_m",start subnode={d1l2},end subnode={d2l2},C-=0.5, marrow=<]\ar[d,"\cdots","i_1"',start subnode={d1r1},end subnode={d2r1},C=0.5, marrow = >]\ar[d,"i_n",start subnode={d1r2},end subnode={d2r2},C-=0.5, marrow = >] &\rightArrowedManyDots{} \\[\zxWRow]
\leftArrowedManyDots{}\zxGateMulti[
additional code={
\node[
inner sep=0pt,
at={($(zxMainNode.north west)+(1mm,0mm)$)},
zx subnode={d2l1}
] {};
\node[
inner sep=0pt,
at={($(zxMainNode.north west)+(2mm,0mm)$)},
zx subnode={d2l2}
] {};
\node[
inner sep=0pt,
at={($(zxMainNode.north east)+(-2mm,0mm)$)},
zx subnode={d2r1}
] {};
\node[
inner sep=0pt,
at={($(zxMainNode.north east)+(-1mm,0mm)$)},
zx subnode={d2r2}
] {};
}%
]{1}{2}{D_2}&\rightArrowedManyDots{}
\end{ZX} \right\rrbracket
\begin{aligned}
    \mathrel=
    \int \dd{s_{\bar\imath}}
    \dd{s_{\bar\jmath}}
%
%
%
%
&\subscripts
	{q_{\bar\imath}}
	{\mel{s_{\bar\imath}}
	{\llbracket {D}_1\rrbracket}{s_{\bar\jmath}}}{q_{\bar\jmath}}
\\
\otimes
&\subscripts
	{q_{\bar\jmath}}
	{\mel{s_{\bar\jmath}}
	{\llbracket {D}_2\rrbracket}{s_{\bar\imath}}}{q_{\bar\imath}}
\end{aligned}
$}
\end{equation}
where~$\bar\imath = i_1, \ldots, i_n$ and $\bar\jmath = j_1, \ldots, j_m$ are compound subscripts representing multiple copies of the symbol being subscripted, one for each item in the respective list. Notice that the contractions along $\bar\imath$ can be done in any basis across those modes (and similarly for $\bar\jmath$). This has an interpretation as a type of partial trace on ${\llbracket {D}_1\rrbracket} \otimes {\llbracket {D}_2\rrbracket}$ over the contracted modes, introducing generalized input-output relationship between the two operators. In fact, when $D_2$ represents the identity operation, the diagram exactly corresponds to partial trace on the operator $\llbracket D_1 \rrbracket$, which can be confirmed from the following equation:
\begin{align}
    &\left\llbracket
     \begin{ZX}[circuit]
    \leftArrowedManyDots{}\zxGateMulti[
    additional code={
    \node[
    inner sep=0pt,
    at={($(zxMainNode.south west)+(0.3mm,0.5mm)$)},
    zx subnode={d1l1}
    ] {};
    \node[
    inner sep=0pt,
    at={($(zxMainNode.south west)+(2mm,0mm)$)},
    zx subnode={d1l2}
    ] {};
    \node[
    inner sep=0pt,
    at={($(zxMainNode.south east)+(-2mm,0mm)$)},
    zx subnode={d1r1}
    ] {};
    \node[
    inner sep=0pt,
    at={($(zxMainNode.south east)+(-0.3mm,0.5mm)$)},
    zx subnode={d1r2}
    ] {};
    }%
    ]{1}{1}{D} \ar[d, start subnode={d1l1},end subnode={L},C=0.5, marrow=<,"A"' left]{}\ar[d,"A" right,start subnode={d1r2},end subnode={R},C-=0.5, marrow = >]\rightArrowedManyDots{} \\[\zxWRow]
    &
    \node[circle, zx subnode=L,inner sep = 0pt,xshift = -3mm,yshift = -0.5mm]{};
    \node[circle, zx subnode=R,inner sep = 0pt,xshift = 3mm,yshift = -0.5mm]{};
    \node[circle, zx subnode=C,inner sep = 0pt,xshift = -3.4mm,yshift = -0.5mm]{};
    \node[circle, zx subnode=D,inner sep = 0pt,xshift = 3.4mm,yshift = -0.5mm]{};
    \ar[start subnode={C},end subnode={D}, marrow=<]
    \end{ZX} \right\rrbracket
    \\ = &\iint_{\mathbb{R}^2} \dd{s_1}\dd{s_2} \subscripts{q_1}{\mel{s_1}{\llbracket {D}\rrbracket}{s_2}}{q_2}
    \subscripts{q_2}{\braket{s_2}{s_1}}{q_1}
    \\ = &\iint_{\mathbb{R}^2} \dd{s_1}\dd{s_2} \delta(s_2-s_1)\subscripts{q_1}{\mel{s_1}{\llbracket {D}\rrbracket}{s_2}}{q_2}
    \\ = &\int_{\mathbb{R}} \dd{s} \subscripts{q}{\mel{s}{\llbracket {D}\rrbracket}{s}}{q}
    \\ = &\Tr_A\left(\llbracket D \rrbracket\right)
    \end{align}
where $A$ denotes the qumode to be traced out.
Compositon of three or more diagrams can be defined by recursively applying the above interpretation.


In some cases, the contraction rule may not function properly because of infinite non-converging integrals. For example, the rule suggests that $\llbracket\begin{ZX}\zxZ{}\rar[marrow=<]&[\zxwCol]\zxZ{}\end{ZX}\rrbracket=\subscripts{p}{\braket{0}{0}}{p}=\int_\mathbb{R}1/\sqrt{2\pi}\dd t=\delta(0)$, which is technically infinite. Nevertheless, we will treat it as a global scalar that can be ignored (for instance, when discussing operator equivalence for corresponding diagrams). Issues like this are ubiquitous when working with CV systems, as mentioned in subsection~\ref{ssec:CV_quad}, and solving them is beyond the scope of this work. Our purpose is to introduce a toolbox that can be the starting point for further exploration.


\subsection{Graphical representation of CV quantum gates} \label{ssec:resp_gates}

By combining spiders defined above, one can obtain graphical representation of a wide variety of quantum gates. In this subsection, we show diagrams of the CV quantum gates listed in Table.~\ref{tab:table_gate}. We verify the consistency of the diagrammatic representations in Appendix \ref{sec:consistency}.

\subsubsection{Displacement gate}

The displacement gate has a linear Hamiltonian and thus can be decomposed into two displacements, one each with respect to $q$ and $p$. With this observation, we obtain the following representation of the displacement gate:
\begin{equation}
\label{eq:Dgate}
    \begin{ZX}
        \zxN{}&[\zxWCol] \zxBox{D(\alpha)} \lar[inarrow=>] \rar[inarrow=<] & [\zxWCol] \zxN{}
    \end{ZX}
    \coloneqq
    \begin{ZX}
        \zxN{}&[\zxWCol] \zxX{\sqrt{2}~\Re(\alpha)x} \lar[inarrow=>] \rar[marrow=<] &[\zxWCol] \zxZ{\sqrt{2}~\Im(\alpha)x}\rar[inarrow=<]&[\zxWCol]\zxN{}
    \end{ZX}
\end{equation}
Since any two displacement gates commute with each other up to an irrelevant global phase factor, the order of the $q$-spider and $p$-spider in the diagram above can be reversed.
In Sec.~\ref{ssec:gauss_univ}, we provide diagrams that are equivalent to the definitions of the quantum gates given in this chapter with their derivations based upon rewrite rules we define in the following sections.

\subsubsection{Phase rotation gate}

A phase rotation uses a quadratic Hamiltonian and can be decomposed into three shear gates. For this reason, the diagram corresponding to a phase rotation gate is described as three quadratic spiders:
\begin{equation}
    \begin{ZX}
        \zxN{}&[\zxWCol] \zxBox{R(\theta)} \lar[inarrow=>] \rar[inarrow=<] & [\zxWCol] \zxN{}
    \end{ZX}
    \coloneqq
    \begin{ZX}
        \zxN{}&[\zxWCol] \zxX{\frac{\tan(\theta/2)}{2}x^2} \lar[inarrow=>] \rar[inarrow=<] &[\zxWCol] \zxZ-{\frac{\sin\theta}{2}x^2} \rar[marrow=<]&[\zxWCol] \zxX{\frac{\tan(\theta/2)}{2}x^2} \rar[marrow=<]&[\zxWCol]\zxN{}
    \end{ZX}\label{def:rot_sp}
\end{equation}
%
Note that this definition is not formally defined where $\theta$ is an odd multiple of $\pi$ since $\tan{\frac{\theta}{2}}$ is infinite in that case. However, as a $\pi$ rotation gate is equivalent to a squared Fourier operation in Table~\ref{tab:table_gen}, we can employ its diagram as the definition of $\pi$-rotation diagram.


\subsubsection{1-mode squeezing gate}

Unlike the phase rotation gate, the 1-mode squeezing gate requires at least four shear gates for its decomposition, as is shown in the following diagram:
\begin{equation}
    \begin{ZX}
        \zxN{}&[\zxWCol] \zxBox{Sq(\tau)} \lar[inarrow=>] \rar[inarrow=<] & [\zxWCol] \zxN{}
    \end{ZX}
    \coloneqq
    \begin{ZX}
        \zxN{}&[\zxWCol] \zxX{\frac{\tau(1-\tau)}{4}x^2} \lar[inarrow=>] \rar[inarrow=<] &[\zxWCol] \zxZ-{\frac{1}{\tau}x^2} \rar[marrow=<]&[\zxWCol] \zxX{\frac{(\tau-1)}{4}x^2} \rar[marrow=<]&[\zxWCol] \zxZ{x^2} \rar[marrow=<]&[\zxWCol]\zxN{}
    \end{ZX}\label{eq:def_sq_diag}
\end{equation}
By setting $\widehat{Sq}(\tau)\coloneqq\hat{S}(r)$ with $\tau=e^{-r}$ one obtains the squeezing gate defined in Table~\ref{tab:table_gate}. However, here we allow $\tau$ to be negative to represent general cases. In particular, when $\tau=-1$, this is equivalent to $\hat{F}^2$---i.e.,~the square of Fourier transform. Thus, we can expand the definition of rotation spiders for odd multiples of $\pi$ using a 1-mode squeezing gate, namely
\begin{equation}
\begin{split}
\begin{ZX}
    \zxN{}&[\zxWCol] \zxBox{R((2n+1)\pi)} \lar[inarrow=>] \rar[inarrow=<] & [\zxWCol] \zxN{}
    \end{ZX}
    \coloneqq&
    \begin{ZX}
        \zxN{}&[\zxWCol] \zxBox{Sq(-1)} \lar[inarrow=>] \rar[inarrow=<] & [\zxWCol] \zxN{}
    \end{ZX}\\
    =&
    \begin{ZX}
        \zxN{}&[\zxWCol] \zxX-{\frac{1}{2}x^2} \lar[inarrow=>] \rar[inarrow=<] &[\zxWCol] \zxZ{x^2} \rar[marrow=<]&[\zxWCol] \zxX-{\frac{1}{2}x^2} \rar[marrow=<]&[\zxWCol] \zxZ{x^2} \rar[marrow=<]&[\zxWCol]\zxN{}
    \end{ZX}
\end{split}
\end{equation}
for all $n\in \mathbb{Z}$. 

\subsubsection{Controlled-sum gate}

The controlled-sum (CSUM) gate, also called a quantum nondemolition (QND) gate in the context of experimental operations for historical reasons, is called \emph{unbiased} when its gain $g$ is equal to~$1$. In that case, the unbiased CSUM gate $e^{-i\hat{q}_1\hat{p}_2}$ can be represented as a combination of two spiders as follows:
\begin{equation}
    \begin{ZX}
    \zxN{}\ar[drrr,-N,outarrow=<]&[\zxwCol]&[\zxwCol]&[\zxWCol]&[\zxwCol]&[\zxwCol]\\
    &&&\zxZ{} \ar[rr,marrow=<]&&\zxN{}\\[\zxWRow]
    \zxN{}\ar[rr,marrow=<] &&\zxX{}\ar[ru,s.,marrow=<]\ar[drrr,N-,inarrow=<]\\
    &&&&&\zxN{}
    \end{ZX}\label{eq:csum}
\end{equation}
With the fact that the diagram for the unbiased CSUM gate is interpreted as a CNOT gate in the original qubit ZX~calculus, one can naturally see the correspondence between the CSUM gate for CV systems and the CNOT gate for qubits. Another way of understanding the diagrams above is to interpret the wire in the middle of the two spiders as if it conveyed information from the first mode to the second mode. From this perspective, each wire of the diagram can be regarded as the flow of information, 
where the upper and lower wires correspond to the first (control) mode and the second (target) mode, respectively.

As for the general CSUM gate with gain~$g$, the effects of this gain can be equivalently represented  
by
squeezing one of the modes. Specifically, it can be represented by the following diagram:
\begin{equation}
    \begin{ZX}
    \zxN{}\ar[r,marrow=<]&[\zxwCol,1mm]\zxBox{Sq(g^{-1})}\ar[drr,-N,inarrow=<]&[\zxwCol]&[\zxWCol]&[\zxwCol,1mm]&[\zxwCol,1mm]\\[\zxZeroRow,-4mm]
    &&&\zxZ{} \ar[r,marrow=<]&\zxBox{Sq(g)}\ar[r,marrow=<]&\zxN{}\\[\zxWRow,4mm]
    \zxN{}\ar[rr,marrow=<] &&\zxX{}\ar[ru,s.,marrow=<]\ar[drrr,N-,inarrow=<]\\
    &&&&&\zxN{}
    \end{ZX}\label{eq:csum-gen}
\end{equation}
Substituting $g=-1$ into the diagram above yields the inverse of \eqref{eq:csum}. In the following sections, we will show that the inverse of the CSUM gate is equal to its own conjugation (which will be defined later).

\subsubsection{Controlled-Z gate}

As the Fourier transform interchanges $q$ and $p$ eigenstates (up to a sign), the controlled-Z~(CZ) operation can be represented using Fourier diagram and two $q$-spiders. For unbiased ($g=1$) CZ gate, its diagrammatic representation is given as
\begin{equation}
\begin{ZX}
&[\zxwCol]&[\zxwCol]&[\zxWCol]&[\zxWCol]&[\zxwCol]&[\zxwCol]\\
\zxN{}\ar[rr,marrow=<] &&\zxZ{}\ar[rd,s.,marrow=<]\ar[urrrr,N-,inarrow=<]\\[\zxWRow]
&&&\zxF{} \ar[rd,s.,marrow=<]&&\zxN{}\\[\zxWRow]
&&&&\zxZ{} \ar[rr,marrow=<]&&\zxN{}\\
\zxN{}\ar[urrrr,-N,outarrow=<]
\end{ZX}  
\end{equation}
Like for the CSUM gate, a general diagrammatic representation of the CZ gate is given by simply inserting a squeezing gate before and after the CZ gate, namely
\begin{equation}
    \begin{ZX}
        &[\zxwCol,2mm]&[\zxwCol,2mm]&[\zxWCol]&[\zxWCol]\zxBox{Sq(g)}\ar[rr,inarrow=<]&[\zxwCol]&[\zxwCol]\\[\zxZeroRow,-2mm]
        \zxN{}\ar[r,marrow=<] &\zxBox{Sq(g^{-1})}\ar[r,marrow=<]&\zxZ{}\ar[rd,s.,marrow=<]\ar[urr,N-,marrow=<]\\[\zxZeroRow,2mm]
        &&&\zxF{} \ar[rd,s.,marrow=<]&&\zxN{}\\[\zxWRow]
        &&&&\zxZ{} \ar[rr,marrow=<]&&\zxN{}\\
        \zxN{}\ar[urrrr,-N,outarrow=<]
        \end{ZX}
\end{equation}

The controlled-Z gate can also be regarded as a CSUM gate being converted via appropriate Fourier transforms. In addition, note that the controlled-Z gate is symmetric and thus invariant under mode permutation. These behaviors are also reflected in the diagrammatic representation, which will be proven via rewrite rules we define later.

\subsubsection{Beamsplitter gate}

Displacement, phase rotation, squeezing and CSUM gates form a universal gate set for Gaussian operations.
Hence, a graphical representation of any Gaussian gate can be obtained by combining the diagrams defined above \cite{lloyd_quantum_1999}. One such decomposition for the beamsplitter operation~$\exp(-i\theta(\hat{q}_1\hat{p} -\hat{p}_1\hat{q}_2))$ is given by 
\begin{equation}
  \begin{aligned}
BS(\theta)\equiv& \exp(\theta(\hat{q}_1\hat{p}_2-\hat{p}_1\hat{q}_2))\\
=&\widehat{Sq}_1\left(\frac{1}{\tan\theta}\right)\cdot \mathrm{CS}_{1,2}\cdot\widehat{Sq}_1\left(\frac{\sin^2\theta}{\cos\theta}\right)\\
&\cdot \widehat{Sq}_2\left(\frac{1}{\cos\theta}\right)\cdot \mathrm{CS}_{2,1}^\dagger\cdot\widehat{Sq}_1\left(\frac{1}{\tan\theta}\right)
  \end{aligned}
\end{equation}
up to irrelevant global phase factor, as is confirmed in \ref{ssec:bs_decom}. 
Thus, the beamsplitter gate can be represented by the following diagram.

\begin{equation}
\label{eq:BStheta}
\begin{ZX}
\zxN{}\ar[r,marrow=<]&[\zxwCol,2mm]\zxBox{Sq\left(\frac{1}{\tan\theta}\right)}\ar[drr,-N,inarrow=<]&[\zxWCol,1mm]&[\zxwCol]&[\zxwCol,1mm]&[\zxwCol,1mm]&[\zxwCol]&[\zxwCol,1mm]&[\zxwCol,1mm]\\[\zxZeroRow,-2mm]
&&&\zxZ{} \ar[r,marrow=<]&\zxBox{Sq\left(\frac{\sin^2\theta}{\cos\theta}\right)}\ar[drr,-N,inarrow=<]&\zxN{}&\\[\zxZeroRow,-2mm]
&&&&&&\zxX{} \ar[r,marrow=<]&\zxBox{Sq\left(\frac{1}{\tan\theta}\right)}\ar[r,marrow=<]&\zxN{}\\
\zxN{}\ar[rr,marrow=<]&&\zxX{}\ar[ruu,s.,marrow=<]\ar[drr,N-,outarrow=<]\\[\zxZeroRow]
&&&&\zxBox{Sq\left(\frac{1}{\cos\theta}\right)}\ar[r,marrow=<]&\zxZ{}\ar[ruu,s.,marrow=<]\ar[drrr,N-,inarrow=<]\\
&&&&&&&&\zxN{}
\end{ZX}
\end{equation}
When $\theta = \frac{\pi}{4}$, the beamsplitter is often called a \emph{balanced beamsplitter} or a \emph{half-splitter} because its 
reflectance and transmittance are equal.
In that case, its diagram can be represented more compactly as follows:
\begin{equation}
\begin{ZX}
\zxN{}\ar[drr,-N,outarrow=<]&[\zxWCol,2mm]&[\zxwCol]&[\zxwCol,1mm]&[\zxwCol,1mm]&[\zxwCol,1mm]&[\zxWCol,2mm]\\[\zxZeroRow]
&&\zxZ{} \ar[r,marrow=<]&\zxBox{Sq\left(\frac{1}{\sqrt{2}}\right)}\ar[drr,-N,inarrow=<]&\zxN{}&\\[\zxZeroRow]
&&&&&\zxX{} \ar[r,marrow=<]&\zxN{}\\
\zxN{}\ar[r,marrow=<]&\zxX{}\ar[ruu,s.,marrow=<]\ar[drr,N-,outarrow=<]\\[\zxZeroRow]
&&&\zxBox{Sq(\sqrt{2})}\ar[r,marrow=<]&\zxZ{}\ar[ruu,s.,marrow=<]\ar[drr,N-,inarrow=<]\\[\zxZeroRow]
&&&&&&\zxN{}
\end{ZX}
\end{equation}

\subsubsection{Cubic phase gate}

By definition, the cubic phase gate (CPG) is represented by a single $q$-spider with a cubic phase function:
\begin{equation}
    \begin{ZX}
        \zxN{}&[\zxwCol,1mm] \zxBox{CPG(\gamma)} \lar[inarrow=>] \rar[inarrow=<] & [\zxwCol,1mm] \zxN{}
    \end{ZX}
    \coloneqq
    \begin{ZX}
        \zxN{}&[\zxwCol,1mm] \zxZ{\gamma x^3} \lar[inarrow=>] \rar[inarrow=<] &[\zxwCol,1mm]\zxN{}
    \end{ZX}
\end{equation}
In our graphical model, the non-Gaussian feature of quantum process lies not so much in the complexity of the appearance of a diagram but rather in how they are transformed by rewrite rules. In the following arguments, we will also discuss the distinct property of Gaussian and non-Gaussian properties of diagrams.

\section{Properties of CV ZX~calculus}

In this section, first we list the rewrite rules that constitute the axioms of our CV graphical calculus. Each rule depicts the corresponding equations of linear maps over CV modes. Consequently, we
derive basic properties of our graphical model and various equations utilized in CV quantum information processing protocols based upon the rules.

\subsection{Basic rewrite rules} \label{ssec:CV_basic_rules}

Before describing individual rewrite rules, we shall define a conjugate diagram. 

\begin{dfn}[Conjugate diagram]
    Let $D$ be a proper diagram. The \emph{conjugate diagram of $D$}, denoted by $D^\dagger$, is the diagram obtained by inverting directions of all arrows and multiplying $-1$ to all phase functions in $D$ (this operation interchanges Fourier diagram and inverse Fourier diagram.)
\end{dfn}

Note that conjugation is compatible with parallelizing and combining diagrams---i.e.,~$(D_1\otimes D_2)^\dagger=D_1^\dagger \otimes D_2^\dagger$ and $(D_2\circ D_1)^\dagger=D_1^\dagger \circ D_2^\dagger$. This, together with the fact that $\llbracket G\rrbracket^\dagger =\llbracket G^\dagger\rrbracket$ for each generator $G$ in Table~\ref{tab:table_gen}, it is easy to see that $\llbracket D\rrbracket^\dagger =\llbracket D^\dagger\rrbracket$ holds for any proper diagram $D$.

In our calculus model, we employ a meta-rule that a rewrite rule is conserved under conjugation. More specifically, for each rewrite rule claiming $D_1 = D_2$, we accept $D_1^\dagger = D_2^\dagger$ as rewrite rule, as well. Thus, whenever two diagrams are shown to be equal via sequential application of rewrite rules, its conjugative counterpart also holds in the same manner. Also, as we are only interested in the quantum dynamics, and global scalars do not affect processes for that purpose, so we will ignore closed diagrams that appear in the derivations.The proof of soundness for each rewrite rule is given in Appendix~\ref{sec:rewrite_sound}.

\subsubsection{Identity rule}

The identity rule~$(\hyperlink{li:id}{id})$ states
\hypertarget{li:id}{}\begin{gather}
\begin{ZX}
   \zxN{}\ar[r,marrow=<]&[\zxwCol,1mm]\zxZ{a}\ar[r,marrow=<]&[\zxwCol,1mm]\zxN{}
\end{ZX}=\begin{ZX}
   \zxN{}\ar[r,marrow=<]&[\zxwCol,1mm]\zxX{b}\ar[r,marrow=<]&[\zxwCol,1mm]\zxN{}
\end{ZX}\overset{(\hyperlink{li:id}{id})}{=}
\begin{ZX}
    \zxN{}\ar[r,marrow=<]&[\zxwCol,5mm]\zxN{}
    \end{ZX}
\end{gather}
where $a$ and $b$ denotes arbitrary constant-valued functions. Note that when the constants are zero, these diagrams are denoted by blank spiders.

\subsubsection{Fusion rule}

The fusion rule~$(\hyperlink{li:fusion}{f})$ states
\hypertarget{li:fusion}{}\begin{gather}
\begin{ZX}
   \leftArrowedManyDots{n}\zxZ{f(x)}\ar[dd,C=1,start anchor=south,end anchor=north]\ar[dd,C-=1,start anchor=south,end anchor=north]\rightArrowedManyDots{m}\\
   &\ldots\\
   \leftArrowedManyDots{n'}\zxZ{g(x)}\rightArrowedManyDots{m'}
\end{ZX}\overset{(\hyperlink{li:fusion}{f})}{=}\begin{ZX}
   \leftArrowedManyDots{n+n'}\zxZ{(f+g)(x)} \rightArrowedManyDots{m+m'}
\end{ZX}\\
   \begin{ZX}
   \leftArrowedManyDots{n}\zxX{f(x)}\ar[dd,C=1,start anchor=south,end anchor=north]\ar[dd,C-=1,start anchor=south,end anchor=north]\rightArrowedManyDots{m}\\
   &\ldots\\
   \leftArrowedManyDots{n'}\zxX{g(x)}\rightArrowedManyDots{m'}
\end{ZX}\overset{(\hyperlink{li:fusion}{f})}{=}\begin{ZX}
   \leftArrowedManyDots{n+n'}\zxX{(f+g)(x)} \rightArrowedManyDots{m+m'}
\end{ZX}
\end{gather}
where the undirected lines between the spiders represent an arbitrary number of wires (except zero), with arbitrary directions.

\subsubsection{Bialgebra rule}

The bialgebra rule~$(\hyperlink{li:bialgebra}{b})$ has two forms:
\hypertarget{li:bialgebra}{}\begin{equation}
\begin{ZX}
   \zxN{}\ar[dr,-N, outarrow=<]&[\zxwCol,1mm]&[\zxwCol,2mm]&[\zxwCol,1mm]\zxN{}\\[\zxWRow]
   &  \zxZ{}\ar[r,marrow=<]&\zxX{}\ar[dr,N-, inarrow=<]\ar[ur,N-, inarrow=<]\\[\zxWRow]
   \zxN{}\ar[ur,-N, outarrow=<]&&&\zxN{}
\end{ZX}
\overset{(\hyperlink{li:bialgebra}{b})}{=}
\begin{ZX}
   &[\zxwCol,2mm]&[\zxwCol,1mm]\zxN{}&[\zxwCol,1mm]&[\zxwCol,2mm]\\
   \zxN{}\ar[r,marrow=<]&\zxX{}\ar[drr,s.,numarrow={<}{0.25},numarrow={<}{0.75}]\ar[rr,o'={-=.2,L=.25},marrow=<]&&\zxZ{}\ar[r,marrow=<]&\zxN{}\\[\zxWRow,.7mm]
   \zxN{}\ar[r,marrow=<]&\zxX{}\ar[rr,o.={-=.2,L=.25},marrow=<]\ar[urr,s.,numarrow={<}{0.25},numarrow={<}{0.75}]&&\zxZ{}\ar[r,marrow=<]&\zxN{}
\end{ZX}
\end{equation}
Conjugating the above rule gives the following equation:
\begin{equation}
\begin{ZX}
   \zxN{}\ar[dr,-N, outarrow=<]&[\zxwCol,1mm]&[\zxwCol,2mm]&[\zxwCol,1mm]\zxN{}\\[\zxWRow]
   &  \zxX{}\ar[r,marrow=<]&\zxZ{}\ar[dr,N-, inarrow=<]\ar[ur,N-, inarrow=<]\\[\zxWRow]
   \zxN{}\ar[ur,-N, outarrow=<]&&&\zxN{}
\end{ZX}
\overset{(\hyperlink{li:bialgebra}{b})}{=}
\begin{ZX}
   &[\zxwCol,2mm]&[\zxwCol,1mm]\zxN{}&[\zxwCol,1mm]&[\zxwCol,2mm]\\
   \zxN{}\ar[r,marrow=<]&\zxZ{}\ar[drr,s.,numarrow={<}{0.25},numarrow={<}{0.75}]\ar[rr,o'={-=.2,L=.25},marrow=<]&&\zxX{}\ar[r,marrow=<]&\zxN{}\\[\zxWRow,.7mm]
   \zxN{}\ar[r,marrow=<]&\zxZ{}\ar[rr,o.={-=.2,L=.25},marrow=<]\ar[urr,s.,numarrow={<}{0.25},numarrow={<}{0.75}]&&\zxX{}\ar[r,marrow=<]&\zxN{}
\end{ZX}
\end{equation}
Much of the power of the graphical calculus is contained in this rule. (Recall the importance of the bialgebra property in quantum mechanics discussed in Sec.~\ref{subsec:catQM}.)

\subsubsection{Fourier rule}

The Fourier rule~$(\hyperlink{li:fourier}{F})$ is
\hypertarget{li:fourier}{}\begin{equation}\label{rule:fo}
    \begin{ZX}\\[\zxHRow]
\zxN{}\rar[inarrow=<]&[\zxWCol]\zxF{}\rar[inarrow=<]&[\zxWCol]\zxN{}\\[\zxHRow]
\end{ZX}\overset{(\hyperlink{li:fourier}{F})}{=}\begin{ZX}
    \zxN{}&[\zxWCol] \zxBox{R\left(-\frac{\pi}{2}\right)} \lar[inarrow=>] \rar[inarrow=<] & [\zxWCol]
\end{ZX}
\end{equation}
Conjugation gives
\begin{equation}\label{rule:fo_dag}
    \begin{ZX}\\[\zxHRow]
\zxN{}\rar[inarrow=<] &[\zxWCol]\zxFdag{}\rar[inarrow=<]&[\zxwCol]\zxN{}\\[\zxHRow]
\end{ZX}\overset{(\hyperlink{li:fourier}{F})}{=}\begin{ZX}
    \zxN{}&[\zxWCol] \zxBox{R\left(\frac{\pi}{2}\right)} \lar[inarrow=>] \rar[inarrow=<] & [\zxWCol]
\end{ZX}
\end{equation}
and composing this with the original gives
\begin{equation}
    \begin{ZX}\\[\zxHRow]
\zxN{}\rar[inarrow=<] &[\zxWCol]\zxFsq{}\rar[inarrow=<]&[\zxwCol]\zxN{}\\[\zxHRow]
\end{ZX}\overset{(\hyperlink{li:fourier}{F})}{=}\begin{ZX}
    \zxN{}&[\zxWCol] \zxBox{R(\pi)} \lar[inarrow=>] \rar[inarrow=<] & [\zxWCol]
\end{ZX}=\begin{ZX}
    \zxN{}&[\zxWCol] \zxBox{Sq(-1)} \lar[inarrow=>] \rar[inarrow=<] & [\zxWCol]
\end{ZX}
\end{equation}

\subsubsection{Copy rule}

The copy rule~$(\hyperlink{li:copy}{c})$ has two forms:
\hypertarget{li:copy}{}\begin{equation}
    \begin{ZX}
   \zxN{}\ar[dr,-N, outarrow=<]&[\zxwCol,1mm]&[\zxwCol,1mm]\\
   \makebox[0pt][l]{\scalebox{0.8}{$\cvdotsCenterMathline$}}& \zxZ{f(x)}\ar[r,marrow=<]&\zxX{}\\
   \zxN{}\ar[ur,-N, outarrow=<]
\end{ZX}\overset{(\hyperlink{li:copy}{c})}{=}\begin{ZX}
   \zxN{}\ar[r,marrow=<]&[\zxwCol,1mm]\zxX{}\\
   &\makebox[0pt][l]{\scalebox{0.8}{$\cvdotsCenterMathline$}}&\\
   \zxN{}\ar[r,marrow=<]&\zxX{}
\end{ZX}
\end{equation}
and
\begin{equation}
    \begin{ZX}
   \zxN{}\ar[dr,-N, outarrow=<]&[\zxwCol,1mm]&[\zxwCol,1mm]\\
   \makebox[0pt][l]{\scalebox{0.8}{$\cvdotsCenterMathline$}}& \zxX{f(x)}\ar[r,marrow=<]&\zxZ{}\\
   \zxN{}\ar[ur,-N, outarrow=<]
\end{ZX}\overset{(\hyperlink{li:copy}{c})}{=}\begin{ZX}
   \zxN{}\ar[r,marrow=<]&[\zxwCol,1mm]\zxZ{}\\
   &\makebox[0pt][l]{\scalebox{0.8}{$\cvdotsCenterMathline$}}&\\
   \zxN{}\ar[r,marrow=<]&\zxZ{}
\end{ZX}
\end{equation}
where dotted lines denote arbitrary number of wires and diagrams (and so on in the following equations).

\subsubsection{Displacement rule}


The displacement rule~$(\hyperlink{li:disp}{d})$ also has two forms:
\hypertarget{li:disp}{}
\begin{equation}
    \begin{ZX}
   \zxN{}\ar[dr,-N, outarrow=<]&[\zxwCol,1mm]&[\zxwCol,1mm]\zxX{ax}\ar[r,marrow=<]&[\zxwCol,1mm]\zxN{}\\
   \makebox[0pt][l]{\scalebox{0.8}{$\cvdotsCenterMathline$}}&  \zxZ{f(x)}\ar[ur,N-,inarrow=<]\ar[dr,N-,inarrow=<]&\makebox[0pt][r]{\scalebox{0.8}{$\cvdotsCenterMathline$}}\\
   \zxN{}\ar[ur,-N, outarrow=<]&&\zxX{ax}\ar[r,marrow=<]&\zxN{}
\end{ZX}\overset{(\hyperlink{li:disp}{d})}{=}\begin{ZX}
   \zxN{}\ar[r,marrow=<]&[\zxwCol,1mm]\zxX{ax}\ar[dr,-N, marrow=<]&[\zxwCol,1mm]&[\zxwCol,2mm]\zxN{}\\
   &\makebox[0pt][l]{\scalebox{0.8}{$\cvdotsCenterMathline$}}&  \zxZ{f(x+a)}\ar[ru,N-,marrow=<]\ar[rd,N-,marrow=<]&\makebox[0pt][r]{\scalebox{0.8}{$\cvdotsCenterMathline$}}\\
   \zxN{}\ar[r,marrow=<]&\zxX{ax}\ar[ur,-N, marrow=<]&&\zxN{}
\end{ZX}
\end{equation}
and
\begin{equation}
    \begin{ZX}
   \zxN{}\ar[dr,-N, outarrow=<]&[\zxwCol,1mm]&[\zxwCol,1mm]\zxZ{ax}\ar[r,marrow=<]&[\zxwCol,1mm]\zxN{}\\
   \makebox[0pt][l]{\scalebox{0.8}{$\cvdotsCenterMathline$}}&  \zxX{f(x)}\ar[ur,N-,inarrow=<]\ar[dr,N-,inarrow=<]&\makebox[0pt][r]{\scalebox{0.8}{$\cvdotsCenterMathline$}}\\
   \zxN{}\ar[ur,-N, outarrow=<]&&\zxZ{ax}\ar[r,marrow=<]&\zxN{}
\end{ZX}\overset{(\hyperlink{li:disp}{d})}{=}\begin{ZX}
   \zxN{}\ar[r,marrow=<]&[\zxwCol,1mm]\zxZ{ax}\ar[dr,-N, marrow=<]&[\zxwCol,1mm]&[\zxwCol,2mm]\zxN{}\\
   &\makebox[0pt][l]{\scalebox{0.8}{$\cvdotsCenterMathline$}}&  \zxX{f(x+a)}\ar[ru,N-,marrow=<]\ar[rd,N-,marrow=<]&\makebox[0pt][r]{\scalebox{0.8}{$\cvdotsCenterMathline$}}\\
   \zxN{}\ar[r,marrow=<]&\zxZ{ax}\ar[ur,-N, marrow=<]&&\zxN{}
\end{ZX}
\end{equation}
In the case that the diagram has no inputs (or no outputs), the linear spider disappears after acting on the central spider function.

\subsubsection{Antipode rule}\label{ssec:anti}

The antipode rule~$(\hyperlink{li:anti}{a})$ states
\hypertarget{li:anti}{}\begin{equation}
\begin{ZX}
    \zxN{}&[\zxWCol] \zxFsq{} \lar[inarrow=>] \rar[inarrow=<] & [\zxWCol] \zxN{}
\end{ZX}
\overset{(\hyperlink{li:anti}{a})}{=}
\begin{ZX}
   \zxN{}&[\zxwCol]&[\zxwCol,2mm]&[\zxwCol]\\
   && \zxZ{} \ar[ull,N-, inarrow=>] \ar[dl,s.={L=.4}, marrow=>]\\
   &\zxX{}\ar[drr,N-, inarrow=<]&\\
   &&&\zxN{}
\end{ZX}
\end{equation}
The left-side diagram of the eqaution is called \emph{antipode} for its algebraic action to the phase space. We will investigate in its useful properties in the following sections.

%
%
%

\subsubsection{Squeezing rule}

The squeezing rule~$(\hyperlink{li:sq}{s})$ also has two forms:
\hypertarget{li:sq}{}\begin{equation}
    \begin{ZX}
   \zxN{}\ar[dr,-N, outarrow=<]&[\zxwCol,1mm]&[\zxwCol,1mm]\zxBox{Sq\left(\tau\right)}\ar[r,marrow=<]&[\zxwCol,1mm]\zxN{}\\
   \makebox[0pt][l]{\scalebox{0.8}{$\cvdotsCenterMathline$}}&  \zxZ{f(x)}\ar[ur,N-,numarrow={<}{0.4}]\ar[dr,N-,numarrow={<}{0.4}]&\makebox[0pt][r]{\scalebox{0.8}{$\cvdotsCenterMathline$}}\\
   \zxN{}\ar[ur,-N, outarrow=<]&&\zxBox{Sq\left(\tau\right)}\ar[r,marrow=<]&\zxN{}
\end{ZX}\overset{(\hyperlink{li:sq}{s})}{=}\begin{ZX}
   \zxN{}\ar[r,marrow=<]&[\zxwCol,1mm]\zxBox{Sq(\tau)}\ar[dr,-N, numarrow={<}{0.6}]&[\zxwCol,1mm]&[\zxwCol,2mm]\zxN{}\\
   &\makebox[0pt][l]{\scalebox{0.8}{$\cvdotsCenterMathline$}}&  \zxZ{f(\tau x)}\ar[ru,N-,marrow=<]\ar[rd,N-,marrow=<]&\makebox[0pt][r]{\scalebox{0.8}{$\cvdotsCenterMathline$}}\\
   \zxN{}\ar[r,marrow=<]&\zxBox{Sq(\tau)}\ar[ur,-N, numarrow={<}{0.6}]&&\zxN{}
\end{ZX}
\end{equation}
and
\begin{equation}
    \begin{ZX}
   \zxN{}\ar[dr,-N, outarrow=<]&[\zxwCol,1mm]&[\zxwCol,1mm]\zxBox{Sq(\tau)}\ar[r,marrow=<]&[\zxwCol,1mm]\zxN{}\\
   \makebox[0pt][l]{\scalebox{0.8}{$\cvdotsCenterMathline$}}&  \zxX{f(x)}\ar[ur,N-,numarrow={<}{0.4}]\ar[dr,N-,numarrow={<}{0.4}]&\makebox[0pt][r]{\scalebox{0.8}{$\cvdotsCenterMathline$}}\\
   \zxN{}\ar[ur,-N, outarrow=<]&&\zxBox{Sq(\tau)}\ar[r,marrow=<]&\zxN{}
\end{ZX}\overset{(\hyperlink{li:sq}{s})}{=}\begin{ZX}
   \zxN{}\ar[r,marrow=<]&[\zxwCol,1mm]\zxBox{Sq\left(\tau\right)}\ar[dr,-N, numarrow={<}{0.6}]&[\zxwCol,1mm]&[\zxwCol,2mm]\zxN{}\\
   &\makebox[0pt][l]{\scalebox{0.8}{$\cvdotsCenterMathline$}}&  \zxX{f\left(\frac{x}{\tau}\right)}\ar[ru,N-,marrow=<]\ar[rd,N-,marrow=<]&\makebox[0pt][r]{\scalebox{0.8}{$\cvdotsCenterMathline$}}\\
   \zxN{}\ar[r,marrow=<]&\zxBox{Sq(\tau)}\ar[ur,-N, numarrow={<}{0.6}]&&\zxN{}
\end{ZX}
\end{equation}
In the case that the diagram has no inputs (or no outputs), the squeezing gate disappears after acting on the spider function.

\subsubsection{Quadratic rule}

The quadratic rule~$(\hyperlink{li:quad}{q})$ is
\hypertarget{li:quad}{}
\begin{align}
&\begin{ZX}
        \zxN{}&[\zxWCol] \zxZ{ax^2} \lar[inarrow=>] \rar[inarrow=<] &[\zxWCol] \zxX{bx^2} \rar[marrow=<]&[\zxWCol] \zxZ{cx^2} \rar[marrow=<]&[\zxWCol]\zxN{}
    \end{ZX}\\
\overset{(\hyperlink{li:quad}{q})}{=}&
\begin{ZX}
        \zxN{}&[\zxWCol] \zxX{\frac{bc}{4abc+a+c}x^2} \lar[inarrow=>] \rar[inarrow=<] &[\zxWCol] \zxZ{(4abc+a+c)x^2} \rar[marrow=<]&[\zxWCol] \zxX{\frac{ab}{4abc+a+c}x^2} \rar[marrow=<]&[\zxWCol]\zxN{}
    \end{ZX}
\end{align}
which can also be written
\begin{align}
    &
\begin{ZX}
        \zxN{}&[\zxWCol] \zxX{ax^2} \lar[inarrow=>] \rar[inarrow=<] &[\zxWCol] \zxZ{bx^2} \rar[marrow=<]&[\zxWCol] \zxX{cx^2} \rar[marrow=<]&[\zxWCol]\zxN{}
    \end{ZX}\\
\overset{(\hyperlink{li:quad}{q})}{=}&
\begin{ZX}
        \zxN{}&[\zxWCol] \zxZ{\frac{bc}{4abc+a+c}x^2} \lar[inarrow=>] \rar[inarrow=<] &[\zxWCol] \zxX{(4abc+a+c)x^2} \rar[marrow=<]&[\zxWCol] \zxZ{\frac{ab}{4abc+a+c}x^2} \rar[marrow=<]&[\zxWCol]\zxN{}
    \end{ZX}
\end{align}
Note that $4abc+a+c\neq0$ must be satisfied whenever applying the quadratic rule.

\subsubsection{Inversion rule}

The inversion rule~$(\hyperlink{li:inv}{inv})$ can be used to change spider types:
\hypertarget{li:inv}{}\begin{equation}
    \begin{ZX}
   \zxN{}\ar[r,marrow=<]&[\zxwCol,1mm]\zxF{}\ar[dr,-N, marrow=<]&[\zxwCol,1mm]&[\zxwCol,1mm]\zxFdag{}\ar[r,marrow=<]&[\zxwCol,1mm]\zxN{}\\
   &\makebox[0pt][l]{\scalebox{0.8}{$\cvdotsCenterMathline$}}&  \zxZ{f(x)}\ar[ur,N-,marrow=<]\ar[dr,N-,,marrow=<]&\makebox[0pt][r]{\scalebox{0.8}{$\cvdotsCenterMathline$}}\\
  \zxN{}\ar[r,marrow=<]&\zxF{}\ar[ur,-N, marrow=<]&&\zxFdag{}\ar[r,marrow=<]&\zxN{}
\end{ZX}\overset{(\hyperlink{li:inv}{inv})}{=}\begin{ZX}
   \zxN{}\ar[dr,-N, marrow=<]&[\zxwCol,1mm]&[\zxwCol,2mm]\zxN{}\\[\zxWRow]
   \makebox[0pt][l]{\scalebox{0.8}{$\cvdotsCenterMathline$}}&  \zxX{f(x)}\ar[ru,N-,marrow=<]\ar[rd,N-,marrow=<]&\makebox[0pt][r]{\scalebox{0.8}{$\cvdotsCenterMathline$}}\\[\zxWRow]
   \zxN{}\ar[ur,-N, marrow=<]&&\zxN{}
\end{ZX}
\end{equation}
Sandwiching the above equation between $\hat{F}^{\dagger}$ and $\hat{F}$, the following equivalent form can also be derived:
\begin{equation}
    \begin{ZX}
   \zxN{}\ar[r,marrow=<]&[\zxwCol,1mm]\zxFdag{}\ar[dr,-N, marrow=<]&[\zxwCol,1mm]&[\zxwCol,1mm]\zxF{}\ar[r,marrow=<]&[\zxwCol,1mm]\zxN{}\\
   &\makebox[0pt][l]{\scalebox{0.8}{$\cvdotsCenterMathline$}}&  \zxX{f(x)}\ar[ur,N-,marrow=<]\ar[dr,N-,marrow=<]&\makebox[0pt][r]{\scalebox{0.8}{$\cvdotsCenterMathline$}}\\
  \zxN{}\ar[r,marrow=<]&\zxFdag{}\ar[ur,-N, marrow=<]&&\zxF{}\ar[r,marrow=<]&\zxN{}
\end{ZX}\overset{(\hyperlink{li:inv}{inv})}{=}\begin{ZX}
   \zxN{}\ar[dr,-N, marrow=<]&[\zxwCol,1mm]&[\zxwCol,2mm]\zxN{}\\[\zxWRow]
   \makebox[0pt][l]{\scalebox{0.8}{$\cvdotsCenterMathline$}}&  \zxZ{f(x)}\ar[ru,N-,inarrow=<]\ar[rd,N-,inarrow=<]&\makebox[0pt][r]{\scalebox{0.8}{$\cvdotsCenterMathline$}}\\[\zxWRow]
   \zxN{}\ar[ur,-N, marrow=<]&&\zxN{}
\end{ZX}
\end{equation}


\subsection{Derived rewrite rules}

The following rewrite rules can be derived using the basic ones.

\subsubsection{Commutative and associative property of displacement gate}
Direct calculation using the Baker-Campbell-Hausdorff formula shows that the displacement gates are commutative and hence associative---i.e. $\hat{D}(\alpha_1)\hat{D}(\alpha_2)\sim\hat{D}(\alpha_2)\hat{D}(\alpha_1)\sim\hat{D}(\alpha_1+\alpha_2)$ up to an irrelevant global phase. This fact can be confirmed using our graphical calculus. First, using displacement rule, one obtains the following equation:
\begin{subequations}
\begin{align}
    \begin{ZX}
       \zxN{}\ar[r,marrow=<] &[\zxwCol,1mm] \zxZ{ax} \ar[r,marrow=<] 
       &[\zxwCol,1mm]\zxX{bx} \ar[r,marrow=<]&[\zxwCol,1mm]\zxN{}
    \end{ZX}
\overset{(\hyperlink{li:disp}{d})}{=}&
    \begin{ZX}
       \zxN{}\ar[r,marrow=<] &[\zxwCol,1mm]\zxX{bx} \ar[r,marrow=<] 
       &[\zxwCol,1mm]\zxZ{a(x+b)} \ar[r,marrow=<]&[\zxwCol,1mm]\zxN{}
    \end{ZX}\\
\overset{(\hyperlink{li:fusion}{f})}{=}&
    \begin{ZX}
       \zxN{}\ar[r,marrow=<] &[\zxwCol,1mm]\zxX{bx} \ar[r,marrow=<] 
       &[\zxwCol,1mm]\zxZ{ax} \ar[r,marrow=<]&[\zxwCol,1mm]\zxZ{ab}\ar[r,marrow=<]&[\zxwCol,1mm]\zxN{}
    \end{ZX}\\
\overset{(\hyperlink{li:id}{id})}{=}&
    \begin{ZX}
       \zxN{}\ar[r,marrow=<] &[\zxwCol,1mm]\zxX{bx} \ar[r,marrow=<] 
       &[\zxwCol,1mm]\zxZ{ax} \ar[r,marrow=<]&[\zxwCol,1mm]\zxN{}
    \end{ZX}
\end{align}\label{eq:d_com}
\end{subequations}
By using this commutation rule, the associative property can be confirmed as is shown below.
\begin{align}
    &\scalebox{0.9}{\begin{ZX}
        \zxN{}&[\zxWCol] \zxBox{D(\alpha_1)} \lar[inarrow=>] \rar[marrow=<] &[\zxWCol] \zxBox{D(\alpha_2)}  \rar[inarrow=<] & [\zxWCol] \zxN{}
    \end{ZX}}\\
    \overset{\eqref{eq:Dgate}}{=}&\scalebox{0.9}{\begin{ZX}
        \zxN{}&[\zxWCol] \zxX{\sqrt{2}~\Re(\alpha_1)x} \lar[inarrow=>] \rar[marrow=<] &[\zxWCol] \zxZ{\sqrt{2}~\Im(\alpha_1)x}\rar[marrow=<]&[\zxWCol] \zxX{\sqrt{2}~\Re(\alpha_2)x} \rar[marrow=<] &[\zxWCol] \zxZ{\sqrt{2}~\Im(\alpha_2)x}\rar[inarrow=<]&[\zxWCol]\zxN{}
    \end{ZX}}\\
    \overset{\eqref{eq:d_com}}{=}&\scalebox{0.9}{\begin{ZX}
        \zxN{}&[\zxWCol] \zxX{\sqrt{2}~\Re(\alpha_1)x} \lar[inarrow=>] \rar[marrow=<]&[\zxWCol] \zxX{\sqrt{2}~\Re(\alpha_2)x} \rar[marrow=<] &[\zxWCol] \zxZ{\sqrt{2}~\Im(\alpha_1)x}\rar[marrow=<]&[\zxWCol] \zxZ{\sqrt{2}~\Im(\alpha_2)x}\rar[inarrow=<]&[\zxWCol]\zxN{}
    \end{ZX}}\\
    \overset{(\hyperlink{li:fusion}{f})}{=}&\scalebox{0.9}{\begin{ZX}
        \zxN{}&[\zxWCol] \zxX{\sqrt{2}~\Re(\alpha_1+\alpha_2)x} \lar[inarrow=>] \rar[marrow=<]&[\zxWCol] \zxZ{\sqrt{2}~\Im(\alpha_1\alpha_2)x}\rar[inarrow=<]&[\zxWCol]\zxN{}
    \end{ZX}}\\
    \overset{\eqref{eq:Dgate}}{=}&\scalebox{0.9}{\begin{ZX}
        \zxN{}&[\zxWCol] \zxBox{D(\alpha_1+\alpha_2)} \lar[inarrow=>] \rar[inarrow=<]&[\zxWCol]\zxN{}
    \end{ZX}}
\end{align}

\subsubsection{Self-conjugateness of antipode}
As is mentioned in subsection~\ref{ssec:anti}, the antipode diagram represents the square of Fourier transform, which is self-conjugate---i.e. $(\hat{F}^2)^\dagger = \hat{F}^2$. The same property holds for the diagram as well:
\begin{equation}
    \begin{ZX}
       \zxN{}&[\zxwCol]&[\zxwCol,2mm]&[\zxwCol]\\
       && \zxZ{} \ar[ull,N-, inarrow=>] \ar[dl,s.={L=.4}, marrow=>]\\
       &\zxX{}\ar[drr,N-, inarrow=<]&\\
       &&&\zxN{}
    \end{ZX}
    =
    \left(\begin{ZX}
       \zxN{}&[\zxwCol]&[\zxwCol,2mm]&[\zxwCol]\\
       && \zxZ{} \ar[ull,N-, inarrow=>] \ar[dl,s.={L=.4}, marrow=>]\\
       &\zxX{}\ar[drr,N-, inarrow=<]&\\
       &&&\zxN{}
    \end{ZX}\right)^{\dagger}=
    \begin{ZX}
       \zxN{}&[\zxwCol]&[\zxwCol,2mm]&[\zxwCol]\\
       && \zxX{} \ar[ull,N-, inarrow=>] \ar[dl,s.={L=.4}, marrow=>]\\
       &\zxZ{}\ar[drr,N-, inarrow=<]&\\
       &&&\zxN{}
    \end{ZX}\label{eq:anti_self_conj}
    \end{equation}
This equation can be derived in the following way. First, we have
\begin{align}
    \begin{ZX}
                \zxN{}&[\zxWCol,5mm] \zxN{} \lar[marrow=>]
            \end{ZX}
    \overset{\eqref{eq:anti_self}}{=}&
    \begin{ZX}
        \zxN{}&[\zxWCol] \zxBox{Sq(-1)} \lar[inarrow=>] \rar[marrow=<] &[\zxWCol] \zxBox{Sq(-1)}\rar[inarrow=<]&[\zxWCol]\zxN{}
    \end{ZX}
    \\
    \overset{(\hyperlink{li:fourier}{F})}{=}&
    \begin{ZX}
        \zxN{}&[\zxWCol] \zxFsq{} \lar[inarrow=>] \rar[marrow=<] &[\zxWCol] \zxFsq{}\rar[inarrow=<]&[\zxWCol]\zxN{}
    \end{ZX}\\
    \\
    \overset{(\hyperlink{li:anti}{a})}{=}&
    \begin{ZX}
       \zxN{}&[\zxwCol]&[\zxwCol,2mm]&[\zxwCol]\\
       && \zxZ{} \ar[ull,N-, inarrow=>] \ar[dl,s.={L=.4}, marrow=>]\\
       &\zxX{}\ar[dr,s.={L=.4}, marrow=<]\\
       &&\zxZ{}\ar[dl,s.={L=.4}, marrow=>]&\\
       &\zxX{}\ar[drr,N-, inarrow=<]\\
       &&&\zxN{}
    \end{ZX}\label{eq:id_anti}
    \end{align}
in which we used multiplicativity of squeezing gate to be shown in Appendix~\ref{ssec:sq_assoc}. Consequently,
    \begin{align}
    \begin{ZX}
       \zxN{}&[\zxwCol]&[\zxwCol,2mm]&[\zxwCol]\\
       && \zxX{} \ar[ull,N-, inarrow=>] \ar[dl,s.={L=.4}, marrow=>]\\
       &\zxZ{}\ar[drr,N-, inarrow=<]&\\
       &&&\zxN{}
    \end{ZX}
    \overset{\eqref{eq:id_anti}}{=}&\begin{ZX}
       \zxN{}&[\zxwCol]&[\zxwCol,2mm]&[\zxwCol]\\
       && \zxX{} \ar[ull,N-, inarrow=>] \ar[dl,s.={L=.4}, marrow=>]\\
       &\zxZ{}\ar[dr,s.={L=.4}, marrow=<]\\
       && \zxZ{}\ar[dl,s.={L=.4}, marrow=>]\\
       &\zxX{}\ar[dr,s.={L=.4}, marrow=<]\\
       &&\zxZ{}\ar[dl,s.={L=.4}, marrow=>]&\\
       &\zxX{}\ar[drr,N-, inarrow=<]\\
       &&&\zxN{}
    \end{ZX}\\
    \overset{\mathclap{(\hyperlink{li:fusion}{f},\hyperlink{li:id}{id})}}{=}&\begin{ZX}
       \zxN{}&[\zxwCol]&[\zxwCol,2mm]&[\zxwCol]\\
       && \zxX{} \ar[ull,N-, inarrow=>] \ar[dl,s.={L=.4}, marrow=>]\\
       &\zxX{}\ar[dr,s.={L=.4}, marrow=<]\\
       &&\zxZ{}\ar[dl,s.={L=.4}, marrow=>]&\\
       &\zxX{}\ar[drr,N-, inarrow=<]\\
       &&&\zxN{}
    \end{ZX}\\
    \overset{\mathclap{(\hyperlink{li:fusion}{f},\hyperlink{li:id}{id})}}{=}&\begin{ZX}
       \zxN{}&[\zxwCol]&[\zxwCol,2mm]&[\zxwCol]\\
       && \zxZ{} \ar[ull,N-, inarrow=>] \ar[dl,s.={L=.4}, marrow=>]\\
       &\zxX{}\ar[drr,N-, inarrow=<]&\\
       &&&\zxN{}
    \end{ZX}
    \end{align}
holds.

\subsubsection{Antipode reversal}

The antipode reversal rule~$(\hyperlink{li:rev}{rev})$ is written in two forms:
\hypertarget{li:rev}{}\begin{align}
    \begin{ZX}
   \leftArrowedManyDots{}\zxZ{f(x)}\rightArrowedManyDots{}\\
   &\zxFsq{} \uar[inarrow=>] \dar[inarrow=<]\\
   \leftArrowedManyDots{}\zxX{g(x)}\rightArrowedManyDots{}
\end{ZX}
\overset{(\hyperlink{li:rev}{rev})}{=}
\begin{ZX}
   \leftArrowedManyDots{}\zxZ{f(x)}\dar[inarrow=>] \rightArrowedManyDots{}\\[\zxWRow]
   \leftArrowedManyDots{}\zxX{g(x)}\rightArrowedManyDots{}
\end{ZX}
\end{align}
\hypertarget{li:rev}{}\begin{align}
    \begin{ZX}
   \leftArrowedManyDots{}\zxX{f(x)}\rightArrowedManyDots{}\\
   &\zxFsq{} \uar[inarrow=>] \dar[inarrow=<]\\
   \leftArrowedManyDots{}\zxZ{g(x)}\rightArrowedManyDots{}
\end{ZX}
\overset{(\hyperlink{li:rev}{rev})}{=}
\begin{ZX}
   \leftArrowedManyDots{}\zxX{f(x)}\dar[inarrow=>] \rightArrowedManyDots{}\\[\zxWRow]
   \leftArrowedManyDots{}\zxZ{g(x)}\rightArrowedManyDots{}
\end{ZX}
\end{align}
The antipode reversal rule claims that the squared Fourier diagram between two spiders can 
be deleted by reversing its input and output. The first equation can be derived as follows:
\begin{equation}
\begin{ZX}
   \leftArrowedManyDots{}\zxZ{f(x)}\rightArrowedManyDots{}\\
   &\zxFsq{} \uar[inarrow=>] \dar[inarrow=<]\\
   \leftArrowedManyDots{}\zxX{g(x)}\rightArrowedManyDots{}
\end{ZX}\overset{(\hyperlink{li:anti}{a})}{=}
\begin{ZX}
   \leftArrowedManyDots{}\zxZ{f(x)}\rightArrowedManyDots{}\\
   && \zxZ{} \ar[ul,-N, inarrow=>] \ar[dll,s.={L=.4}, marrow=>]\\
   \zxX{}\ar[dr,-N, inarrow=<]\\
   \leftArrowedManyDots{}\zxX{g(x)}\rightArrowedManyDots{}
\end{ZX}
\overset{(\hyperlink{li:fusion}{f})}{=}
\begin{ZX}
   \leftArrowedManyDots{}\zxZ{f(x)}\dar[inarrow=>] \rightArrowedManyDots{}\\[\zxWRow]
   \leftArrowedManyDots{}\zxX{g(x)}\rightArrowedManyDots{}
\end{ZX}
\end{equation}
The second equation can also be derived in the same way by using Eq.~\eqref{eq:anti_self_conj}.

\subsubsection{Quadrature copy rule}
By combining the copy rule~$(\hyperlink{li:copy}{c})$ with the displacement rule~$(\hyperlink{li:disp}{d})$, one obtains a generalized copy rule represented in two forms:
\hypertarget{li:q_copy}{}\begin{equation}
    \begin{ZX}
   \zxN{}\ar[dr,-N, outarrow=<]&[\zxwCol,1mm]&[\zxwCol,1mm]\\
   \makebox[0pt][l]{\scalebox{0.8}{$\cvdotsCenterMathline$}}& \zxZ{f(x)}\ar[r,marrow=<]&\zxX{ax}\\
   \zxN{}\ar[ur,-N, outarrow=<]
\end{ZX}\overset{(\hyperlink{li:q_copy}{qc})}{=}\begin{ZX}
   \zxN{}\ar[r,marrow=<]&[\zxwCol,1mm]\zxX{ax}\\
   &\makebox[0pt][l]{\scalebox{0.8}{$\cvdotsCenterMathline$}}&\\
   \zxN{}\ar[r,marrow=<]&\zxX{ax}
\end{ZX}
\end{equation}
and
\begin{equation}
    \begin{ZX}
   \zxN{}\ar[dr,-N, outarrow=<]&[\zxwCol,1mm]&[\zxwCol,1mm]\\
   \makebox[0pt][l]{\scalebox{0.8}{$\cvdotsCenterMathline$}}& \zxX{f(x)}\ar[r,marrow=<]&\zxZ{ax}\\
   \zxN{}\ar[ur,-N, outarrow=<]
\end{ZX}\overset{(\hyperlink{li:q_copy}{qc})}{=}\begin{ZX}
   \zxN{}\ar[r,marrow=<]&[\zxwCol,1mm]\zxZ{ax}\\
   &\makebox[0pt][l]{\scalebox{0.8}{$\cvdotsCenterMathline$}}&\\
   \zxN{}\ar[r,marrow=<]&\zxZ{ax}
\end{ZX}
\end{equation}
The first equation can be derived in the following way:
\begin{align}
    \begin{ZX}
   \zxN{}\ar[dr,-N, outarrow=<]&[\zxwCol,1mm]&[\zxwCol,1mm]\\
   \makebox[0pt][l]{\scalebox{0.8}{$\cvdotsCenterMathline$}}& \zxZ{f(x)}\ar[r,marrow=<]&\zxX{ax}\\
   \zxN{}\ar[ur,-N, outarrow=<]
\end{ZX}\overset{(\hyperlink{li:fusion}{f})}{=}&
\begin{ZX}
   \zxN{}\ar[dr,-N, outarrow=<]&[\zxwCol,1mm]&[\zxwCol,1mm]&[\zxwCol,1mm]\\
   \makebox[0pt][l]{\scalebox{0.8}{$\cvdotsCenterMathline$}}& \zxZ{f(x)}\ar[r,marrow=<]&\zxX{ax}\ar[r,marrow=<]&\zxX{}\\
   \zxN{}\ar[ur,-N, outarrow=<]
\end{ZX}\\
\overset{(\hyperlink{li:disp}{d})}{=}&
\begin{ZX}
   \zxN{}\ar[r,marrow=<]&[\zxwCol,1mm]\zxX{ax}\ar[dr,-N, outarrow=<]&[\zxwCol,1mm]&[\zxwCol,1mm]\\
   &\makebox[0pt][l]{\scalebox{0.8}{$\cvdotsCenterMathline$}}& \zxZ{f(x+a)}\ar[r,marrow=<]&\zxX{}\\
   \zxN{}\ar[r,marrow=<]&\zxX{ax}\ar[ur,-N, outarrow=<]
\end{ZX}\\
\overset{(\hyperlink{li:copy}{c})}{=}&
\begin{ZX}
   \zxN{}\ar[r,marrow=<]&[\zxwCol,1mm]\zxX{ax}\ar[r,marrow=<]&[\zxwCol,1mm]\zxX{}\\
   &\makebox[0pt][l]{\scalebox{0.8}{$\cvdotsCenterMathline$}}\\
   \zxN{}\ar[r,marrow=<]&\zxX{ax}\ar[r,marrow=<]&\zxX{}
\end{ZX}\\
\overset{(\hyperlink{li:fusion}{f})}{=}&
\begin{ZX}
   \zxN{}\ar[r,marrow=<]&[\zxwCol,1mm]\zxX{ax}\\
   &\makebox[0pt][l]{\scalebox{0.8}{$\cvdotsCenterMathline$}}\\
   \zxN{}\ar[r,marrow=<]&\zxX{ax}
\end{ZX}
\end{align}
and the same applies to the second one. The orientation of wires can also be reversed with an additional factor of $-1$. For example, 
\begin{align}
    \begin{ZX}
&[\zxwCol,1mm]&[\zxwCol,1mm]\zxX{ax}\\
   \zxN{}\ar[r,marrow=<]&\zxZ{f(x)}\ar[ur,N-, inarrow=<]\ar[dr,N-, inarrow=<]\\
   &&\zxN{}
\end{ZX}\overset{(\hyperlink{li:fusion}{f})}{=}&\begin{ZX}
   \zxN{}\ar[dr,-N, outarrow=<]&[\zxwCol,1mm]&[\zxwCol,1mm]\\
   & \zxZ{f(x)}\ar[r,marrow=<]&\zxX{ax}\\
   \zxZ{}\ar[ur,s, outarrow=<]\ar[drr,N-, marrow=<]\\
   &&\zxN{}\end{ZX}\\
\overset{(\hyperlink{li:q_copy}{qc})}{=}&\begin{ZX}
   \zxN{}\ar[r,marrow=<]&[\zxwCol,1mm]\zxX{ax}&[\zxwCol,1mm]\\
   & \zxX{ax}\\
   \zxZ{}\ar[ur,s, outarrow=<]\ar[drr,N-, marrow=<]\\
   &&\zxN{}\end{ZX}\\
\overset{\eqref{eq:anti_self_conj}}{=}&\begin{ZX}
   \zxN{}\ar[r,marrow=<]&[\zxwCol,1mm]\zxX{ax}&[\zxwCol,1mm]\zxX{ax}\ar[r,marrow=<]&[\zxwCol,1mm]\zxBox{Sq(-1)}\ar[r,marrow=<]&[\zxwCol,1mm]\zxN{}\end{ZX}\\
\overset{(\hyperlink{li:sq}{s})}{=}&\begin{ZX}
   \zxN{}\ar[r,marrow=<]&[\zxwCol,1mm]\zxX{ax}&[\zxwCol,1mm]\zxX{-ax}\ar[r,marrow=<]&[\zxwCol,1mm]\zxN{}\end{ZX}
\end{align}
holds.

\subsubsection{Hopf rule}

The Hopf rule~$(\hyperlink{li:hopf}{h})$ is
\hypertarget{li:hopf}{}\begin{align}
\begin{ZX}
   \zxN{}\ar[r,marrow=<]&[\zxwCol,2mm]\zxX{}\ar[rrr,o'={-=.2,L=.25},marrow=<]\ar[rrr,o.={-=.2,L=.25},marrow=>]&[\zxwCol,1mm]\zxN{}&[\zxwCol,1mm]&[\zxwCol,1mm]\zxZ{}\ar[r,marrow=<]&[\zxwCol,2mm]\zxN{}
\end{ZX}
\overset{(\hyperlink{li:hopf}{h})}{=}
\begin{ZX}
   \zxN{}\ar[r,marrow=<]&[\zxwCol,2mm]\zxX{}&[\zxwCol,2mm]\zxZ{}\ar[r,marrow=<]&[\zxwCol,2mm]\zxN{}
\end{ZX}
\end{align}
which 
can be derived from basic rewrite rules as follows:
\begin{align}
&\begin{ZX}
   \zxN{}\ar[r,marrow=<]&[\zxwCol,2mm]\zxX{}\ar[rr,o'={-=.2,L=.25},marrow=<]\ar[rr,o.={-=.2,L=.25},marrow=>]&[\zxwCol,1mm]\zxN{}&[\zxwCol,1mm]\zxZ{}\ar[r,marrow=<]&[\zxwCol,2mm]\zxN{}
\end{ZX}\\
\overset{(\hyperlink{li:rev}{rev})}{=}&
\begin{ZX}
   \zxN{}\ar[r,marrow=<]&[\zxwCol,2mm]\zxX{}\ar[rr,o'={-=.2,L=.25},marrow=<]\ar[dr,N-,outarrow=<]&[\zxwCol,1mm]\zxN{}&[\zxwCol,1mm]\zxZ{}\ar[r,marrow=<]&[\zxwCol,2mm]\zxN{}\\[\zxZeroRow]
   &&\zxFsq{}\ar[ur, -N, inarrow=<]
\end{ZX}\\
\overset{(\hyperlink{li:anti}{a})}{=}&
\begin{ZX}
   \zxN{}\ar[r,marrow=<]&[\zxwCol,2mm]\zxX{}\ar[rr,marrow=<]\ar[rrr,o'={-=.2,L=.25},marrow=<]&&[\zxwCol,1mm]\zxZ{}\ar[dl,s.={L=.4}, marrow=>]&[\zxwCol,1mm]\zxZ{}\ar[r,marrow=<]&[\zxwCol,2mm]\zxN{}\\
   &&\zxX{}\ar[urr, -N, inarrow=<]
\end{ZX}\\
=&
\begin{ZX}
   &[\zxwCol,2mm]&[\zxwCol,1mm]\zxN{}&[\zxwCol,1mm]&[\zxwCol,2mm]\\
   \zxN{}\ar[r,marrow=<]&\zxX{}\ar[drr,s.,numarrow={<}{0.25},numarrow={<}{0.75}]\ar[rr,o'={-=.2,L=.25},marrow=<]&&\zxZ{}\ar[r,marrow=<]&\zxN{}\\[\zxwRow]
   &\zxX{}\ar[rr,o.={-=.2,L=.25},marrow=<]\ar[urr,s.,numarrow={<}{0.25},numarrow={<}{0.75}]&&\zxZ{}\\
   &&\zxN{}
\end{ZX}\\
\overset{(\hyperlink{li:fusion}{f})}{=}&
\begin{ZX}
   &[\zxwCol,2mm]&[\zxwCol,1mm]\zxN{}&[\zxwCol,1mm]&[\zxwCol,2mm]\\
   \zxN{}\ar[r,marrow=<]&\zxX{}\ar[drr,s.,numarrow={<}{0.25},numarrow={<}{0.75}]\ar[rr,o'={-=.2,L=.25},marrow=<]&&\zxZ{}\ar[r,marrow=<]&\zxN{}\\[\zxwRow]
   \zxX{}\ar[r,marrow=<]&\zxX{}\ar[rr,o.={-=.2,L=.25},marrow=<]\ar[urr,s.,numarrow={<}{0.25},numarrow={<}{0.75}]&&\zxZ{}\ar[r,marrow=<]&\zxZ{}\\
   &&\zxN{}
\end{ZX}\\
\overset{(\hyperlink{li:bialgebra}{b})}{=}&
\begin{ZX}
   \zxN{}\ar[dr,-N, outarrow=<]&[\zxwCol,1mm]&[\zxwCol,2mm]&[\zxwCol,1mm]\zxN{}\\
   &  \zxZ{}\ar[r,marrow=<]&\zxX{}\ar[dr,N-, inarrow=<]\ar[ur,N-, inarrow=<]\\
   \zxX{}\ar[ur,-N, outarrow=<]&&&\zxZ{}
\end{ZX}\\
\overset{(\hyperlink{li:fusion}{f})}{=}&
\begin{ZX}
   &[\zxwCol,1mm]&[\zxwCol,2mm]&[\zxwCol,1mm]\zxN{}\\
   &\zxX{}\ar[urr,N-, inarrow=<]\ar[dr,s., marrow=<]&&\\
   \zxN{}\ar[dr,-N, outarrow=<]&&\zxX{}\ar[r,inarrow=<]&\zxZ{}\\
   &  \zxZ{}\ar[ur,s., marrow=<]\\
   \zxX{}\ar[ur,-N, outarrow=<]&&&
\end{ZX}\\
\overset{(\hyperlink{li:copy}{c})}{=}&\begin{ZX}
   \zxN{}\ar[dr,-N, outarrow=<]&[\zxwCol,1mm]&[\zxwCol]&[\zxwCol]&[\zxwCol,1mm]\zxN{}\\
   &  \zxZ{}\ar[r,marrow=<]&\zxZ{}&\zxX{}\ar[ur,N-, inarrow=<]\ar[dr,N-, inarrow=<]\\
   \zxX{}\ar[ur,-N, outarrow=<]&&&&\zxZ{}
\end{ZX}\\
\overset{(\hyperlink{li:fusion}{f})}{=}&\begin{ZX}
   \zxN{}\ar[dr,-N, outarrow=<]&[\zxwCol,1mm]&[\zxwCol]&[\zxwCol,1mm]\zxN{}\\
   &  \zxZ{}&\zxZ{}\ar[ur,N-, inarrow=<]\ar[dr,N-, inarrow=<]\\
   \zxX{}\ar[ur,-N, outarrow=<]&&&\zxZ{}
\end{ZX}\\
\overset{(\hyperlink{li:anti}{a})}{=}&
\begin{ZX}
   \zxN{}\ar[r,marrow=<]&[\zxwCol,1mm]\zxFsq{}\ar[r,marrow=<]&[\zxwCol,1mm]\zxX{}&[\zxwCol]\zxZ{}\ar[r,marrow=<]&[\zxwCol,1mm]\zxFsq{}\ar[r,marrow=<]&[\zxwCol,1mm]\zxN{}
\end{ZX}\\
\overset{(\hyperlink{li:sq}{s})}{=}&
\begin{ZX}
   \zxN{}\ar[r,marrow=<]&[\zxWCol]\zxX{}&[\zxWCol]\zxZ{}\ar[r,marrow=<]&[\zxwCol,2mm]\zxN{}
\end{ZX}
\end{align}

Using the Hopf rule, one can
show that conjugation of the CSUM diagram yields its own inverse. In fact,
\begin{align}
    \begin{ZX}
    \zxN{}\ar[drr,-N,outarrow=<]&[\zxwCol,1mm]&[\zxwCol]&[\zxWCol]&[\zxwCol]&[\zxwCol,1mm]\\
    &&\zxZ{} \ar[r,marrow=<]&\zxZ{}\ar[urr,N-,inarrow=<]\ar[rd,s.,marrow=<]&\zxN{}\\[\zxWRow]
    \zxN{}\ar[r,marrow=<] & \zxX{}\ar[ru,s.,marrow=<]\ar[rrr,o.={-=.2,L=.25},marrow=<]&&&\zxX{}\ar[r,marrow=<]&\zxN{}
    \end{ZX}
    \overset{(\hyperlink{li:fusion}{f})}{=}&
    \begin{ZX}
    &[\zxwCol,1mm]&[\zxwCol]&[\zxWCol]\zxN{}\\
    \zxN{}\ar[r,marrow=<] &\zxZ{}\ar[rd,(,marrow=>]\ar[rd,),marrow=<]\ar[urr,N-,inarrow=<]\\[\zxWRow]
    &&\zxX{} \ar[r,marrow=<]&\zxN{}\\
    \zxN{}\ar[urr,-N,outarrow=<]
    \end{ZX}\\
    \overset{(\hyperlink{li:fusion}{f})}{=}&
    \begin{ZX}
    &[\zxwCol,1mm]&&[\zxWCol]&&[\zxWCol]\zxN{}\\
    \zxN{}\ar[r,marrow=<] &\zxZ{}\ar[rd,N-,marrow=<]\ar[urrrr,N-,inarrow=<]\\[\zxwRow]
    &&\zxZ{}\ar[r,o'={-=.2,L=.25},marrow=<]\ar[r,o.={-=.2,L=.25},marrow=>]&\zxX{}\ar[rd,-N,marrow=<]\\[\zxwRow]
    &&&&\zxX{} \ar[r,marrow=<]&\zxN{}\\
    \zxN{}\ar[urrrr,-N,outarrow=<]
    \end{ZX}\\
    \overset{(\hyperlink{li:hopf}{h})}{=}&
    \begin{ZX}
    &[\zxwCol,1mm]&&[\zxWCol]&&[\zxWCol]\zxN{}\\
    \zxN{}\ar[r,marrow=<] &\zxZ{}\ar[rd,N-,marrow=<]\ar[urrrr,N-,inarrow=<]\\[\zxwRow]
    &&\zxZ{}&\zxX{}\ar[rd,-N,marrow=<]\\[\zxwRow]
    &&&&\zxX{} \ar[r,marrow=<]&\zxN{}\\
    \zxN{}\ar[urrrr,-N,outarrow=<]
    \end{ZX}\\
    \overset{(\hyperlink{li:fusion}{f})}{=}&
    \begin{ZX}
    \zxN{}\ar[r,marrow=<]&[\zxwCol,5mm]\zxN{}\\[\zxwRow,3mm]
    \zxN{}\ar[r,marrow=<]&\zxN{}
    \end{ZX}
\end{align}
and
\begin{align}
    \begin{ZX}
    \zxN{}\ar[r,marrow=<] &[\zxwCol,1mm]\zxZ{}\ar[rd,s.,marrow=<]\ar[rrr,o'={-=.2,L=.25},marrow=<]&[\zxwCol]&[\zxwCol,1mm]&[\zxwCol]\zxZ{}\ar[r,marrow=<]&[\zxwCol,1mm]\zxN{}\\[\zxWRow]
    &&\zxX{} \ar[r,marrow=<]&\zxX{}\ar[drr,N-,inarrow=<]\ar[ru,s.,marrow=<]&\zxN{}\\
    \zxN{}\ar[urr,-N,outarrow=<]&&[\zxwCol]&[\zxWCol]&[\zxwCol]&[\zxwCol,1mm]
    \end{ZX}=&
    \begin{ZX}
    \zxN{}\ar[r,marrow=<]&[\zxwCol,5mm]\zxN{}\\[\zxwRow,3mm]
    \zxN{}\ar[r,marrow=<]&\zxN{}
    \end{ZX}
\end{align}
holds.

\subsubsection{Rotation rule}

The rotation rule~$(\hyperlink{li:rot}{rot})$ is
\hypertarget{li:rot}{}\begin{equation}
    \begin{ZX}
    \zxN{}\ar[r,marrow=<]&[\zxwCol,1mm]\zxBox{R(\theta)}\ar[r,marrow=<]&[\zxwCol,1mm] \zxZ{}
    \end{ZX}
    \overset{(\hyperlink{li:rot}{rot})}{=}\begin{ZX}
    \zxN{}\ar[r,marrow=<]&[\zxwCol,1mm]\zxZ{-\frac{\tan\theta}{2}x^2}
    \end{ZX}
\end{equation}
which holds for any value of~$\theta$ that is not an 
odd multiple of $\frac{\pi}{2}$. (In this exceptional case, the Fourier rule can be applied instead.) The rotation rule can be shown in the following way:
\begin{align}
    &\begin{ZX}
    \zxN{}\ar[r,marrow=<]&[\zxwCol]\zxBox{R(\theta)}\ar[r,marrow=<]&[\zxwCol] \zxZ{}
    \end{ZX}\\
\overset{\mathclap{\eqref{def:rot_sp}}}{=}&
\begin{ZX}
        \zxN{}\rar[marrow=<]&[\zxwCol] \zxX{\frac{\tan(\theta/2)}{2}x^2}  \rar[marrow=<] &[\zxwCol] \zxZ-{\frac{\sin\theta}{2}x^2} \rar[marrow=<]&[\zxwCol] \zxX{\frac{\tan(\theta/2)}{2}x^2} \rar[marrow=<]&[\zxwCol]\zxZ{}
    \end{ZX}\\
\overset{(\hyperlink{li:copy}{c})}{=}&
\begin{ZX}
        \zxN{}\rar[marrow=<]&[\zxwCol] \zxX{\frac{\tan(\theta/2)}{2}x^2} \rar[marrow=<] &[\zxwCol] \zxZ-{\frac{\sin\theta}{2}x^2} \rar[marrow=<]&[\zxwCol]\zxZ{}
    \end{ZX}\\
\overset{\eqref{eq:sq_cos}}{=}&
\begin{ZX}
        \zxN{}\ar[r,marrow=<]&[\zxwCol]\zxZ{-\frac{\tan\theta}{2}x^2}\rar[marrow=<]&[\zxwCol]\zxX{\frac{\cos\theta\tan(\theta/2)}{2}x^2}\ar[r,marrow=<]&[\zxwCol] \zxBox{Sq(\cos\theta)} \rar[marrow=<]&[\zxwCol]\zxZ{}
    \end{ZX}\\
\overset{(\hyperlink{li:sq}{s})}{=}&
\begin{ZX}
        \zxN{}\ar[r,marrow=<]&[\zxwCol]\zxZ{-\frac{\tan\theta}{2}x^2}\rar[marrow=<]&[\zxwCol]\zxX{\frac{\cos\theta\tan(\theta/2)}{2}x^2}\ar[r,marrow=<]&[\zxwCol]\zxZ{}
    \end{ZX}\\
\overset{(\hyperlink{li:copy}{c})}{=}&
\begin{ZX}
        \zxN{}\ar[r,marrow=<]&[\zxwCol]\zxZ{-\frac{\tan\theta}{2}x^2}\rar[marrow=<]&[\zxwCol]\zxZ{}
    \end{ZX}\\
\overset{(\hyperlink{li:fusion}{f})}{=}&\begin{ZX}
    \zxN{}\ar[r,marrow=<]&[\zxwCol]\zxZ{-\frac{\tan\theta}{2}x^2}
    \end{ZX}
\end{align}
in which we used the following property
\begin{subequations}
    \begin{align}
    &\begin{ZX}
        \zxN{}&[\zxwCol] \zxBox{Sq(\cos\theta)} \lar[inarrow=>] \rar[inarrow=<] & [\zxwCol] \zxN{}
    \end{ZX}\\
    =&\begin{ZX}
        \zxN{}\ar[r,marrow=<]&[\zxwCol]\zxX{-\frac{\cos\theta\tan(\theta/2)}{2}x^2}\ar[r,marrow=<]&[\zxwCol]\zxZ{\frac{\tan\theta}{2}x^2}\rar[marrow=<]&[\zxwCol] \zxX{\frac{\tan(\theta/2)}{2}x^2}\rar[marrow=<] &[\zxwCol] \zxZ-{\frac{\sin\theta}{2}x^2} \rar[marrow=<]&[\zxwCol]\zxN{}
    \end{ZX}
\end{align}\label{eq:sq_cos}
\end{subequations}
to be proven in Lemma~\ref{thm:sq_eq}. Since conjugating the rotation gate yields its own inverse, one also obtains
\begin{equation}
    \begin{ZX}
        \zxZ{}\ar[r,marrow=<]&[\zxwCol,1mm]\zxBox{R(-\theta)}\ar[r,marrow=<]&[\zxwCol,1mm] \zxN{}
    \end{ZX}
        \overset{(\hyperlink{li:rot}{rot})^\dagger}{=}\begin{ZX}
            \zxZ{\frac{\tan\theta}{2}x^2}\ar[r,marrow=<]&[\zxwCol,1mm]\zxN{}
        \end{ZX}
\end{equation}
and thus
\begin{equation}
    \begin{ZX}
        \zxZ{}\ar[r,marrow=<]&[\zxwCol,1mm]\zxBox{R(\theta)}\ar[r,marrow=<]&[\zxwCol,1mm] \zxN{}
        \end{ZX}
        =\begin{ZX}
            \zxZ-{\frac{\tan\theta}{2}x^2}\ar[r,marrow=<]&[\zxwCol,1mm]\zxN{}
        \end{ZX}\label{eq:rot_rule_conj}
\end{equation}
as well.

The rotation rule comes with another useful form of equality as is shown below:
\begin{align}
    &\begin{ZX}
    \zxN{}\ar[r,marrow=<]&[\zxwCol,1mm]\zxBox{R(\theta)}\ar[r,marrow=<]&[\zxwCol,1mm] \zxX{}
    \end{ZX}\\
    \overset{(\hyperlink{li:inv}{inv})}{=}&
    \begin{ZX}
        \zxN{}\ar[r,marrow=<]&[\zxwCol,1mm]\zxBox{R(\theta)}\ar[r,marrow=<]&[\zxwCol,1mm] \zxF{}\ar[r,marrow=<]&[\zxwCol,1mm] \zxZ{}
        \end{ZX}
    \\
    \overset{(\hyperlink{li:fourier}{f})}{=}&\begin{ZX}
        \zxN{}\ar[r,marrow=<]&[\zxwCol,1mm]\zxBox{R(\theta)}\ar[r,marrow=<]&[\zxwCol,1mm] \zxBox{R(-\pi/2)}\ar[r,marrow=<]&[\zxwCol,1mm] \zxZ{}
    \end{ZX}
    \\
    \overset{\eqref{eq:rot_assoc}}{=}&\begin{ZX}
        \zxN{}\ar[r,marrow=<]&[\zxwCol,1mm] \zxBox{R(-\pi/2)}\ar[r,marrow=<]&[\zxwCol,1mm]\zxBox{R(\theta)}\ar[r,marrow=<]&[\zxwCol,1mm] \zxZ{}
    \end{ZX}
    \\
    \overset{(\hyperlink{li:fourier}{f})}{=}&\begin{ZX}
        \zxN{}\ar[r,marrow=<]&[\zxwCol,1mm] \zxF{}\ar[r,marrow=<]&[\zxwCol,1mm]\zxBox{R(\theta)}\ar[r,marrow=<]&[\zxwCol,1mm] \zxZ{}
    \end{ZX}
    \\
    \overset{(\hyperlink{li:rot}{rot})}{=}&\begin{ZX}
        \zxN{}\ar[r,marrow=<]&[\zxwCol,1mm] \zxF{}\ar[r,marrow=<]&[\zxwCol,1mm]\zxZ{-\frac{\tan\theta}{2}x^2}
    \end{ZX}
    \\
    \overset{(\hyperlink{li:inv}{inv})}{=}&\begin{ZX}
        \zxN{}\ar[r,marrow=<]&[\zxwCol,1mm]\zxX{-\frac{\tan\theta}{2}x^2}
    \end{ZX}
\end{align}
and
\begin{align}
        &\begin{ZX}
    \zxN{}\ar[r,marrow=<]&[\zxwCol,1mm]\zxBox{R(\theta)}\ar[r,marrow=<]&[\zxwCol,1mm] \zxX{}
    \end{ZX}\\\overset{(\hyperlink{li:inv}{inv})}{=}&
    \begin{ZX}
        \zxN{}\ar[r,marrow=<]&[\zxwCol,1mm]\zxBox{R(\theta)}\ar[r,marrow=<]&[\zxwCol,1mm] \zxF{}\ar[r,marrow=<]&[\zxwCol,1mm] \zxZ{}
        \end{ZX}
    \\
    \overset{(\hyperlink{li:fourier}{f})}{=}&\begin{ZX}
        \zxN{}\ar[r,marrow=<]&[\zxwCol,1mm]\zxBox{R(\theta)}\ar[r,marrow=<]&[\zxwCol,1mm] \zxBox{R(-\pi/2)}\ar[r,marrow=<]&[\zxwCol,1mm] \zxZ{}
    \end{ZX}
    \\
    \overset{\eqref{eq:rot_assoc}}{=}&\begin{ZX}
        \zxN{}\ar[r,marrow=<]&[\zxwCol,1mm]\zxBox{R(\theta-\pi/2)}\ar[r,marrow=<]&[\zxwCol,1mm] \zxZ{}
    \end{ZX}
    \\
    \overset{(\hyperlink{li:rot}{rot})}{=}&\begin{ZX}
        \zxN{}\ar[r,marrow=<]&[\zxwCol,1mm]\zxZ{-\frac{\tan(\theta-\pi/2)}{2}x^2}
    \end{ZX}
    \\
    =&\begin{ZX}
        \zxN{}\ar[r,marrow=<]&[\zxwCol,1mm]\zxZ{\frac{\cot\theta}{2}x^2}
    \end{ZX}
\end{align}
thus one obtains
\begin{equation}
    \begin{ZX}
        \zxN{}\ar[r,marrow=<]&[\zxwCol,1mm]\zxX{-\frac{\tan\theta}{2}x^2}
    \end{ZX}
    =
    \begin{ZX}
        \zxN{}\ar[r,marrow=<]&[\zxwCol,1mm]\zxZ{\frac{\cot\theta}{2}x^2}
    \end{ZX}\label{eq:quad_state_id}
\end{equation}
for any value of~$\theta$ that is not a
multiple of $\frac{\pi}{2}$.
\subsubsection{Symmetry of Controlled-Z gate}

As is shown in Table.~\ref{tab:table_gate}, controlled-Z gate has a symmetric Hamiltonian over two modes and thus it is invariant under swapping operation of the inputs. This fact can also be easiry verified with graphical calculus:
\begin{align}
&\begin{ZX}
&[\zxwCol]&[\zxwCol]&[\zxWCol]&[\zxWCol]&[\zxwCol]&[\zxwCol]\\
\zxN{}\ar[rr,marrow=<] &&\zxZ{}\ar[rd,s.,marrow=<]\ar[urrrr,N-,inarrow=<]\\[\zxWRow]
&&&\zxF{} \ar[rd,s.,marrow=<]&&\zxN{}\\[\zxWRow]
&&&&\zxZ{} \ar[rr,marrow=<]&&\zxN{}\\
\zxN{}\ar[urrrr,-N,outarrow=<]
\end{ZX}\\
\overset{(\hyperlink{li:inv}{inv})}{=}&
\begin{ZX}
&[\zxwCol]&[\zxwCol]&[\zxWCol]&[\zxWCol]&[\zxwCol,1mm]&[\zxwCol,1mm]\\
\zxN{}\ar[rr,marrow=<] &&\zxZ{}\ar[rrd,s.,marrow=<]\ar[urrrr,N-,inarrow=<]\\[\zxWRow,5mm]
&&&&\zxX{} \ar[r,marrow=<]&\zxF{}\ar[r,marrow=<]&\zxN{}\\[\zxZeroRow,-6mm]
\zxN{}\ar[rr,inarrow=<]&&\zxFdag{}\ar[urr,-N,marrow=<]
\end{ZX}\\
\overset{(\hyperlink{li:rev}{rev})}{=}&
\begin{ZX}
\zxN{}\ar[drrrr,-N,inarrow=<]&[\zxwCol,2mm]&[\zxwCol]&[\zxWCol]&[\zxwCol]&[\zxwCol,1mm]\\
&&&&\zxZ{}\ar[r,marrow=<]&\zxN{}\\
&&&\zxFsq{}\ar[ru,s.,marrow=<]\\[\zxZeroRow]
\zxN{}\ar[r,marrow=<]&\zxFdag{}\ar[r,marrow=<]&\zxX{}\ar[dr,N-,outarrow=<]\ar[ru,s.,marrow=<]\\[\zxZeroRow,-3mm]
&&&\zxF{}\ar[rr,marrow=<]&&\zxN{}
\end{ZX}\\
\overset{\phantom{(inv)}}{=}&  
\begin{ZX}
\zxN{}\ar[drrrrr,-N,inarrow=<]&[\zxwCol,2mm]&[\zxwCol]&[\zxWCol]&[\zxwCol]&[\zxwCol]\\
&&&&&\zxZ{}\ar[rr,marrow=<]&&\zxN{}\\[\zxZeroRow,-1mm]
&&&&\zxF{}\ar[ru,s.,marrow=<]\\[\zxZeroRow,-3mm]
&&&\zxF{}\ar[ru,s.,marrow=<]\\[\zxZeroRow,-3mm]
\zxN{}\ar[r,marrow=<]&\zxFdag{}\ar[r,marrow=<]&\zxX{}\ar[dr,N-,outarrow=<]\ar[ru,N-,outarrow=<]\\[\zxZeroRow,-3mm]
&&&\zxF{}\ar[rrrr,marrow=<]&&&&\zxN{}
\end{ZX}\\
\overset{(\hyperlink{li:inv}{inv})}{=}&
\begin{ZX}
\zxN{}\ar[drrrr,-N,inarrow=<]&[\zxwCol]&[\zxwCol]&[\zxWCol]&[\zxWCol]&[\zxwCol]&[\zxwCol]\\
&&&&\zxZ{}\ar[rr,marrow=<]&&\zxN{}\\[\zxWRow]
&&&\zxF{} \ar[ru,s.,marrow=<]\\[\zxWRow]
\zxN{}\ar[rr,marrow=<]&&\zxZ{}\ar[drrrr,N-,outarrow=<]\ar[ru,s.,marrow=<]\\
&&&&&&\zxN{}
\end{ZX}
\end{align}
This derivation reflects the symmetric property of the controlled-Z gate in 
that its diagram remains identical under a vertical flip.

\subsubsection{Rotation self-loop removal}

For any $\theta$ that is not multiple of $\pi$,
\begin{equation}
    \begin{ZX}
    &\zxBox{R(\theta)}\ar[d,C,marrow=>]\ar[d,C-,marrow=<]&\\
    &\zxX{}\ar[rd]&\\
    \zxN{}\ar[ru]&\cdots&\zxN{}
    \end{ZX}=\begin{ZX}
    &\zxX{(\tan\frac{\theta}{2})x^2}\ar[rd]&\\
    \zxN{}\ar[ru]&\cdots&\zxN{}
    \end{ZX}\label{eq:rot_loop-1}
\end{equation}
and
\begin{equation}
    \begin{ZX}
    &\zxBox{R(\theta)}\ar[d,C,marrow=>]\ar[d,C-,marrow=<]&\\
    &\zxZ{}\ar[rd]&\\
    \zxN{}\ar[ru]&\cdots&\zxN{}
    \end{ZX}=\begin{ZX}
    &\zxZ{-(\tan\frac{\theta}{2})x^2}\ar[rd]&\\
    \zxN{}\ar[ru]&\cdots&\zxN{}
    \end{ZX}\label{eq:rot_loop-2}
\end{equation}
holds. Eq.~\eqref{eq:rot_loop-1} can be shown as
\begin{align}
    &\begin{ZX}
    &\zxBox{R(\theta)}\ar[d,C,marrow=>]\ar[d,C-,marrow=<]&\\
    &\zxX{}\ar[rd]&\\
    \zxN{}\ar[ru]&\cdots&\zxN{}
    \end{ZX}\\
    \overset{(\hyperlink{li:rot}{rot})}{=}&\begin{ZX}
   \zxX{\frac{\tan(\theta/2)}{2}x^2} \ar[dr,C,marrow=>] \rar[inarrow=<] &\zxZ-{\frac{\sin\theta}{2}x^2} \rar[marrow=<]&\zxX{\frac{\tan(\theta/2)}{2}x^2}\ar[dl,C-,marrow=<]\\
    &\zxX{}\ar[rd]&\\
    \zxN{}\ar[ru]&\cdots&\zxN{}
    \end{ZX}\\
    \overset{(\hyperlink{li:fusion}{f})}{=}&\begin{ZX}
    &\zxZ-{\frac{\sin\theta}{2}x^2}\ar[d,C,marrow=>]\ar[d,C-,marrow=<]&\\
    &\zxX{(\tan\frac{\theta}{2})x^2}\ar[rd]&\\
    \zxN{}\ar[ru]&\cdots&\zxN{}
    \end{ZX}\\
    \overset{(\hyperlink{li:fusion}{f})}{=}&\begin{ZX}
        &\zxX{(\tan\frac{\theta}{2})x^2}\rar[marrow=<]\ar[rd]&[\zxwCol]\zxX{}\ar[rrr,o'={-=.2,L=.25},marrow=<]\ar[rrr,o.={-=.2,L=.25},marrow=>]&[\zxwCol]\zxN{}&[\zxwCol]&[\zxwCol]\zxZ{}\ar[r,marrow=<]&[\zxwCol]\zxZ-{\frac{\sin\theta}{2}x^2}\\
        \zxN{}\ar[ru]&\cdots&\zxN{}
        \end{ZX}\\
    \overset{(\hyperlink{li:hopf}{h})}{=}&\begin{ZX}
        &\zxX{(\tan\frac{\theta}{2})x^2}\rar[marrow=<]\ar[rd]&[\zxwCol]\zxX{}&[\zxwCol]\zxN{}&[\zxwCol]&[\zxwCol]\zxZ{}\ar[r,marrow=<]&[\zxwCol]\zxZ-{\frac{\sin\theta}{2}x^2}\\
        \zxN{}\ar[ru]&\cdots&\zxN{}
        \end{ZX}\\
    \overset{(\hyperlink{li:fusion}{f})}{=}&\begin{ZX}
    &\zxX{(\tan\frac{\theta}{2})x^2}\ar[rd]&\\
    \zxN{}\ar[ru]&\cdots&\zxN{}
    \end{ZX}
\end{align}
in which we neglected scalar term \begin{ZX}\zxZ-{\frac{\sin\theta}{2}x^2}\end{ZX} 
in the final step. Note this is legal because $\llbracket\begin{ZX}\zxZ-{\frac{\sin\theta}{2}x^2}\end{ZX}\rrbracket=\int_\mathbb{R}\dd{x}\exp(-i\frac{\sin\theta}{2}x^2)$ is a nonzero value in our standard interpretation. 

\subsection{Completeness for 1-mode Gaussian unitary gates}\label{ssec:gauss_univ}

Since any Gaussian operation can be represented by its symplectic action to the canonical operators, one can perfectly simulate arbitrary Gaussian quantum dynamics in polynomial time classically \cite{bartlett_efficient_2002}. Now the same question arises for the graphical calculus: when given two Gaussian diagrams representing the same process, is it possible to find a way to convert one into the other? Though it is hard to find such a transformation in the general case, we can give a straightforward
proof for when the diagrams are 
limited to 1-mode Gaussian unitary gates. We start with the following lemma.

\begin{lem}\label{thm:sq_eq}
    Let $k$ and $\tau$ be an arbitrary nonzero real numbers. Then, we have the following expansions of a squeezing operator with squeezing factor~$\tau$:
    \begin{align}
        \begin{ZX}
        \zxN{}&[\zxWCol] \zxBox{Sq(\tau)} \lar[inarrow=>] \rar[inarrow=<] & [\zxWCol] \zxN{}
        \end{ZX}=&
        \begin{ZX}
        \zxN{}&[\zxWCol] \zxX{\frac{\tau(1-\tau)}{4k}x^2} \lar[inarrow=>] \rar[inarrow=<] &[\zxWCol] \zxZ-{\frac{k}{\tau}x^2} \rar[marrow=<]&[\zxWCol] \zxX{\frac{\tau-1}{4k}x^2} \rar[marrow=<]&[\zxWCol] \zxZ{kx^2} \rar[marrow=<]&[\zxWCol]\zxN{}
    \end{ZX}\label{eq:sq_k_X}\\
    =&\begin{ZX}
        \zxN{}&[\zxWCol] \zxZ{\frac{\tau-1}{4k\tau^2}x^2} \lar[inarrow=>] \rar[inarrow=<] &[\zxWCol] \zxX{-k\tau x^2} \rar[marrow=<]&[\zxWCol] \zxZ{\frac{1-\tau}{4k\tau}x^2} \rar[marrow=<]&[\zxWCol] \zxX{kx^2} \rar[marrow=<]&[\zxWCol]\zxN{}
    \end{ZX}\label{eq:sq_k_Z}
    \end{align}
\end{lem}
\begin{widetext}
\begin{prf*}
The top equation can be shown as
\begin{align}
    \begin{ZX}
        \zxN{}&[\zxWCol] \zxBox{Sq(\tau)} \lar[inarrow=>] \rar[inarrow=<] & [\zxWCol] \zxN{}
    \end{ZX}
    \coloneqq&
    \begin{ZX}
        \zxN{}&[\zxWCol] \zxX{\frac{\tau(1-\tau)}{4}x^2} \lar[inarrow=>] \rar[inarrow=<] &[\zxWCol] \zxZ-{\frac{1}{\tau}x^2} \rar[marrow=<]&[\zxWCol] \zxX{\frac{\tau-1}{4}x^2} \rar[marrow=<]&[\zxWCol] \zxZ{x^2} \rar[marrow=<]&[\zxWCol]\zxN{}
    \end{ZX}\\
    \overset{(\hyperlink{li:fusion}{f})}{=}&\begin{ZX}
        \zxN{}&[\zxWCol] \zxX{\frac{\tau(1-\tau)}{4}x^2} \lar[inarrow=>] \rar[inarrow=<] &[\zxWCol] \zxZ-{\frac{1}{\tau}x^2} \rar[marrow=<]&[\zxWCol] \zxX{\frac{\tau-1}{4}x^2} \rar[marrow=<]&[\zxWCol] \zxZ{(1-k)x^2} \rar[marrow=<]&[\zxWCol] \zxZ{kx^2} \rar[marrow=<]&[\zxWCol]\zxN{}
    \end{ZX}\\
    \overset{(\hyperlink{li:quad}{q})}{=}&\begin{ZX}
        \zxN{}&[\zxWCol] \zxX{\frac{\tau(1-\tau)}{4}x^2} \lar[inarrow=>] \rar[inarrow=<] &[\zxWCol] \zxX{\frac{\tau(\tau-1)(k-1)}{4k}x^2} \rar[marrow=<]&[\zxWCol] \zxZ-{\frac{k}{\tau}x^2} \rar[marrow=<]&[\zxWCol] \zxX{\frac{\tau-1}{4k}x^2} \rar[marrow=<]&[\zxWCol] \zxZ{kx^2} \rar[marrow=<]&[\zxWCol]\zxN{}
    \end{ZX}\\
    \overset{(\hyperlink{li:fusion}{f})}{=}&\begin{ZX}
        \zxN{}&[\zxWCol] \zxX{\frac{\tau(1-\tau)}{4k}x^2} \lar[inarrow=>] \rar[inarrow=<] &[\zxWCol] \zxZ-{\frac{k}{\tau}x^2} \rar[marrow=<]&[\zxWCol] \zxX{\frac{\tau-1}{4k}x^2} \rar[marrow=<]&[\zxWCol] \zxZ{kx^2} \rar[marrow=<]&[\zxWCol]\zxN{}
    \end{ZX}
\end{align}
in which we applied quadratic rule under the constraint $k\neq 0$. 
To obtain the second form, we make use of the first. The statement of the lemma prescribes $k$. Now, let $k'$ be a nonzero real number such that $(1-4k'k)\tau\neq1$, which always exists. Using $k'$ in the the result above, we have
\begin{align}
    \begin{ZX}
        \zxN{}&[\zxWCol] \zxBox{Sq(\tau)} \lar[inarrow=>] \rar[inarrow=<] & [\zxWCol] \zxN{}
    \end{ZX}
    \coloneqq&
    \begin{ZX}
        \zxN{}&[\zxWCol] \zxX{\frac{\tau(1-\tau)}{4{k'}}x^2} \lar[inarrow=>] \rar[inarrow=<] &[\zxWCol] \zxZ-{\frac{{k'}}{\tau}x^2} \rar[marrow=<]&[\zxWCol] \zxX{\frac{\tau-1}{4{k'}}x^2} \rar[marrow=<]&[\zxWCol] \zxZ{{k'}x^2} \rar[marrow=<]&[\zxWCol]\zxN{}
    \end{ZX}\\
    \overset{(\hyperlink{li:fusion}{f})}{=}&
    \begin{ZX}
        \zxN{}&[\zxWCol] \zxX{\frac{\tau(1-\tau)}{4{k'}}x^2} \lar[inarrow=>] \rar[inarrow=<] &[\zxWCol] \zxZ-{\frac{{k'}}{\tau}x^2} \rar[marrow=<]&[\zxWCol] \zxX{\frac{\tau-1}{4{k'}}x^2} \rar[marrow=<]&[\zxWCol] \zxZ{{k'}x^2} \rar[marrow=<]&[\zxWCol] \zxX-{{k}x^2} \rar[marrow=<]&[\zxWCol] \zxX{{k}x^2} \rar[marrow=<]&[\zxWCol]\zxN{}
    \end{ZX}\\
    \overset{(\hyperlink{li:quad}{q})}{=}&\begin{ZX}
        \zxN{}&[\zxWCol] \zxX{\frac{\tau(1-\tau)}{4{k'}}x^2} \lar[inarrow=>] \rar[inarrow=<] &[\zxWCol] \zxZ-{\frac{{k'}}{\tau}x^2} \rar[marrow=<]&[\zxWCol] \zxZ{\frac{4{k'}^2{k}}{(4{k'}{k}-1)\tau+1}x^2} \rar[marrow=<]&[\zxWCol] \zxX{\frac{(1-4{k'}{k})\tau-1}{4{k'}}x^2} \rar[marrow=<]&[\zxWCol] \zxZ{\frac{{k'}(1-\tau)}{(4{k'}{k}-1)\tau+1}x^2} \rar[marrow=<]&[\zxWCol] \zxX{{k}x^2} \rar[marrow=<]&[\zxWCol]\zxN{}
    \end{ZX}\\
    \overset{(\hyperlink{li:fusion}{f})}{=}&\begin{ZX}
        \zxN{}&[\zxWCol] \zxX{\frac{\tau(1-\tau)}{4{k'}}x^2} \lar[inarrow=>] \rar[inarrow=<] &[\zxWCol] \zxZ{\frac{{k'}(\tau-1)}{\tau(4{k'}{k}\tau-\tau+1)}x^2} \rar[marrow=<]&[\zxWCol] \zxX{\frac{(1-4{k'}{k})\tau-1}{4{k'}}x^2} \rar[marrow=<]&[\zxWCol] \zxZ{\frac{{k'}(1-\tau)}{(4{k'}{k}-1)\tau+1}x^2} \rar[marrow=<]&[\zxWCol] \zxX{{k}x^2} \rar[marrow=<]&[\zxWCol]\zxN{}
    \end{ZX}\\
    \overset{(\hyperlink{li:quad}{q})}{=}&\begin{ZX}
        \zxN{}&[\zxWCol] \zxZ{\frac{\tau-1}{4{k}\tau^2}x^2} \lar[inarrow=>] \rar[inarrow=<] &[\zxWCol] \zxX{-{k}\tau x^2} \rar[marrow=<]&[\zxWCol] \zxZ{\frac{(\tau-1)^2}{4{k}\tau(4{k'}{k}\tau-\tau+1)}x^2} \rar[marrow=<]&[\zxWCol] \zxZ{\frac{{k'}(1-\tau)}{(4{k'}{k}-1)\tau+1}x^2} \rar[marrow=<]&[\zxWCol] \zxX{{k}x^2} \rar[marrow=<]&[\zxWCol]\zxN{}
    \end{ZX}
    \\
    \overset{(\hyperlink{li:fusion}{f})}{=}&\begin{ZX}
        \zxN{}&[\zxWCol] \zxZ{\frac{\tau-1}{4{k}\tau^2}x^2} \lar[inarrow=>] \rar[inarrow=<] &[\zxWCol] \zxX{-{k}\tau x^2} \rar[marrow=<]&[\zxWCol] \zxZ{\frac{1-\tau}{4{k}\tau}x^2} \rar[marrow=<]&[\zxWCol] \zxX{{k}x^2} \rar[marrow=<]&[\zxWCol]\zxN{}
    \end{ZX}\label{eq:sq_k_prf}
\end{align}
where application of quadratic rule is legal since $k\neq 0$, $k'\neq0$, and $(1-4k'k)\tau\neq1$.
\hfill \qed
\end{prf*}
\end{widetext}
We can also go in the other direction, as shown in the following lemma.
\begin{lem}
\label{thm:sq}
Let $a,b,c,d$ be 
    nonzero real values such that
    $4abc+a+c=4bcd+b+d=0$.
    Then, $4ab+1\neq 0$, and both of the following hold:
    \begin{gather}
        \begin{ZX}\zxN{}&[\zxwCol]\zxX{ax^2}\lar[inarrow=>]\rar[marrow=<]&[\zxwCol]\zxZ{bx^2}\rar[marrow=<]&[\zxwCol]\zxX{cx^2}\rar[marrow=<]&[\zxwCol]\zxZ{dx^2}\rar[inarrow=<]&[\zxwCol]\zxN{}\end{ZX}
        =\begin{ZX}\zxN{}&[\zxWCol] \zxBox{Sq(4ab+1)} \lar[inarrow=>] \rar[inarrow=<] & [\zxWCol] \zxN{}\end{ZX}\\
        \begin{ZX}\zxN{}&[\zxwCol]\zxZ{ax^2}\lar[inarrow=>]\rar[marrow=<]&[\zxwCol]\zxX{bx^2}\rar[marrow=<]&[\zxwCol]\zxZ{cx^2}\rar[marrow=<]&[\zxwCol]\zxX{dx^2}\rar[inarrow=<]&[\zxwCol]\zxN{}\end{ZX}
        =\begin{ZX}\zxN{}&[\zxWCol] \zxBox{Sq\left(\frac{1}{4ab+1}\right)} \lar[inarrow=>] \rar[inarrow=<] & [\zxWCol] \zxN{}\end{ZX}
    \end{gather}
\end{lem}
\begin{prf*}
    By solving $4abc+a+c=4bcd+b+d=0$ for $a,b$, we obtain
\begin{subequations}
\begin{align}
	 a=&-c(4cd+1),\\
        b=&-
        d (4cd+1)^{-1}.
\end{align}
\end{subequations}
Let us define $\tau=4cd+1$.
Then, these coefficients can be written as follows:
\begin{subequations}
\begin{align}
	a=&-c \tau = \frac{\tau^{-1}-1}{4d\tau^{-2}} = \frac{\tau(1-\tau^{-1})}{4d}\\*
        b=&-\frac{d}{\tau}=d\tau^{-1}\\*
        c=& \frac{\tau-1}{4d}=\frac{1-\tau^{-1}}{4d\tau^{-1}} = \frac{\tau-1}{4d} 
\end{align}%
\label{eq:coeff_abc}%
\end{subequations}%
By comparing Eqs.~\eqref{eq:coeff_abc} with Eqs.~\eqref{eq:sq_k_X} and~\eqref{eq:sq_k_Z}, we obtain the result.
\hfill\qed
\end{prf*}%
\noindent Finally, we have the main theorem for representing Gaussian unitary gates.
\begin{thm}
    Any diagram written as a (finite) chain of quadratic spiders such as
   \begin{equation}
    \begin{ZX}
        \zxN{}&[\zxWCol] \zxX{a_1x^2} \lar[marrow=>] \rar[marrow=<] &[\zxWCol] \zxZ{a_2x^2}\rar[marrow=<]&[\zxWCol] \zxX{a_3x^2} \rar[marrow=<]&[\zxWCol] \zxZ{a_4x^2} \rar[marrow=<]&[\zxWCol] \zxN{}\end{ZX} \cdots \begin{ZX}
        \zxN{}&[\zxWCol] \zxX{a_Nx^2} \lar[marrow=>] \rar[marrow=<] &[\zxWCol] \zxN{}\end{ZX}\label{eq:quad_chain}
\end{equation}
 can be rewritten into at most four quadratic spiders, corresponding to one of the diagrams listed in Table~\ref{tab:table_gaussian_class}.
\end{thm}

\begin{prf*}
    Without loss of generality, we can assume all coefficients are nonzero. Consider a generalized form of the diagram composed of a squeezing part and a chain part as shown here:
    \begin{equation}
    \begin{ZX}
        \zxN{}\rar[marrow=<]&[\zxWCol]\zxBox{Sq(\tau)}\rar[marrow=<]&[\zxWCol] \zxX{a_1x^2}  \rar[marrow=<] &[\zxWCol] \zxZ{a_2x^2}\rar[marrow=<]&[\zxWCol] \zxX{a_3x^2} \rar[marrow=<]&[\zxWCol] \zxZ{a_4x^2}  \rar[inarrow=<]& \zxN{}\end{ZX}  \cdots \begin{ZX}
        \zxN{}\rar[outarrow=<]& \zxX{a_Nx^2} \rar[marrow=<] &[\zxWCol] \zxN{}\end{ZX}\label{eq:sq_quad_chain}
\end{equation}
    By applying the quadratic rule repeatedly to the chain part of \eqref{eq:sq_quad_chain}, the colours of the spiders are reversed (three at a time), and thus they can be merged with neighboring spiders. Based on this approach, the number of spiders of the chain spiders in \eqref{eq:sq_quad_chain} can be reduced to at most three by repeating the following procedures.
    \begin{enumerate}
        \item \label{enu:algo-1}If the first three spiders of the chain part are convertible---i.e.,~$4a_1a_2a_3+a_1+a_3 \neq 0$, then the quadratic rule can be applied to these three, giving
\begin{align}
    &\scalebox{0.9}{$\begin{ZX}
    \zxN{}\rar[marrow=<]&[\zxWCol]\zxBox{Sq(\tau)}\rar[marrow=<]&[\zxWCol] \zxX{a_1x^2}  \rar[marrow=<] &[\zxWCol] \zxZ{a_2x^2}\rar[marrow=<]&[\zxWCol] \zxX{a_3x^2} \rar[marrow=<]&[\zxWCol] \zxZ{a_4x^2}  \rar[inarrow=<]& \zxN{}\end{ZX}  \cdots \begin{ZX}
    \zxN{}\rar[outarrow=<]& \zxX{a_Nx^2} \rar[marrow=<] &[\zxWCol] \zxN{}\end{ZX}$}
    \\\overset{(\hyperlink{li:quad}{q})}{=}
    &\scalebox{0.9}{$\begin{ZX}
    \zxN{}\rar[marrow=<]&[\zxWCol]\zxBox{Sq(\tau)}\rar[marrow=<]&[\zxWCol] \zxZ{a_1'x^2}  \rar[marrow=<] &[\zxWCol] \zxX{a_2'x^2}\rar[marrow=<]&[\zxWCol] \zxZ{a_3'x^2} \rar[marrow=<]&[\zxWCol] \zxZ{a_4x^2}  \rar[inarrow=<]& \zxN{}\end{ZX}  \cdots \begin{ZX}
    \zxN{}\rar[outarrow=<]& \zxX{a_Nx^2}\rar[marrow=<] &[\zxWCol] \zxN{}\end{ZX}$}
    \\\overset{(\hyperlink{li:fusion}{f})}{=}
    &\scalebox{0.9}{$\begin{ZX}
    \zxN{}\rar[marrow=<]&[\zxWCol]\zxBox{Sq(\tau)}\rar[marrow=<]&[\zxWCol] \zxZ{a_1'x^2}  \rar[marrow=<] &[\zxWCol] \zxX{a_2'x^2}\rar[marrow=<]&[\zxWCol] \zxZ{(a_3'+a_4)x^2} \rar[inarrow=<]& \zxN{}\end{ZX}  \cdots \begin{ZX}
    \zxN{}\rar[outarrow=<]& \zxX{a_Nx^2} \rar[marrow=<] &[\zxWCol] \zxN{}\end{ZX}$}
    \end{align}
        where $a_1',a_2',a_3'$ are determined by the quadratic rule. Note the number of spiders in the chain part is reduced by one in the procedure above.
        \item
        \begin{enumerate}
        \item Otherwise,
        \label{enu:algo-2} check whether $4a_2a_3a_4+a_2+a_4\neq 0$ holds or not. If it holds, then the quadratic rule can be applied to the second, third and fourth spiders, and thus the number of spiders can be reduced in the same way as \ref{enu:algo-1}.
        \item If both \ref{enu:algo-1} and \ref{enu:algo-2} fails, then $4a_1a_2a_3+a_1+a_3 =4a_2a_3a_4+a_2+a_4= 0$ holds and thus, due to Lemma.~\ref{thm:sq}, there exists a nonzero real value $\tau'$ such that
        \begin{equation}
        \begin{ZX}\zxN{}&[\zxwCol]\zxX{a_1x^2}\lar[inarrow=>]\rar[marrow=<]&[\zxwCol]\zxZ{a_2x^2}\rar[marrow=<]&[\zxwCol]\zxX{a_3x^2}\rar[marrow=<]&[\zxwCol]\zxZ{a_4x^2}\rar[inarrow=<]&[\zxwCol]\zxN{}\end{ZX}
        =\begin{ZX}\zxN{}&[\zxWCol] \zxBox{Sq(\tau')} \lar[inarrow=>] \rar[inarrow=<] & [\zxWCol] \zxN{}\end{ZX}
    \end{equation}
    is satisfied. In this case, the diagram can be rewritten as
    \begin{align}
        &\begin{ZX}
        \zxN{}\rar[marrow=<]&[\zxWCol]\zxBox{Sq(\tau)}\rar[marrow=<]&[\zxWCol] \zxX{a_1x^2}  \rar[marrow=<] &[\zxWCol] \zxZ{a_2x^2} \rar[inarrow=<]& \zxN{}\end{ZX}  \cdots \begin{ZX}
        \zxN{}\rar[outarrow=<]& \zxX{a_Nx^2} \rar[marrow=<] &[\zxWCol] \zxN{}\end{ZX}
        \\\overset{(\hyperlink{li:quad}{q})}{=}
        &\begin{ZX}
        \zxN{}\rar[marrow=<]&[\zxWCol]\zxBox{Sq(\tau)}\rar[marrow=<]&[\zxWCol]\zxBox{Sq(\tau')}\rar[marrow=<]&[\zxWCol] \zxX{a_5x^2} \rar[inarrow=<]& \zxN{}\end{ZX}  \cdots \begin{ZX}
        \zxN{}\rar[outarrow=<]& \zxX{a_Nx^2}\rar[marrow=<] &[\zxWCol] \zxN{}\end{ZX}
        \\\overset{\eqref{eq:sq_assoc}}{=}
        &\begin{ZX}
        \zxN{}\rar[marrow=<]&[\zxWCol]\zxBox{Sq(\tau\tau')}\rar[marrow=<]&[\zxWCol]\zxX{a_5x^2} \rar[inarrow=<]& \zxN{}\end{ZX}  \cdots \begin{ZX}
        \zxN{}\rar[outarrow=<]& \zxX{a_Nx^2}\rar[marrow=<] &[\zxWCol] \zxN{}\end{ZX}
        \end{align}
\end{enumerate}
and reduction can be performed anyway.
    \end{enumerate}
    Once the number of spiders in the chain part is reduced to 3 or less, then the whole diagram looks like either of the following two:
    \begin{gather}
        \begin{ZX}
        \zxN{}\rar[marrow=<]&[\zxWCol]\zxBox{Sq(\tau)}\rar[marrow=<]&[\zxWCol] \zxZ{c_1x^2}  \rar[marrow=<] &[\zxWCol] \zxX{c_2x^2}\rar[marrow=<]&[\zxWCol] \zxZ{c_3x^2} \rar[inarrow=<]&[\zxWCol] \zxN{}\end{ZX}\label{eq:sq_quad_6_1}
        \\
        \begin{ZX}
        \zxN{}\rar[marrow=<]&[\zxWCol]\zxBox{Sq(\tau)}\rar[marrow=<]&[\zxWCol] \zxX{c_1x^2}  \rar[marrow=<] &[\zxWCol] \zxZ{c_2x^2}\rar[marrow=<]&[\zxWCol] \zxX{c_3x^2} \rar[inarrow=<]&[\zxWCol] \zxN{}\end{ZX}\label{eq:sq_quad_6_2}
    \end{gather}
    where we can assume $c_1=0$ holds if and only if $c_1=c_2=c_3=0$ (and the diagram is equivalent to single squeezing gate). When $c_1\neq 0$, then Eq.~\eqref{eq:sq_quad_6_1} can be written as
    \begin{widetext}
    \begin{align}
    &\begin{ZX}
        &\zxN{}\rar[marrow=<]&[\zxWCol]\zxBox{Sq(\tau)}\rar[marrow=<]&[\zxWCol] \zxZ{c_1x^2}  \rar[marrow=<] &[\zxWCol] \zxX{c_2x^2}\rar[marrow=<]&[\zxWCol] \zxZ{c_3x^2} \rar[inarrow=<]&[\zxWCol] \zxN{}\end{ZX}\\
        \overset{\eqref{eq:sq_k_X}}{=}&
        \begin{ZX}
        \zxN{}&[\zxWCol] \zxX{\frac{\tau(\tau-1)}{4c_1}x^2} \lar[inarrow=>] \rar[marrow=<] &[\zxWCol] \zxZ{\frac{c_1}{\tau}x^2} \rar[marrow=<]&[\zxWCol] \zxX{\frac{1-\tau}{4c_1}x^2} \rar[marrow=<]&[\zxWCol] \zxZ{-c_1x^2} \rar[marrow=<]&[\zxWCol] \zxZ{c_1x^2}  \rar[marrow=<] &[\zxWCol] \zxX{c_2x^2}\rar[marrow=<]&[\zxWCol] \zxZ{c_3x^2} \rar[inarrow=<]&[\zxWCol] \zxN{}
    \end{ZX}\\
        \overset{(\hyperlink{li:fusion}{f})}{=}&
        \begin{ZX}
        \zxN{}&[\zxWCol] \zxX{\frac{\tau(\tau-1)}{4c_1}x^2} \lar[inarrow=>] \rar[marrow=<] &[\zxWCol] \zxZ{\frac{c_1}{\tau}x^2} \rar[marrow=<]&[\zxWCol] \zxX{\left(\frac{1-\tau}{4c_1}+c_2\right)x^2} \rar[marrow=<]&[\zxWCol] \zxZ{c_3x^2} \rar[inarrow=<]&[\zxWCol] \zxN{}
    \end{ZX}
    \end{align}
    \end{widetext}
    and so is Eq.~\eqref{eq:sq_quad_6_2}. It is easy to see that the last diagram is either (1)~equivalent to a squeezing gate or (2)~convertible by the quadratic rule, thereby reducing 
the number of spiders
to at most 3, which must coincide to one of the diagrams specified in Table~\ref{tab:table_gaussian_class} (as the list is complete for 1-mode Gaussian operations).
\hfill \qed
\end{prf*}

This theorem assures any chain of quadratic spiders can be uniquely rewritten into standard form given in Table.~\ref{tab:table_gaussian_class}. As linear phase functions can be driven out to the edge of the diagram using displacement rule~(\hyperlink{li:disp}{d}), this result immediately implies completeness of our rewrite rules over 1-mode Gaussian diagrams written as sequence of at most quadratic spiders. Also, 1-mode spiders are interpreted as unitary operations that are well-defined on the CV Hilbert space. Thus, soundness is 
assured by
the argument above, and we
conclude that the ZX graphical calculus provides a complete characterization of 1-mode Gaussian gates.

By combining this theorem with Eq.~\eqref{eq:quad_state_id}, one can observe that completeness of the model is not only limited to unitary operations but extends to infinitely-squeezed 1-mode Gaussian states and projections as well. Extension of the result to multi-mode diagrams awaits further exploration. Recent work has 
shown that Gaussian quantum mechanics can be treated within a graphical calculus by an axiomization based on the Heisenberg picture~\cite{booth_complete_2024,booth_graphical_2024}. In these works, the authors avoid problems with infinite squeezing by confining the target of 
their investigations
to Gaussian processes and utilizing a
graphical calculus of affine transformations for 
these cases. Although 
that work is applicable only to Gaussian processes and has yet to deal with general CV processes, for the purpose of establishing a CV ZX calculus, it is a promising 
avenue of future research
to integrate these authors' methodologies with our CV ZX calculus model.

\begin{table}[b]
\caption{\label{tab:table_gaussian_class}Complete classification of 1-mode Gaussian unitary gates. Each coefficient is nonzero.}
\begin{ruledtabular}
\begin{tabular}{cc}
Diagram & Representation matrix\\
\colrule
$$\begin{ZX}\zxN{}&[\zxWCol, 5mm] \zxN{}\lar[marrow=>] \end{ZX}$$&$\mat{1&0\\0&1}$\\
$$\begin{ZX}\zxN{}&[\zxWCol] \zxZ{cx^2} \lar[inarrow=>] \rar[marrow=<]&[\zxWCol]\zxN{}\end{ZX}$$&$\mat{1&0\\2c&1}$\\
$$\begin{ZX}\zxN{}&[\zxWCol] \zxX{cx^2} \lar[inarrow=>] \rar[marrow=<]&[\zxWCol]\zxN{}\end{ZX}$$&$\mat{1&2c\\0&1}$\\
$$\begin{ZX}\zxN{}&[\zxWCol] \zxZ{c_1x^2} \lar[inarrow=>] \rar[marrow=<]&[\zxWCol]\zxX{c_2x^2}\rar[marrow=<]&[\zxWCol]\zxN{}\end{ZX}$$&$\mat{1&2c_2\\2c_1&1+4c_1c_2}$\\
$$\begin{ZX}\zxN{}&[\zxWCol]\zxX{c_1x^2}\lar[inarrow=>]\rar[marrow=<]&[\zxWCol]\zxZ{c_2x^2}\rar[marrow=<]&[\zxWCol]\zxN{}\end{ZX}$$ &$\mat{1+4c_1c_2&2c_2\\2c_1&1}$\\
$$\begin{ZX}\zxN{}&[\zxwCol]\zxZ{c_1x^2}\lar[inarrow=>]\rar[marrow=<]&[\zxwCol]\zxX-{\frac{c_1+c_2}{4c_1c_2}x^2}\rar[marrow=<]&[\zxwCol]\zxZ{c_2x^2}\rar[marrow=<]&[\zxwCol]\zxN{}\end{ZX}$$&$\mat{-\frac{c_2}{c_1}&-\frac{c_1+c_2}{c_1c_2}\\0&-\frac{c_1}{c_2}}$\\
$$\begin{ZX}\zxN{}&[\zxwCol]\zxX{c_1x^2}\lar[inarrow=>]\rar[marrow=<]&[\zxwCol]\zxZ-{\frac{c_1+c_2}{4c_1c_2}x^2}\rar[marrow=<]&[\zxwCol]\zxX{c_2x^2}\rar[marrow=<]&[\zxwCol]\zxN{}\end{ZX}$$ &$\mat{-\frac{c_1}{c_2}&0\\-\frac{c_1+c_2}{c_1c_2}&-\frac{c_2}{c_1}}$\\
\begin{ZX}\zxN{}&[\zxWCol]\zxZ{c_1x^2}\lar[inarrow=>]\rar[marrow=<]&[\zxWCol]\zxX{c_2x^2}\rar[marrow=<]&[\zxWCol]\zxZ{c_3x^2}\rar[marrow=<]&[\zxWCol]\zxN{}\end{ZX}&$\mat{1+4c_2c_3&2c_2\\8c_1c_2c_3+2c_1+2c_3&1+4c_1c_2}$\\
\begin{ZX}\zxN{}&[\zxWCol] \zxBox{Sq(\tau)} \lar[inarrow=>] \rar[inarrow=<] & [\zxWCol] \zxN{}\end{ZX}& $\mat{\tau&0\\0&\frac{1}{\tau}}$
\end{tabular}
\end{ruledtabular}
\end{table}

\section{Examples of graphical calculus}


In this section, we provide with several examples to see how our diagrammatic calculus 
serves as a comprehensive tool for graphical reasoning about CV quantum information processing.

\subsection{ Measurement-induced universal squeezer}

\begin{figure}[b]
\includegraphics[width=0.7\linewidth]{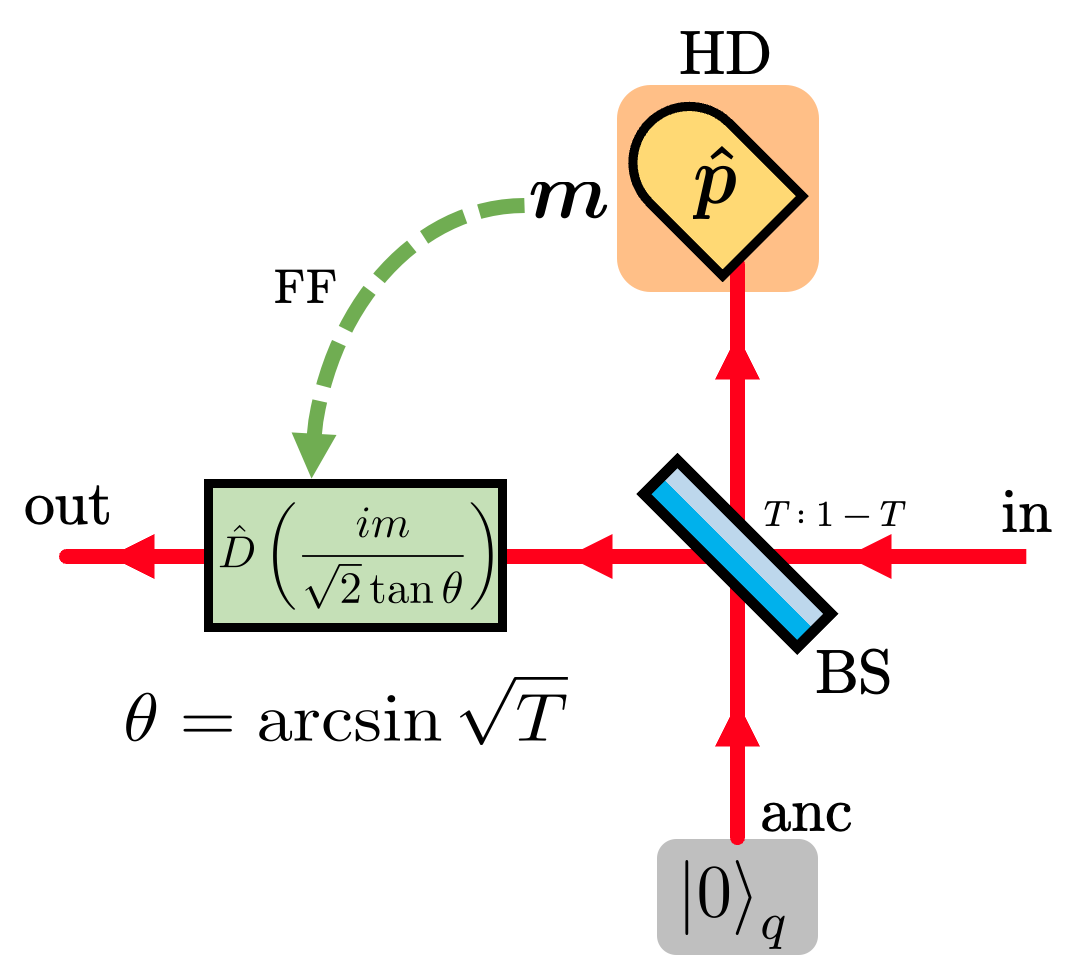}
\caption{\label{fig:uni_sq} Schematic illustration of measurement-induced universal squeezer \cite{filip_measurement-induced_2005}. The measurement outcome of homodyne detection is linearly feedforwarded to subsequent displacement operation. The acronyms are as follows: HD = homodyne detection, BS = beamsplitter, FF = feedforward, anc = ancillary input.}
\end{figure}

The measurement-induced universal squeezer is a device that implements, via measurement, a
squeezing gate with the aid of a squeezed state used as an ancilla 
and 
an unbalanced beamsplitter~%
\cite{filip_measurement-induced_2005,miyata_experimental_2014}. Its (idealised) experimental architecture is illustrated in Fig.~\ref{fig:uni_sq}. Deriving the function of this device is a good example of our graphical formalism.

In this setup, the transmittance~$T$ of the beamsplitter determines the squeezing parameter~$r$ acting on the target mode, where the squeezing gain $e^{-r}$ is equal to $\sqrt{T}=\sin\theta$. This procedure can be treated within the framework of our graphical calculus.

Recall the representation of $\ket 0_q$ [Eq.~\eqref{eq:zeroq}], a beamsplitter [Eq.~\eqref{eq:BStheta}], a displacement gate [Eq.~\eqref{eq:Dgate}], and a measurement in the $p$ basis with outcome~$m$ (which is just projection onto the bra~$\subscripts{p}{\bra{m}}{}$) [first line of Table~\ref{tab:table_gen}, arity~$0 \leftarrow 1$, with $f(x) = -mx$]. Combining these, we can write the diagram corresponding to Fig.~\ref{fig:uni_sq}:
\begin{equation}
    \scalebox{0.9}{
    \begin{ZX}
        [
            execute at end picture={
                \node[draw, rounded corners, fill=orange!50!white, name = hom,
                node on layer=background,
                fit=(\zxGetNameAbsoluteNode{1}{1}),
                label = {[label distance=1mm]above:\footnotesize{HD}},
                inner sep=1mm
                ]{};
                \node[draw, rounded corners, fill=gray!50!white,
                node on layer=background,
                fit=(\zxGetNameAbsoluteNode{6}{9}),
                label = {[label distance=1mm]above:\footnotesize{$\ket{0}_q$}},
                inner sep=1mm
                ]{};
                \draw [color=blue!30!green, ultra thick, dotted,arrows=-Stealth] (hom) to [bend right=40] node[left,black, label={[label distance=1mm,color=black]left:\scriptsize{FF}}]{} (ff);
            }]
    \zxZ{-mx}\ar[r,marrow=<]&[\zxwCol]\zxBox{Sq\left(\frac{1}{\tan\theta}\right)}\ar[drr,-N,inarrow=<]&[\zxwCol]&&[\zxwCol]&[\zxwCol]&[\zxwCol]&[\zxwCol]&[\zxwCol]\\[\zxZeroRow,-2mm]
    &&&\zxZ{} \ar[r,marrow=<]&\zxBox{Sq\left(\frac{\sin^2\theta}{\cos\theta}\right)}\ar[drr,-N,inarrow=<]&\zxN{}&\\[\zxZeroRow,-2mm]
    &&&&&&\zxX{} \ar[r,marrow=<]&\zxBox{Sq\left(\frac{1}{\tan\theta}\right)}\ar[r,marrow=<,"\mathrm{in}" above right]&\zxN{}\\
    \zxN{}\ar[r,marrow=<,"\mathrm{out}" above left]&\zxBox[a=ff]{D\left(\frac{im}{\sqrt{2}\tan\theta}\right)}\ar[r,marrow=<]&\zxX{}\ar[ruu,s.,marrow=<]\ar[drr,N-,inarrow=<]\\[\zxZeroRow]
    &&&&\zxBox{Sq\left(\frac{1}{\cos\theta}\right)}\ar[r,marrow=<]&\zxZ{}\ar[ruu,s.,marrow=<]\ar[drrr,N-,marrow=<,"\mathrm{anc}" near end]\\
    &&&&&&&&\zxX{}
    \end{ZX}}
    \end{equation}
    Now we can use rewrite rules to simplify this down to its essential effect. Starting from the diagram above, we have the following chain of equalities:
    \begin{align}
    &\scalebox{0.9}{\begin{ZX}
    \zxZ{-mx}\ar[r,marrow=<]&[\zxwCol]\zxBox{Sq\left(\frac{1}{\tan\theta}\right)}\ar[drr,-N,inarrow=<]&[\zxwCol]&&[\zxwCol]&[\zxwCol]&[\zxwCol]&[\zxwCol]&[\zxwCol]\\[\zxZeroRow,-2mm]
    &&&\zxZ{} \ar[r,marrow=<]&\zxBox{Sq\left(\frac{\sin^2\theta}{\cos\theta}\right)}\ar[drr,-N,inarrow=<]&\zxN{}&\\[\zxZeroRow,-2mm]
    &&&&&&\zxX{} \ar[r,marrow=<]&\zxBox{Sq\left(\frac{1}{\tan\theta}\right)}\ar[r,marrow=<]&\zxN{}\\
    \zxN{}\ar[r,outarrow=<]&\zxZ{\frac{m}{\tan\theta}}\ar[r,marrow=<]&\zxX{}\ar[ruu,s.,marrow=<]\ar[drr,N-,inarrow=<]\\[\zxZeroRow]
    &&&&\zxBox{Sq\left(\frac{1}{\cos\theta}\right)}\ar[r,marrow=<]&\zxZ{}\ar[ruu,s.,marrow=<]\ar[drrr,N-,inarrow=<]\\
    &&&&&&&&\zxX{}
    \end{ZX}}\label{eq:uni_sq_diagram}\\
    \overset{(\hyperlink{li:copy}{c})}{=}&\scalebox{0.9}{
    \begin{ZX}
    \zxZ{-mx}\ar[r,marrow=<]&[\zxwCol]\zxBox{Sq\left(\frac{1}{\tan\theta}\right)}\ar[drr,-N,inarrow=<]&[\zxwCol]&&[\zxwCol]&[\zxwCol]&[\zxwCol]&[\zxwCol]&[\zxwCol]\\[\zxZeroRow,-2mm]
    &&&\zxZ{} \ar[r,marrow=<]&\zxBox{Sq\left(\frac{\sin^2\theta}{\cos\theta}\right)}\ar[drr,-N,inarrow=<]&\zxN{}&\\[\zxZeroRow,-2mm]
    &&&&&&\zxX{} \ar[r,marrow=<]&\zxBox{Sq\left(\frac{1}{\tan\theta}\right)}\ar[r,marrow=<]&\zxN{}\\
    \zxN{}\ar[r,outarrow=<]&\zxZ{\frac{m}{\tan\theta}}\ar[r,marrow=<]&\zxX{}\ar[ruu,s.,marrow=<]\ar[drr,N-,inarrow=<]&&&\zxX{}\ar[ru,s.,marrow=<]\\[\zxZeroRow]
    &&&&\zxBox{Sq\left(\frac{1}{\cos\theta}\right)}\ar[r,marrow=<]&\zxX{}
    \end{ZX}}
    \\
    \overset{(\hyperlink{li:sq}{s},\hyperlink{li:fusion}{f})}{=}&\scalebox{0.9}{
    \begin{ZX}
    \zxZ{-mx}\ar[r,marrow=<]&[\zxwCol]\zxBox{Sq\left(\frac{1}{\tan\theta}\right)}\ar[drr,-N,inarrow=<]&[\zxwCol]&&[\zxwCol]&[\zxwCol]&[\zxwCol]&[\zxwCol]&[\zxwCol]\\[\zxZeroRow,-2mm]
    &&&\zxZ{} \ar[r,marrow=<]&\zxBox{Sq\left(\frac{\sin^2\theta}{\cos\theta}\right)}\ar[dr,-N,numarrow={<}{0.8}]&\zxN{}&\\[\zxZeroRow,-2mm]
    &&&&&\zxX{} \ar[r,marrow=<]&\zxBox{Sq\left(\frac{1}{\tan\theta}\right)}\ar[r,marrow=<]&\zxN{}\\
    \zxN{}\ar[r,outarrow=<]&\zxZ{\frac{m}{\tan\theta}}\ar[r,marrow=<]&\zxX{}\ar[ruu,s.,marrow=<]
    \end{ZX}}
    \\
    \overset{(\hyperlink{li:id}{id})}{=}&\scalebox{0.9}{
    \begin{ZX}
    \zxZ{-mx}\ar[r,marrow=<]&[\zxwCol]\zxBox{Sq\left(\frac{1}{\tan\theta}\right)}\ar[dr,-N,inarrow=<]&[\zxwCol]&[\zxwCol]&[\zxwCol]&[\zxwCol,1mm]&[\zxwCol,1mm]&[\zxwCol,1mm]\\
    &&\zxZ{} \ar[r,marrow=<]&\zxBox{Sq\left(\frac{\sin^2\theta}{\cos\theta}\right)}\ar[r,marrow=<]&\zxBox{Sq\left(\frac{1}{\tan\theta}\right)}\ar[r,marrow=<]&\zxN{}\\
    \zxN{}\ar[r,outarrow=<]&\zxZ{\frac{m}{\tan\theta}}\ar[ru,-N,inarrow=<]
    \end{ZX}}
    \\
    \overset{(\hyperlink{li:sq}{s})}{=}&\scalebox{0.9}{
    \begin{ZX}
    &[\zxwCol]\zxZ{-\frac{m}{\tan\theta}x}\ar[dr,-N,inarrow=<]&[\zxwCol]&[\zxwCol]&[\zxwCol]&[\zxwCol,1mm]&[\zxwCol,1mm]&[\zxwCol,1mm]\\
    &&\zxZ{} \ar[r,marrow=<]&\zxBox{Sq\left(\frac{\sin^2\theta}{\cos\theta}\right)}\ar[r,marrow=<]&\zxBox{Sq\left(\frac{1}{\tan\theta}\right)}\ar[r,marrow=<]&\zxN{}\\
    \zxN{}\ar[r,outarrow=<]&\zxZ{\frac{m}{\tan\theta}}\ar[ru,-N,inarrow=<]
    \end{ZX}}
    \\
    \overset{(\hyperlink{li:fusion}{f},\hyperlink{li:id}{id})}{=}&\scalebox{0.9}{
    \begin{ZX}
    \zxN{}\ar[r,marrow=<]&[\zxwCol,1mm]\zxBox{Sq\left(\frac{\sin^2\theta}{\cos\theta}\right)}\ar[r,marrow=<]&[\zxwCol,1mm]\zxBox{Sq\left(\frac{1}{\tan\theta}\right)}\ar[r,marrow=<]&[\zxwCol,1mm]\zxN{}
    \end{ZX}}
    \\
    \overset{\eqref{eq:sq_assoc}}{=}&\scalebox{0.9}{\begin{ZX}
    \zxN{}\ar[r,marrow=<]&[\zxwCol,1mm]\zxBox{Sq\left(\sin\theta\right)}\ar[r,marrow=<]&[\zxwCol,1mm]\zxN{}
    \end{ZX}}
    \end{align}
\subsection{Quantum teleportation}

\begin{figure}[b]
\includegraphics[width=0.7\linewidth]{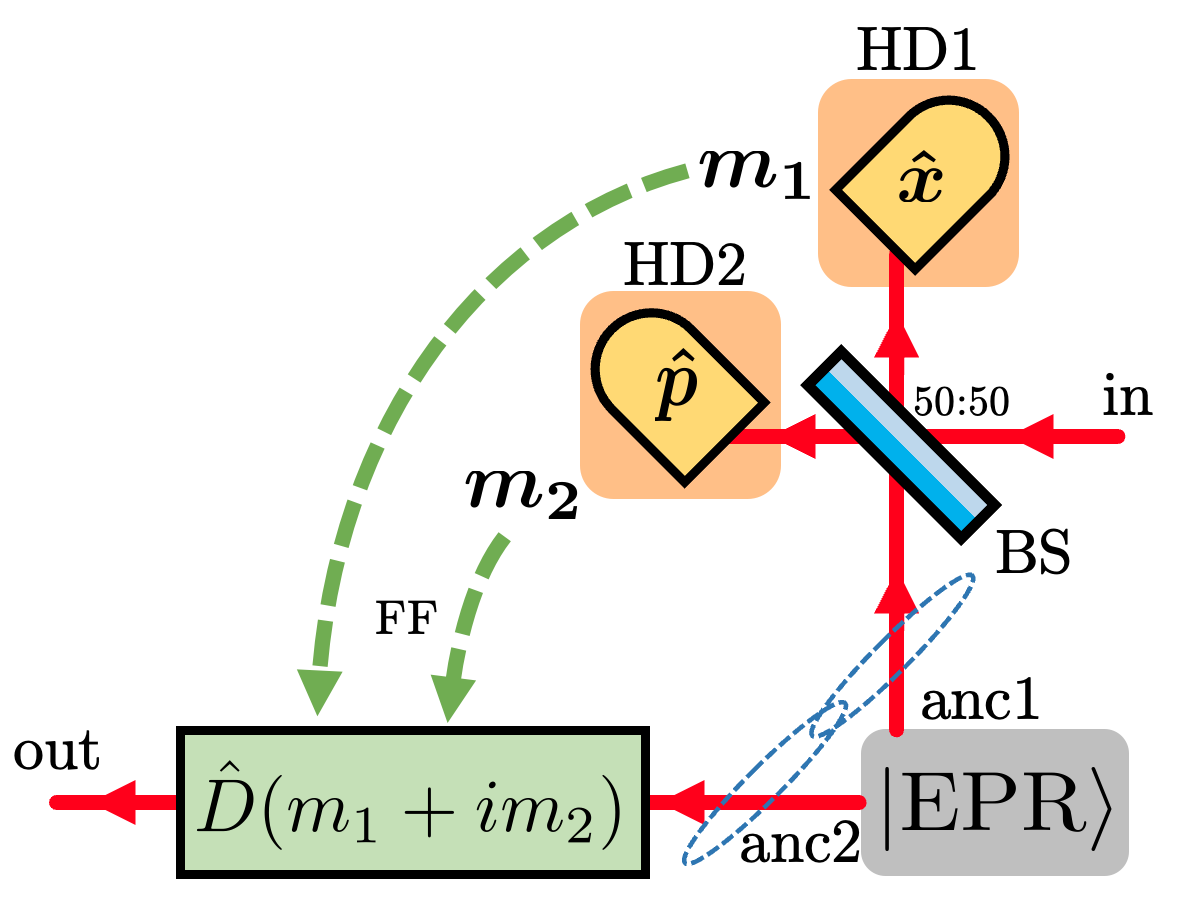}
\caption{\label{fig:q_tele} Schematic illustration of quantum teleportation protocol \cite{braunstein_teleportation_1998}. The measurement outcomes of the two homodyne detections are jointly feedforwarded. The acronyms are as follows: HD = homodyne detection, BS = beamsplitter, FF = feedforward, anc = ancillary input.}
\end{figure}

Among various CV processes, quantum teleportation~\cite{braunstein_teleportation_1998} is regarded as one of the most fundamental
both conceptually and practically and is 
shown Fig.~\ref{fig:q_tele}. In quantum teleportation, performing a joint measurement on an input state and half of an entangled state makes the the input state appear in the remaining system---up to an outcome-dependent displacement that can be corrected by feeding forward the outcomes to the output system and using that data to set the amount of a final displacement.
%

In our graphical calculus, the quantum teleportation experiment in Figure~\ref{fig:q_tele}
can be represented as follows:
\begin{align}
    &\begin{ZX}
        [
        execute at end picture={
        \node[draw, rounded corners, fill=orange!50!white,
        name = hom1,
        node on layer = belownodelayer,
        fit=(\zxGetNameAbsoluteNode{1}{1}),
        label = {[inner sep = 1mm,label distance=1mm]above:\footnotesize{HD1}},
    inner sep=1mm
        ]{};
        \node[draw, rounded corners, fill=orange!50!white,
        name = hom2,
        node on layer = belownodelayer,
        fit=(\zxGetNameAbsoluteNode{4}{1}),
        label = {[fill=white,inner sep = 1mm,label distance=1mm]above:\footnotesize{HD2}},
    inner sep=1mm
        ]{};
        \node[draw, rounded corners, fill=gray!50!white,
        node on layer=background,
        fit=(\zxGetNameAbsoluteNode{5}{6})(\zxGetNameAbsoluteNode{7}{7})(anc1)(anc2),
        label = {[label distance=1mm]below:$\ket{\rm{EPR}}$},
    inner sep=1mm
        ]{};
    \begin{scope}[on background layer]
    \draw [color=blue!30!green, ultra thick, dotted,arrows=-Stealth,node on layer=background] (hom1) to  [bend right=25]  ($(ff)+(-11mm,3mm)$);
    \draw [color=blue!30!green, ultra thick, dotted,arrows=-Stealth] (hom2) to  [bend right=30] node[black, label={[label distance=2mm,color=black]\scriptsize{FF}}]{} ($(ff)+(-11mm,2mm)$);
    \end{scope}
        }]
        \zxX-{m_1x}\ar[rrd,-N,outarrow=<]&[\zxWCol,2mm]&[\zxwCol]&[\zxwCol,1mm]&[\zxwCol,1mm]&[\zxwCol,1mm]&[\zxWCol,2mm]\\[\zxZeroRow]
        &&\zxZ{} \ar[r,marrow=<]&\zxBox{Sq\left(\frac{1}{\sqrt{2}}\right)}\ar[drr,-N,marrow=<]&\zxN{}&\\[\zxZeroRow]
        &&&&&\zxX{} \ar[r,marrow=<,"\mathrm{in}" above right]&\zxN{}\\
        \zxZ-{m_2x}\ar[r,marrow=<]&\zxX{}\ar[ruu,s.,marrow=<]\ar[drr,N-,inarrow=<]\\[\zxZeroRow]
        &&&\zxBox{Sq(\sqrt{2})}\ar[r,marrow=<]&\zxZ{}\ar[ruu,s.,marrow=<]\ar[drr,-N,marrow=<,"\mathrm{anc1}" {above right, name=anc1}]&\\
        &&&&&&\zxZ{}\\
        \zxN{}\ar[rrr,marrow=<,"\mathrm{out}" at start]&&&\zxBox[a=ff]{D(m_1+im_2)}\ar[r]&\zxN{}\ar[urr,-N,outarrow=<,"\mathrm{anc2}" {below right, name=anc2}]&&
        \end{ZX}\\
    \overset{(\hyperlink{li:q_copy}{qc})}{=}&
    \begin{ZX}
    \zxX-{m_1x}\ar[rrd,-N,outarrow=<]&[\zxWCol,2mm]&[\zxwCol]&[\zxwCol,1mm]&[\zxwCol,1mm]&[\zxwCol,1mm]&[\zxWCol,2mm]\\[\zxZeroRow]
    &&\zxZ{} \ar[r,marrow=<]&\zxBox{Sq\left(\frac{1}{\sqrt{2}}\right)}\ar[drr,-N,marrow=<]&\zxN{}&\\[\zxZeroRow]
    &&&&&\zxX{} \ar[r,marrow=<]&\zxN{}\\
    &\zxZ-{m_2x}\ar[ruu,s.,marrow=<]\\[\zxZeroRow]
    &\zxZ-{m_2x}\ar[rr,marrow=<]&&\zxBox{Sq(\sqrt{2})}\ar[r,marrow=<]&\zxZ{}\ar[ruu,s.,marrow=<]\ar[drr,-N,marrow=<]\\
    &&&&&&\zxZ{}\\[\zxwRow,1mm]
    \ar[rr,marrow=<]&&\zxX{\sqrt{2}m_1 x}\rar[marrow=<]&\zxZ{\sqrt{2}m_2 x}\ar[r]&\zxN{}\ar[urr,-N,outarrow=<]&&
    \end{ZX}\\
    \overset{(\hyperlink{li:fusion}{f})}{=}&
    \begin{ZX}
    &[\zxWCol,2mm]\zxX-{m_1x}\ar[r,marrow=<]&[\zxwCol]\zxZ-{m_2x}\ar[r,marrow=<]&[\zxwCol,1mm]\zxBox{Sq\left(\frac{1}{\sqrt{2}}\right)}\ar[drr,-N,marrow=<]&[\zxwCol,1mm]\zxN{}&[\zxwCol,1mm]&[\zxwCol,1mm]\\[\zxZeroRow]
    &&&&&\zxX{} \ar[r,marrow=<]&\zxN{}\\[\zxwRow]
    &\zxZ-{m_2x}\ar[rr,marrow=<]&&\zxBox{Sq(\sqrt{2})}\ar[r,marrow=<]&\zxZ{}\ar[ru,s.,marrow=<]\ar[drr,-N,marrow=<]\\
    &&&&&&\zxZ{}\\[\zxwRow,1mm]
    \ar[rr,marrow=<]&&\zxX{\sqrt{2}m_1 x}\rar[marrow=<]&\zxZ{\sqrt{2}m_2 x}\ar[r]&\zxN{}\ar[urr,-N,outarrow=<]&&
    \end{ZX}\\
    \overset{(\hyperlink{li:q_copy}{qc})}{=}&%
    \begin{ZX}
    &[\zxWCol,2mm]&[\zxwCol]\zxX-{m_1x} \ar[r,marrow=<]&[\zxwCol,1mm]\zxBox{Sq\left(\frac{1}{\sqrt{2}}\right)}\ar[drr,-N,marrow=<]&[\zxwCol,1mm]\zxN{}&[\zxwCol,1mm]&[\zxWCol,2mm]\\
    &&&&&\zxX{} \ar[r,marrow=<]&\zxN{}\\
    &\zxZ-{m_2x}\ar[rr,marrow=<]&&\zxBox{Sq(\sqrt{2})}\ar[r,marrow=<]&\zxZ{}\ar[ru,s.,marrow=<]\ar[drr,-N,marrow=<]\\
    &&&&&&\zxZ{}\\[\zxwRow,1mm]
    \ar[rr,marrow=<]&&\zxX{\sqrt{2}m_1 x}\rar[marrow=<]&\zxZ{\sqrt{2}m_2 x}\ar[r]&\zxN{}\ar[urr,-N,outarrow=<]&&
    \end{ZX}\\
    \overset{(\hyperlink{li:sq}{sq})}{=}&
    \begin{ZX}
    &[\zxWCol,2mm]&[\zxwCol]&[\zxwCol,1mm]\zxX-{\sqrt{2}m_1x}\ar[drr,-N,marrow=<]&[\zxwCol,1mm]\zxN{}&[\zxwCol,1mm]&[\zxWCol,2mm]\\
    &&&&&\zxX{} \ar[r,marrow=<]&\zxN{}\\
    &&&\zxZ-{\sqrt{2}m_2x}\ar[r,marrow=<]&\zxZ{}\ar[ru,s.,marrow=<]\ar[drr,-N,marrow=<]\\
    &&&&&&\zxZ{}\\[\zxwRow,1mm]
    \ar[rr,marrow=<]&&\zxX{\sqrt{2}m_1 x}\rar[marrow=<]&\zxZ{\sqrt{2}m_2 x}\ar[r]&\zxN{}\ar[urr,-N,outarrow=<]&&
    \end{ZX}\\
    \overset{(\hyperlink{li:fusion}{f})}{=}&
    \begin{ZX}
     \zxN{}&[\zxWCol] \zxX{\sqrt{2}m_1x} \lar[marrow=>] \rar[marrow=<] &[\zxWCol] \zxZ{\sqrt{2}m_2x}\rar[marrow=<]&[\zxWCol] \zxZ-{\sqrt{2}m_2x} \rar[marrow=<] &[\zxWCol] \zxX-{\sqrt{2}m_1x}\rar[marrow=<]&[\zxWCol]\zxN{}
    \end{ZX}\\
    \overset{(\hyperlink{li:fusion}{f})}{=}&
    \begin{ZX}
     \zxN{}\rar[marrow=<]&[\zxWCol,5mm] \zxN{}
    \end{ZX}
    \end{align}
where $m_1$ and $m_2$ denotes each homodyne measurement outcome. As the last line of the calculus implies, the whole process of quantum teleportation is equivalent to the identity operation.

\subsection{Cubic state injection}

Cubic state injection is 
one possible way to achieve 
universal quantum computation in an MBQC scheme, as it 
introduces non-Gaussianity of an ancilliary state to the system only with Gaussian operations by implementing the cubic phase gate onto an arbitrary input state. In an MBQC protocol, it is necessary to conduct feedforward operations so that errors induced by previous measurement outcomes are properly corrected. In general, when implementing $n^{\text{th}}$ order operations, one needs $(n-1)^{\text{th}}$ order feedforward to eliminate the effect of projection onto the displaced quadrature eigenstates~\cite{marek_general_2018}.

\begin{figure*}
\includegraphics[width=.8\linewidth]{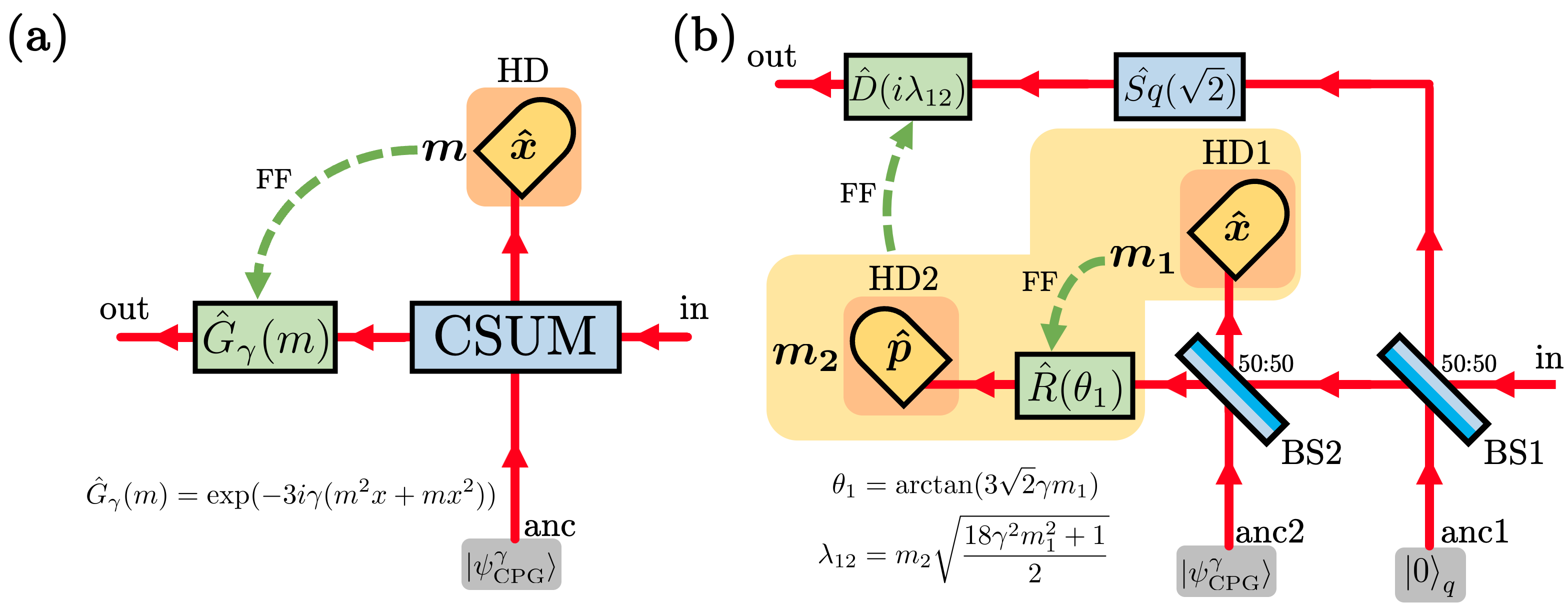}
\caption{\label{fig:cubic_inj}Schematic illustration of (a) simplified cubic state injection and (b) its practical implementation with passive optical components and nonlinear feedforward \cite{miyata_implementation_2016}. Green-colored gates represent feedforwarded operations with tunable parameters and blue-colored gates represent constant opearations. (a) The operation $\hat{G}_\gamma$ is determined by the outcome of homodyne detection. (b) The first homodyne measurement outcome determines the measurement angle of the second homodyne detection. Then the two measurement outcomes are jointly feedforwarded to the displacement operation in the last step. 
The acronyms are as follows: HD = homodyne detection, BS = beamsplitter, FF = feedforward, anc = ancillary input.}
\end{figure*}

The simplest implementation of cubic state injection is based on gate teleportation~\cite{bartlett_quantum_2003}. As illustrated on the left side of Fig.~\ref{fig:cubic_inj}, this consists of a CSUM operation, an ancillary cubic phase state
\begin{equation}
\label{eq:CPSdef}
	\ket{\psi_\mathrm{CPG}^\gamma}\coloneqq e^{i\gamma \hat{q}^3}\ket{0}_p=\frac {1} {\sqrt{2\pi}} \int_\mathbb{R}\dd{s} e^{i\gamma s^3} \ket{s}_q,
\end{equation}
a homodyne measurement, and a tunable Gaussian gate $\hat{G}_\gamma(m)$ defined as
\begin{equation}
    \label{eq:Gdef}
        \hat{G}_\gamma(m)=\exp(-3i\gamma(m^2x+mx^2))
    \end{equation}
where $m$ denotes the homodyne measurement outcome.

In our graphical calculus model, the total process of cubic state injection can be graphically calculated as follows:
\begin{align}
    &\begin{ZX}[
    execute at end picture={
    \node[draw, rounded corners, fill=orange!50!white,
    name = hom,
    node on layer=background,
    fit=(\zxGetNameAbsoluteNode{4}{2}),
    label = {[label distance=1mm]below:\footnotesize{HD}},
    inner sep=1mm
    ]{};
    \node[draw, rounded corners, fill=gray!50!white,
    node on layer=background,
    fit=(\zxGetNameAbsoluteNode{3}{5})(anc),
    label = {[label distance=1mm]below:\footnotesize{$\ket{\psi_\mathrm{CPG}^\gamma}$}},
    inner sep=1mm
    ]{};
    \draw [color=blue!30!green, ultra thick, dotted,arrows=-Stealth] (hom) to node[left,black, label={[label distance=1mm,color=black]left:\scriptsize{FF}}]{} (ff);
    }]
    &[\zxwCol,1mm]&[\zxwCol,1mm]&[\zxwCol]&[\zxwCol,5mm]\zxN{}\\
    \zxN{}\ar[r,marrow=<,"\mathrm{out}" above left]&\zxZ[a=ff]-{3\gamma(mx^2+m^2x)}\ar[r,marrow=<]&\zxZ{}\ar[urr,N-,marrow=<,"\mathrm{in}" at end]\ar[rd,s.,marrow=<]\\[\zxWRow]
    &&&\zxX{}\ar[r,marrow=<, "\mathrm{anc}" {near end, name=anc}]&\zxZ{\gamma x^3}\\
    &\zxX{-mx}\ar[urr,-N,marrow=<]&&&
    \end{ZX}
    \\
    \overset{(\hyperlink{li:fusion}{f})}{=}&
    \begin{ZX}
    &[\zxwCol,1mm]&[\zxwCol,1mm]&[\zxwCol]&[\zxwCol,1mm]\zxN{}\\
    \zxN{}\ar[r,marrow=<]&\zxZ-{3\gamma(mx^2+m^2x)}\ar[r,marrow=<]&\zxZ{}\ar[urr,N-,marrow=<]\ar[dr,N-,marrow=<]\\
    &&&\zxX{-mx}\ar[r,marrow=<]&\zxZ{\gamma x^3}
    \end{ZX}
    \\
    \overset{(\hyperlink{li:disp}{d})}{=}&
    \begin{ZX}
    &[\zxwCol,1mm]&[\zxwCol,1mm]&[\zxwCol]&[\zxwCol,1mm]\zxN{}\\
    \zxN{}\ar[r,marrow=<]&\zxZ-{3\gamma(mx^2+m^2x)}\ar[r,marrow=<]&\zxZ{}\ar[urr,N-,marrow=<]\ar[dr,N-,marrow=<]\\
    &&&\zxZ{\gamma (x+m)^3}
    \end{ZX}
    \\
    \overset{(\hyperlink{li:fusion}{f})}{=}&
    \begin{ZX}
    \zxN{}\ar[r,marrow=<]&[\zxwCol,1mm]\zxZ{\gamma x^3+\gamma m^3}\ar[r,marrow=<]&[\zxwCol,1mm]\zxN{}
    \end{ZX}
    \\
    \overset{(\hyperlink{li:id}{id})}{=}&
    \begin{ZX}
    \zxN{}\ar[r,marrow=<]&[\zxwCol,1mm]\zxZ{\gamma x^3}\ar[r,marrow=<]&[\zxwCol,1mm]\zxN{}
    \end{ZX}
    \end{align}
Thus, by operating the Gaussian feedforward operation of $\hat{G}_\gamma(m)$ after the measurement, one can apply the cubic phase gate of arbitrary $\gamma$ by simply preparing the ancilla~$\ket{\psi_\mathrm{CPG}^\gamma}$ [Eq.~\ref{eq:CPSdef}] (or a suitable approximation thereof).

Although this protocol is straightforward, it has two drawbacks in practice.
First, implementing the CSUM gate is itself
a demanding task 
in optical 
systems \cite{shiozawa_quantum_2018} due to experimental constraints. Second, this scheme requires 
a conditional
shear operation for feedforward, which is also 
challenging
to realize, especially when aimed at ultra-fast computation. In order to overcome these difficulties and 
make the implementation practical,
it is necessary to construct a system with feasible operations by an appropriate decomposition.

Such optical implementation of the 
cubic phase gate has been proposed in Ref.~\cite{miyata_implementation_2016} and recently demonstrated in experiment~\cite{sakaguchi_nonlinear_2023}. In this scheme, as is shown on the right side of Fig.~\ref{fig:cubic_inj}, the CSUM gate is replaced with two beamsplitters and one additional homodyne measurement, and the feedforward is operated by adaptive homodyne measurement and displacement. The rotation angle before the second homodyne measurement $\theta_1$ is determined by the first measurement outcome $m_1$ as
\begin{equation}
    \theta_1 = \arctan(3\sqrt{2}\gamma m_1)\label{eq:def_gamma}
\end{equation}
to cancel out the effect of displacement due to the previous homodyne measurement. Effectively, the rotation operation is equivalent to altering the angle of the measured basis $\hat{q}_{\theta}$ by $-\theta_1$. After obtaining the second measurement outcome $m_2$, the output mode is squeezed by a constant factor of $\sqrt{2}$, and then a feedforwarded displacement follows. The displacement amount in the last step $\lambda_{12}$ is given by the following equation
\begin{equation}
    \lambda_{12}=m_2\sqrt{\frac{18\gamma^2 m_1^2+1}{2}}\label{eq:def_d12}
\end{equation}
so that the unwanted displacement introduced by the two homodyne measurement is eliminated. As the constant squeezing operation can be accomplished by the measurement-induced scheme, the entire cubic phase gate protocol is realizable only with passive optical components and ancilliary states, which is a great advantage for experimental implementation.


Again, one can follow its dynamics using our graphical calculus. The protocol can be understood as follows:
\begin{widetext}
    \begin{align}
        &\begin{ZX}[
            execute at end picture={
            \node[draw, rounded corners, fill=orange!50!white,
            name = hom1,
            node on layer=belownodelayer,
            fit=(\zxGetNameAbsoluteNode{9}{3}),
            label = {[label distance=1mm,name=hd1]below:\footnotesize{HD1}},
        inner sep=1mm
            ]{};
            \node[draw, rounded corners, fill=orange!50!white,
            name = hom2,
            node on layer=belownodelayer,
            fit=(\zxGetNameAbsoluteNode{6}{2}),
            label = {[label distance=1mm]below:\footnotesize{HD2}},
        inner sep=1mm
            ]{};
            \node[draw, rounded corners, fill=myyellow,
            name = hom,
            node on layer=background,
            fit=(hom1)(hom2)(\zxGetNameAbsoluteNode{6}{3})(hd1),
        inner sep=1mm
            ]{};
            \node[draw, rounded corners, fill=gray!50!white,
            node on layer=background,
            fit=(\zxGetNameAbsoluteNode{6}{13})(anc1),
            label = {[label distance=1mm]right:$\ket{0}_q$},
        inner sep=1mm
            ]{};
            \node[draw, rounded corners, fill=gray!50!white,
            node on layer=background,
            fit=(\zxGetNameAbsoluteNode{7}{13})(anc2),
            label = {[label distance=1mm]below:$\ket{\psi_\mathrm{CPG}^\gamma}$},
        inner sep=1mm
            ]{};
            \draw [color=blue!30!green, ultra thick, dotted,arrows=-Stealth,] (hom1) to node[right,black, label={[label distance=1mm,color=black]left:\scriptsize{FF}}]{} (rot);
        \draw [color=blue!30!green, ultra thick, dotted,arrows=-Stealth] (hom) to  [bend left=30] node[left,black, label={[label distance=1mm,color=black]left:\scriptsize{FF}}]{} (ff);
            }]
        \zxN{}\ar[rr,marrow=<, "\mathrm{out}" {pos=0.1}]&&[\zxwCol,1mm]\zxBox[a=ff]{D(i\lambda_{12})}\ar[rrr,marrow=<]&[\zxwCol,1mm]&[\zxwCol,1mm]&[\zxwCol,1mm]\zxBox{Sq(\sqrt{2})}\ar[rrrd,-N,inarrow=<]&[\zxwCol,1mm]&[\zxWCol]&[\zxwCol]&[\zxwCol,1mm]&[\zxwCol,1mm]&[\zxwCol,1mm]&[\zxWCol,2mm]\\[\zxZeroRow]
        &&&&&&&&\zxZ{} \ar[r,marrow=<]&\zxBox{Sq\left(\frac{1}{\sqrt{2}}\right)}\ar[drr,-N,marrow=<]&\zxN{}&\\[\zxZeroRow]
        &&&&&&&&&&&\zxX{} \ar[r,marrow=<, "\mathrm{in}" {pos=0.9}]&\zxN{}\\
        &&&&&&&\zxX{}\ar[ruu,s.,marrow=<]\ar[drr,N-,inarrow=<]\\[\zxZeroRow]
        &&&&&\zxBox{Sq(\sqrt{2})}\ar[r,marrow=<]&\zxZ{}\ar[ur,N-,marrow=<]\ar[rdd,s.,marrow=<]&&&\zxBox{Sq(\sqrt{2})}\ar[r,marrow=<]&\zxZ{}\ar[ruu,s.,marrow=<]\ar[drr,N-,outarrow=<, "\mathrm{anc1}" {pos=0.6, name=anc1}]\\[\zxZeroRow,-2mm]
        &\zxZ-{m_2x}\ar[r,marrow=<]&\zxBox[a=rot]{R(\theta_1)}\ar[r,marrow=<]&\zxX{}\ar[rdd,s.,marrow=<]\ar[urr,N-,outarrow=<]&&&&&&&&&\zxX{}\\
        &&&&&&&\zxX{}\ar[rrrrr,s.,marrow=<, "\mathrm{anc2}" {pos=0.93, name=anc2}]&&&&&\zxZ{\gamma x^3}\\[\zxZeroRow,-2mm]
        &&&&\zxZ{}\ar[r,marrow=<]&\zxBox{Sq\left(\frac{1}{\sqrt{2}}\right)}\ar[urr,-N,inarrow=<]\\[\zxZeroRow]
        &&\zxX-{m_1x}\ar[urr,-N,marrow=<]
        \end{ZX}\\
        \overset{(\hyperlink{li:q_copy}{qc})}{=}
        &
        \begin{ZX}
        \zxN{}\ar[rr,marrow=<]&&[\zxwCol,1mm]\zxZ[a=ff]{\sqrt{2}\lambda_{12}x}\ar[rrr,marrow=<]&[\zxwCol,1mm]&[\zxwCol,1mm]&[\zxwCol,1mm]\zxBox{Sq(\sqrt{2})}\ar[rrrd,-N,inarrow=<]&[\zxwCol,1mm]&[\zxWCol]&[\zxwCol]&[\zxwCol,1mm]&[\zxwCol,1mm]&[\zxwCol,1mm]&[\zxWCol,2mm]\\[\zxZeroRow]
        &&&&&&&&\zxZ{} \ar[r,marrow=<]&\zxBox{Sq\left(\frac{1}{\sqrt{2}}\right)}\ar[drr,-N,marrow=<]&\zxN{}&\\[\zxZeroRow]
        &&&&&&&&&&&\zxX{} \ar[r,marrow=<]&\zxN{}\\
        &&&&&&&\zxX{}\ar[ruu,s.,marrow=<]\ar[drr,N-,inarrow=<]&&&\zxX{}\ar[ru,s.,marrow=<]\\[\zxZeroRow]
        &&&&&\zxBox{Sq(\sqrt{2})}\ar[r,marrow=<]&\zxZ{}\ar[ur,N-,marrow=<]\ar[rdd,s.,marrow=<]&&&\zxBox{Sq(\sqrt{2})}\ar[r,marrow=<]&\zxX{}\\[\zxZeroRow,-2mm]
        &\zxZ-{m_2x}\ar[r,marrow=<]&\zxBox{R(\theta_1)}\ar[r,marrow=<]&\zxX{}\ar[rd,s.,marrow=<]\ar[urr,N-,outarrow=<]\\
        &&&&\zxX{m_1x}&&&\zxX{}\ar[rrrrr,s.,marrow=<]&&&&&\zxZ{\gamma x^3}\\[\zxZeroRow]
        &&&&\zxX-{m_1x}\ar[r,marrow=<]&\zxBox{Sq\left(\frac{1}{\sqrt{2}}\right)}\ar[urr,-N,inarrow=<]
        \end{ZX}\\
        \overset{(\hyperlink{li:sq}{s},\hyperlink{li:fusion}{f})}{=}
        &
        \begin{ZX}
        \zxN{}\ar[rr,marrow=<]&&[\zxwCol,1mm]\zxZ[a=ff]{\sqrt{2}\lambda_{12}x}\ar[rrr,marrow=<]&[\zxwCol,1mm]&[\zxwCol,1mm]&[\zxwCol,1mm]\zxBox{Sq(\sqrt{2})}\ar[rrrd,-N,inarrow=<]&[\zxwCol,1mm]&[\zxWCol]&[\zxwCol]&[\zxwCol,1mm]&[\zxwCol,1mm]&[\zxwCol,1mm]&[\zxWCol,2mm]\\[\zxZeroRow]
        &&&&&&&&\zxZ{} \ar[r,marrow=<]&\zxBox{Sq\left(\frac{1}{\sqrt{2}}\right)}\ar[rr,marrow=<]&&\zxN{}\\
        &&&&&\zxBox{Sq(\sqrt{2})}\ar[r,marrow=<]&\zxZ{}\ar[rdd,s.,marrow=<]\ar[urr,s.,marrow=<]\\[\zxZeroRow,-2mm]
        &\zxZ-{m_2x}\ar[r,marrow=<]&\zxBox{R(\theta_1)}\ar[r,marrow=<]&\zxX{m_1x}\ar[urr,N-,marrow=<]\\
        &&&&&&&\zxX-{\sqrt{2}m_1x}\ar[rrrrr,s.,marrow=<]&&&&&\zxZ{\gamma x^3}
        \end{ZX}\\
        \overset{(\hyperlink{li:fusion}{f})}{=}
        &
        \begin{ZX}
        \zxN{}\ar[rr,marrow=<]&[\zxwCol,1mm]&[\zxwCol,1mm]\zxZ[a=ff]{\sqrt{2}\lambda_{12}x}\ar[rr,marrow=<]&[\zxwCol,1mm]&[\zxwCol,1mm]\zxBox{Sq(\sqrt{2})}\ar[rd,-N,marrow=<]&[\zxwCol,5mm]&[\zxwCol,5mm]\zxBox{Sq\left(\frac{1}{\sqrt{2}}\right)}\ar[r,marrow=<]&[\zxwCol,1mm]\zxN{}&[\zxwCol,1mm]&[\zxwCol,1mm]&[\zxwCol,1mm]\\
        &&&&&\zxZ{}\ar[rd,N-,marrow=<]\ar[ru,N-,marrow=<]&&&&&\\
        &\zxZ-{m_2x}\ar[r,marrow=<]&\zxBox{R(\theta_1)}\ar[r,marrow=<]&\zxX{m_1x}\ar[r,marrow=<]&\zxBox{Sq(\sqrt{2})}\ar[ru,-N,marrow=<]&&\zxX-{\sqrt{2}m_1x}\ar[r,marrow=<]&\zxZ{\gamma x^3}
        \end{ZX}\\
            \overset{(\hyperlink{li:fusion}{f})}{=}
            &
            \begin{ZX}
            \zxN{}\ar[rr,marrow=<]&[\zxwCol,1mm]&[\zxwCol,1mm]\zxZ[a=ff]{\sqrt{2}\lambda_{12}x}\ar[rr,marrow=<]&[\zxwCol,1mm]&[\zxwCol,1mm]\zxBox{Sq(\sqrt{2})}\ar[rd,-N,inarrow=<]&[\zxwCol,5mm]&[\zxwCol,5mm]\zxBox{Sq\left(\frac{1}{\sqrt{2}}\right)}\ar[r,marrow=<]&[\zxwCol,1mm]\zxN{}&[\zxwCol,1mm]&[\zxwCol,1mm]&[\zxwCol,1mm]\\
            &&&&&\zxZ{}\ar[rd,N-,outarrow=<]\ar[ru,N-,outarrow=<]&&&&&\\
            &\zxZ-{m_2x}\ar[r,marrow=<]&\zxBox{R(\theta_1)}\ar[r,marrow=<]&\zxX{m_1x}\ar[r,marrow=<]&\zxBox{Sq(\sqrt{2})}\ar[ru,-N,inarrow=<]&&\zxX-{\sqrt{2}m_1x}\ar[r,marrow=<]&\zxZ{\gamma x^3}
            \end{ZX}\\
            \overset{(\hyperlink{li:sq}{s})}{=}
            &
            \begin{ZX}
            \zxN{}\ar[rr,marrow=<]&[\zxwCol,1mm]&[\zxwCol,1mm]\zxZ[a=ff]{\sqrt{2}\lambda_{12}x}\ar[r]&[\zxwCol,1mm]\zxN{}\ar[rd,-N,inarrow=<]&[\zxwCol,5mm]&[\zxwCol,5mm]\zxN{}\ar[rr]&[\zxwCol,1mm]&[\zxwCol,1mm]\zxN{}\\[\zxwRow]
            &&&&\zxZ{}\ar[rd,N-,outarrow=<]\ar[ru,N-,outarrow=<]\\[\zxwRow]
            &\zxZ-{m_2x}\ar[r,marrow=<]&\zxBox{R(\theta_1)}\ar[r,marrow=<]&\zxX{m_1x}\ar[ru,-N,inarrow=<]&&\zxBox{Sq(\sqrt{2})}\ar[r,marrow=<]&\zxX-{\sqrt{2}m_1x}\ar[r,marrow=<]&\zxZ{\gamma x^3}
            \end{ZX}\\\overset{(\hyperlink{li:sq}{s})}{=}
            &
            \begin{ZX}
            \zxN{}\ar[rr,marrow=<]&[\zxwCol,1mm]&[\zxwCol,1mm]\zxZ[a=ff]{\sqrt{2}\lambda_{12}x}\ar[r]&[\zxwCol,1mm]\zxN{}\ar[rd,-N,inarrow=<]&[\zxwCol,5mm]&[\zxwCol,5mm]\zxN[a=mid,inner sep=-1pt]{}\ar[rr]&[\zxwCol,1mm]&[\zxwCol,1mm]\zxN{}\\[\zxwRow]
            &&&&\zxZ{}\ar[rd,N-,outarrow=<]\ar[ru,N-,outarrow=<]\\[\zxwRow]
            &\zxZ-{m_2x}\ar[r,marrow=<]&\zxBox{R(\theta_1)}\ar[r,marrow=<]&\zxX{m_1x}\ar[ru,-N,inarrow=<]&&\zxX-{2m_1x}\ar[r,marrow=<]&\zxZ{\frac{\gamma}{2\sqrt{2}} x^3}
            \end{ZX}\\\overset{(\hyperlink{li:sq}{s})}{=}
            &
            \begin{ZX}
            \zxN{}\ar[rr,marrow=<]&[\zxwCol,1mm]&[\zxwCol,1mm]\zxZ[a=ff]{\sqrt{2}\lambda_{12}x}\ar[r]&[\zxwCol,1mm]\zxN{}\ar[rd,-N,inarrow=<]&[\zxwCol,5mm]&[\zxwCol,5mm]\zxN[a=mid,inner sep=-1pt]{}\ar[rr]&[\zxwCol,1mm]&[\zxwCol,1mm]\zxN{}\\[\zxwRow]
            &&&&\zxZ{}\ar[rd,N-,outarrow=<]\ar[ru,N-,outarrow=<]\\[\zxwRow]
            &\zxZ-{m_2x}\ar[r,marrow=<]&\zxBox{R(\theta_1)}\ar[r,marrow=<]&\zxX{m_1x}\ar[ru,-N,inarrow=<]&&\zxX-{2m_1x}\ar[r,marrow=<]&\zxZ{\frac{\gamma}{2\sqrt{2}} x^3}
            \end{ZX}\label{eq:disp_rot_com_1}\\\overset{\mathclap{\eqref{eq:rot_meas}}}{=}
            &
            \begin{ZX}
            \zxN{}\ar[rr,marrow=<]&[\zxwCol,1mm]&[\zxwCol,1mm]\zxZ[a=ff]{\sqrt{2}\lambda_{12}x}\ar[r]&[\zxwCol,1mm]\zxN{}\ar[rd,-N,inarrow=<]&[\zxwCol,5mm]&[\zxwCol,5mm]\zxN[a=mid,inner sep=-1pt]{}\ar[rr]&[\zxwCol,1mm]&[\zxwCol,1mm]\zxN{}\\[\zxwRow]
            &&&&\zxZ{}\ar[rd,N-,outarrow=<]\ar[ru,N-,outarrow=<]\\[\zxwRow]
            &&\zxZ{-\frac{\tan\theta_1}{2}x^2-\frac{m_2}{\cos\theta_1}x}\ar[r,marrow=<]&\zxX{m_1x}\ar[ru,-N,inarrow=<]&&\zxX-{2m_1x}\ar[r,marrow=<]&\zxZ{\frac{\gamma}{2\sqrt{2}} x^3}
            \end{ZX}\label{eq:disp_rot_com_2}\\\overset{(\hyperlink{li:disp}{d})}{=}
            &
            \begin{ZX}
            \zxN{}\ar[r,marrow=<]&[\zxwCol,1mm]\zxZ[a=ff]{\sqrt{2}\lambda_{12}x}\ar[rd,-N,inarrow=<]&[\zxwCol,5mm]&[\zxwCol,5mm]\zxN[a=mid,inner sep=-1pt]{}\ar[rr]&[\zxwCol,1mm]&[\zxwCol,1mm]\zxN{}\\[\zxwRow]
            &&\zxZ{}\ar[rd,N-,outarrow=<]\ar[ru,N-,outarrow=<]\\[\zxwRow]
            &\zxZ{-\frac{\tan\theta_1}{2}(x+m_1)^2-\frac{m_2}{\cos\theta_1}(x+m_1)}\ar[ru,-N,inarrow=<]&&\zxZ{\frac{\gamma}{2\sqrt{2}} (x+2m_1)^3}
            \end{ZX}\\\overset{(\hyperlink{li:fusion}{f})}{=}
            &
            \begin{ZX}
            \zxN{}\ar[r,marrow=<]&[\zxwCol,1mm]\zxZ{\frac{\gamma}{2\sqrt{2}} (x+2m_1)^3-\frac{\tan\theta_1}{2}(x+m_1)^2-\frac{m_2}{\cos\theta_1}(x+m_1)+\sqrt{2}\lambda_{12}x}\ar[r,marrow=<]&[\zxwCol,1mm]\zxN{}
            \end{ZX}\\
            =&
            \begin{ZX}
            \zxN{}\ar[r,marrow=<]&[\zxwCol,1mm]\zxZ{\frac{\gamma}{2\sqrt{2}} x^3}\ar[r,marrow=<]&[\zxwCol,1mm]\zxN{}
            \end{ZX}
            \end{align}
\end{widetext}
The last equation directly follows from the definitions (\ref{eq:def_gamma},\ref{eq:def_d12}). 
As can be seen above, the whole process is equivalent to a cubic phase gate with $\gamma'=\frac{\gamma}{\sqrt{2}}$.
Here we used the following commutation property to show the equivalence between \eqref{eq:disp_rot_com_1} and \eqref{eq:disp_rot_com_2}:
\begin{widetext}
\begin{subequations}
\begin{align}
&\begin{ZX}
\zxN{}\ar[r,marrow=<]&[\zxwCol,1mm]\zxBox{R(\theta)}\ar[r,marrow=<]&[\zxwCol,1mm]\zxZ{ax}\ar[r,marrow=<]&[\zxwCol,1mm]\zxN{}
\end{ZX}\\
\overset{\mathclap{\eqref{def:rot_sp}}}{=}&
\begin{ZX}
        \zxN{} \rar[marrow=<]&[\zxwCol,1mm] \zxX{\frac{\tan(\theta/2)}{2}x^2}  \rar[marrow=<] &[\zxwCol,1mm] \zxZ-{\frac{\sin\theta}{2}x^2} \rar[marrow=<]&[\zxwCol,1mm] \zxX{\frac{\tan(\theta/2)}{2}x^2} \ar[r,marrow=<]&[\zxwCol,1mm]\zxZ{ax}\ar[r,marrow=<]&[\zxwCol,1mm]\zxN{}
\end{ZX}\\
\overset{(\hyperlink{li:disp}{d})}{=}&
\begin{ZX}
        \zxN{} \rar[marrow=<]&[\zxwCol,1mm] \zxX{\frac{\tan(\theta/2)}{2}x^2}  \rar[marrow=<] &[\zxwCol,1mm] \zxZ-{\frac{\sin\theta}{2}x^2} \rar[marrow=<]&[\zxwCol,1mm]\zxZ{ax}\ar[r,marrow=<]&[\zxwCol,1mm] \zxX{\frac{\tan(\theta/2)}{2}(x+a)^2} \ar[r,marrow=<]&[\zxwCol,1mm]\zxN{}
\end{ZX}\\
\overset{\mathclap{(\hyperlink{li:fusion}{f},\hyperlink{li:id}{id})}}{=}&
\begin{ZX}
        \zxN{} \rar[marrow=<]&[\zxwCol,1mm] \zxX{\frac{\tan(\theta/2)}{2}x^2}  \rar[marrow=<] &[\zxwCol,1mm] \zxZ-{\frac{\sin\theta}{2}x^2} \rar[marrow=<]&[\zxwCol,1mm]\zxZ{ax}\ar[r,marrow=<]&[\zxwCol,1mm] \zxX{a\tan(\theta/2) x} \ar[r,marrow=<]&[\zxwCol,1mm] \zxX{\frac{\tan(\theta/2)}{2}x^2} \ar[r,marrow=<]&[\zxwCol,1mm]\zxN{}
\end{ZX}\\
\overset{(\hyperlink{li:fusion}{f})}{=}&
\begin{ZX}
        \zxN{} \rar[marrow=<]&[\zxwCol,1mm] \zxX{\frac{\tan(\theta/2)}{2}x^2}  \rar[marrow=<] &[\zxwCol,1mm]\zxZ{ax}\ar[r,marrow=<]&[\zxwCol,1mm] \zxZ-{\frac{\sin\theta}{2}x^2} \rar[marrow=<]&[\zxwCol,1mm] \zxX{a\tan(\theta/2) x} \ar[r,marrow=<]&[\zxwCol,1mm] \zxX{\frac{\tan(\theta/2)}{2}x^2} \ar[r,marrow=<]&[\zxwCol,1mm]\zxN{}
\end{ZX}\\
\overset{(\hyperlink{li:disp}{d})}{=}&\begin{ZX}
        \zxN{} \rar[marrow=<]&[\zxwCol,1mm] \zxX{\frac{\tan(\theta/2)}{2}x^2}  \rar[marrow=<]&[\zxwCol,1mm]\zxZ{ax}\ar[r,marrow=<]&[\zxwCol,1mm] \zxX{a\tan(\theta/2) x} \ar[r,marrow=<] &[\zxwCol,1mm] \zxZ-{\frac{\sin\theta}{2}(x+a\tan(\theta/2))^2} \rar[marrow=<]&[\zxwCol,1mm] \zxX{\frac{\tan(\theta/2)}{2}x^2} \ar[r,marrow=<]&[\zxwCol,1mm]\zxN{}
\end{ZX}\\
\overset{\eqref{eq:d_com}}{=}&\begin{ZX}
        \zxN{} \rar[marrow=<]&[\zxwCol,1mm] \zxX{\frac{\tan(\theta/2)}{2}x^2}  \rar[marrow=<]&[\zxwCol,1mm] \zxX{a\tan(\theta/2) x} \ar[r,marrow=<]&[\zxwCol,1mm]\zxZ{ax}\ar[r,marrow=<] &[\zxwCol,1mm] \zxZ-{\frac{\sin\theta}{2}(x+a\tan(\theta/2))^2} \rar[marrow=<]&[\zxwCol,1mm] \zxX{\frac{\tan(\theta/2)}{2}x^2} \ar[r,marrow=<]&[\zxwCol,1mm]\zxN{}
\end{ZX}\\
\overset{\mathclap{(\hyperlink{li:fusion}{f},\hyperlink{li:id}{id})}}{=}&\begin{ZX}
        \zxN{} \rar[marrow=<]&[\zxwCol,1mm] \zxX{\frac{\tan(\theta/2)}{2}x^2}  \rar[marrow=<]&[\zxwCol,1mm] \zxX{a\tan(\theta/2) x} \ar[r,marrow=<]&[\zxwCol,1mm]\zxZ{ax}\ar[r,marrow=<]&[\zxwCol,1mm]\zxZ-{a\sin\theta\tan(\theta/2)x}\ar[r,marrow=<]  &[\zxwCol,1mm] \zxZ-{\frac{\sin\theta}{2}x^2} \rar[marrow=<]&[\zxwCol,1mm] \zxX{\frac{\tan(\theta/2)}{2}x^2} \ar[r,marrow=<]&[\zxwCol,1mm]\zxN{}
\end{ZX}\\
\overset{(\hyperlink{li:fusion}{f})}{=}&\begin{ZX}
        \zxN{} \rar[marrow=<]&[\zxwCol,1mm] \zxX{a\tan(\theta/2) x}  \rar[marrow=<]&[\zxwCol,1mm] \zxX{\frac{\tan(\theta/2)}{2}x^2} \ar[r,marrow=<]&[\zxwCol,1mm]\zxZ{(a\cos\theta)x}\ar[r,marrow=<]  &[\zxwCol,1mm] \zxZ-{\frac{\sin\theta}{2}x^2} \rar[marrow=<]&[\zxwCol,1mm] \zxX{\frac{\tan(\theta/2)}{2}x^2} \ar[r,marrow=<]&[\zxwCol,1mm]\zxN{}
\end{ZX}\\
\overset{(\hyperlink{li:disp}{d})}{=}&
\begin{ZX}
        \zxN{} \rar[marrow=<]&[\zxwCol,1mm] \zxX{a\tan(\theta/2) x} \ar[r,marrow=<]&[\zxwCol,1mm]\zxZ{(a\cos\theta)x}\ar[r,marrow=<]&[\zxwCol,1mm] \zxX{\frac{\tan(\theta/2)}{2}(x+a\cos\theta)^2}  \rar[marrow=<] &[\zxwCol,1mm] \zxZ-{\frac{\sin\theta}{2}x^2} \rar[marrow=<]&[\zxwCol,1mm] \zxX{\frac{\tan(\theta/2)}{2}x^2} \ar[r,marrow=<]&[\zxwCol,1mm]\zxN{}
\end{ZX}\\
\overset{(\hyperlink{li:fusion}{f})}{=}&
\begin{ZX}
        \zxN{} \rar[marrow=<]&[\zxwCol,1mm] \zxX{a\tan(\theta/2) x} \ar[r,marrow=<]&[\zxwCol,1mm]\zxZ{(a\cos\theta)x}\ar[r,marrow=<]&[\zxwCol,1mm] \zxX{a\cos\theta\tan(\theta/2) x}  \rar[marrow=<] &[\zxwCol,1mm] \zxX{\frac{\tan(\theta/2)}{2}x^2}  \rar[marrow=<] &[\zxwCol,1mm] \zxZ-{\frac{\sin\theta}{2}x^2} \rar[marrow=<]&[\zxwCol,1mm] \zxX{\frac{\tan(\theta/2)}{2}x^2} \ar[r,marrow=<]&[\zxwCol,1mm]\zxN{}
\end{ZX}\\
\overset{\eqref{eq:d_com}}{=}&
\begin{ZX}
        \zxN{} \rar[marrow=<]&[\zxwCol,1mm] \zxX{a\tan(\theta/2) x} \ar[r,marrow=<]&[\zxwCol,1mm] \zxX{a\cos\theta\tan(\theta/2) x}  \rar[marrow=<] &[\zxwCol,1mm]\zxZ{(a\cos\theta)x}\ar[r,marrow=<]&[\zxwCol,1mm] \zxX{\frac{\tan(\theta/2)}{2}x^2}  \rar[marrow=<] &[\zxwCol,1mm] \zxZ-{\frac{\sin\theta}{2}x^2} \rar[marrow=<]&[\zxwCol,1mm] \zxX{\frac{\tan(\theta/2)}{2}x^2} \ar[r,marrow=<]&[\zxwCol,1mm]\zxN{}
\end{ZX}\\
\overset{(\hyperlink{li:fusion}{f})}{=}&
\begin{ZX}
        \zxN{} \rar[marrow=<]&[\zxwCol,1mm] \zxX{(a\sin\theta) x} \ar[r,marrow=<]&[\zxwCol,1mm]\zxZ{(a\cos\theta) x}\ar[r,marrow=<]&[\zxwCol,1mm] \zxX{\frac{\tan(\theta/2)}{2}x^2}  \rar[marrow=<] &[\zxwCol,1mm] \zxZ-{\frac{\sin\theta}{2}x^2} \rar[marrow=<]&[\zxwCol,1mm] \zxX{\frac{\tan(\theta/2)}{2}x^2} \ar[r,marrow=<]&[\zxwCol,1mm]\zxN{}
\end{ZX}\\
\overset{\mathclap{\eqref{def:rot_sp}}}{=}&
\begin{ZX}
        \zxN{} \rar[marrow=<]&[\zxwCol,1mm] \zxX{(a\sin\theta) x} \ar[r,marrow=<]&[\zxwCol,1mm]\zxZ{(a\cos\theta) x}\ar[r,marrow=<]&[\zxwCol,1mm] \zxBox{R(\theta)}\ar[r,marrow=<]&[\zxwCol,1mm]\zxN{}
\end{ZX}
\end{align}
\label{eq:rot_disp_com}
\end{subequations}
\end{widetext}
for arbitrary $\theta,a$, (note that the equation holds when $\theta$ is multiple of $\pi$ as well) and thus
\begin{subequations}
    \begin{align}
    &\begin{ZX}
        \zxZ-{m_2x}\ar[r,marrow=<]&[\zxwCol,1mm]\zxBox{R(\theta_1)}
    \end{ZX}
        \\\overset{(\hyperlink{li:copy}{c})}{=}&
        \begin{ZX}
    \zxZ{}\ar[r,marrow=<]&[\zxwCol,1mm]\zxX-{(m_2\tan\theta)x}\ar[r,marrow=<]&[\zxwCol,1mm]\zxZ{-m_2x}\ar[r,marrow=<]&[\zxwCol,1mm]\zxBox{R(\theta_1)}\ar[r,marrow=<]&[\zxwCol,1mm]\zxN{}
    \end{ZX}\\\overset{\eqref{eq:rot_disp_com}}{=}&
    \begin{ZX}
    \zxZ{}\ar[r,marrow=<]&[\zxwCol,1mm]\zxBox{R(\theta_1)}\ar[r,marrow=<]&[\zxwCol,1mm]\zxZ-{\frac{m_2}{\cos\theta_1}x}\ar[r,marrow=<]&[\zxwCol,1mm]\zxN{}
    \end{ZX}\\\overset{\eqref{eq:rot_rule_conj}}{=}&
    \begin{ZX}
    \zxZ-{\frac{\tan\theta_1}{2}x^2}\ar[r,marrow=<]&[\zxwCol,1mm]\zxZ-{\frac{m_2}{\cos\theta_1}x}\ar[r,marrow=<]&[\zxwCol,1mm]\zxN{}
    \end{ZX}\\\overset{(\hyperlink{li:fusion}{f})}{=}&
    \begin{ZX}
    \zxZ{-\frac{\tan\theta_1}{2}x^2-\frac{m_2}{\cos\theta_1}x}\ar[r,marrow=<]&[\zxwCol,1mm]\zxN{}
    \end{ZX}
    \end{align}
    \label{eq:rot_meas}
    \end{subequations}
holds. 


\section{Discussions}

\subsection{Validity of canonical eigenstates and infinite scalars}\label{ssec:illdef}

In this paper, we introduced a graphical calculus
based on directed graphs and explored its behavior under a certain set of rewrite rules. As we mentioned earlier, diagrams can be interpreted as CV operators by the mapping $\llbracket \cdot \rrbracket \colon D\mapsto \llbracket D\rrbracket$, and each rewrite rule equates two diagrams whose interpretation coincides. However, interpretations of $q$- and $p$-spiders given in Table~\ref{tab:table_gate} are not well-defined (when $n\neq 1$ or $m\neq1$). This is due to the fact that canonical operators $\hat{q},\hat{p}$ are unbounded operators~\cite{wintner_unboundedness_1947} and thus quadrature eigenstates $\{\ket{s}_q\}_{s\in \mathbb{R}},\{\ket{t}_p\}_{t\in \mathbb{R}}$ are not formally defined in the bosonic Hilbert space. One of the most evident example of such trickiness emerges in the most simple diagram:
\begin{equation}
    \llbracket \begin{ZX}
        \zxZ{}
    \end{ZX} \rrbracket = \int_\mathbb{R} \dd{s} 1
\end{equation}
is not defined in $\mathbb{C}$. This diagram can also be evaluated as
\begin{equation}
    \llbracket \begin{ZX}
        \zxZ{}\ar[r,marrow=<]&[\zxwCol,1mm]\zxZ{}
    \end{ZX} \rrbracket = \subscripts{p}{\braket{0}{0}}{p}=\delta(0),
\end{equation}
in which the delta function emerges as Fourier transform of the constant function.
Because of these improper calculation, our graphical model cannot assure soundness in general cases. Even if two diagrams have proper interpretation themselves, the graphical proof of two diagrams being rewritable may include improper transformation and thus leaves ambiguity at the formal level. The reverse is also true: Among our rewrite rules, the identity rule (\hyperlink{li:id}{id}), fusion rule (\hyperlink{li:fusion}{f}), squeezing rule (\hyperlink{li:sq}{s}), displacement rule (\hyperlink{li:disp}{d}), inversion rule (\hyperlink{li:inv}{inv}), and quadratic rule (\hyperlink{li:quad}{q}) do not contain undefinable diagrams when all the inputs and outputs are restricted to a single mode. Since our argument on completeness for 1-mode Gaussian gates, Sec.~\ref{ssec:gauss_univ}, relies only on these rules, the proof is strictly valid.

Generally, it is already known that a compact structure does not exist in the category of infinite-dimensional Hilbert spaces \cite{heunen_categories_2019}, which presents a barrier for naive generalization of the framework of the ZX~calculus for a standard CV system. This issue raises another problem deeply associated with the standard architecture of CV quantum computing. As we did in this paper, the canonical eigenstates are widely used when exploring the CV systems. Just as we emphasized in the introduction, these eigenstates themselves require a sensitive handling, since they are out of the Hilbert space $\mathcal{H}$ and need careful justification for their expected behaviors in the CV system. Nevertheless, quadrature eigenstates are often regarded as infinitely-squeezed vacuum, and so are other ideal processes such as homodyne projection and preparation of perfectly correlated EPR states.
Although these intuitive interpretations may not suffice to ensure formal mathematical rigor, 
this is an issue with all of CV quantum information, and these problems are inherited by our formalism. Nevertheless, the quadrature expansion is quite useful to capture the essence of quantum processes in phase space, and canonical eigenstates serve as a handy tool for analysis of ideal processes with infinitely-squeezed states.

For example, a physical EPR state has uncertainty in the quadrature distribution of the two modes due to its finite correlation. By contrast, an ideal EPR state has perfect correlation, thus the homodyne measurement outcome of the first mode exactly coincides to the displaced amount of the residual quantum state in the second mode. By using this property, one can easily show how the quantum teleportation protocol retrieves the original input state without any disturbance, which is essentially what we did through the graphical calculus.

With this perspective, the graphical calculus model proposed in this paper does not achieve 
the level of rigour used in formal mathematics, but it does so to a standard appropriate for theoretical physics. To wit, it relies on the standard framework of CV quantum computing, and we elucidated how the generalization of the ZX calculus should behave based on it. In order to overcome these ambiguities, an additional framework or structure is needed so that one can handle the infinitely squeezed states in a rigorous way while preserving its expected properties. 

One possible approach to deal with these problems is to introduce infinite objects, such as unbounded scalars, quantum states with infinite norms, and others in a rigorous 
way. Recent research has proposed the 
application of non-standard analysis~\cite{robinson_non-standard_1974}, a mathematical field of logic that formalizes unbounded scalars and infinitesimals, for illustration of infinite-dimensional quantum mechanics~\cite{ozawa_phase_1997, raab_approach_2004} and its categorical
representation~\cite{gogioso_infinite-dimensional_2017,gogioso_quantum_2019}. Using this approach, 
one can 
consider 
linear algebras on infinite-dimensional Hilbert spaces where infinities and infinitesimals can be treated as normal numbers and where 
most features of standard Hilbert spaces are transferred without losing mathematical rigor. Further refinements of this approach 
may help to
formalize 
our graphical model with infinite scalars so that its soundness is strictly guaranteed. The current presentation stands as a useful starting point for a CV ZX calculus, with its more formal analysis---using more sophisticated tools---left to future work. 

\subsection{Representation of physical states and error correcting codes}
As mentioned in Sec.~\ref{ssec:illdef}, our graphical calculus is based on canonical eigenstates which cannot be normalized. This causes another problem in the other direction: Though the model can deal with certain unphysical states, such as infinitely squeezed states and the ideal cubic phase states $\ket{\psi_{\mathrm{CPG}^\gamma}}$ [Eq.~\eqref{eq:CPSdef}], as we illustrated in this paper, it cannot give a diagrammatic representation for any physical (finite-norm) state, including Fock eigenstates, displaced vacuum, and finitely-squeezed states. 

One possible extension of the model to deal with this problem is to admit complex 
phase functions for each spider. Since the vacuum state, for instance, can be expanded with regard to the position basis as
\begin{equation}
    \ket{0}=\int \dd{s}\frac{1}{\pi^{1/4}}\exp\left(-\frac{s^2}{2}\right)\ket{s}_q,
\end{equation}
it can also be regarded as a momentum eigenstate $\ket{0}_p\propto \int \dd{s}\ket{s}_q$ perturbed under imaginary time evolution. With this perspective, one can interpret the following diagram
\begin{equation}
    \begin{ZX}
   \zxN{}\ar[rr,marrow=<]& & [\zxwCol] \zxZ{-i\frac{x^2}{2}}
\end{ZX} = \begin{ZX}
    \zxN{}\ar[r,marrow=<]&[\zxwCol] \zxZ{-i\frac{x^2}{2}} \ar[r,marrow=<]& [\zxwCol] \zxZ{}
 \end{ZX}
\end{equation}
as representing
the vacuum state. Most of the rewrite rules proposed in this paper can be generalized to complex spiders. The properties of non-unitary Gaussian operators can also be studied using extended rewrite rules including the quadratic rule~(\hyperlink{li:quad}{q}) and the diplacement rule~(\hyperlink{li:quad}{q}), which may aassociate our calculus model with graphical symplectic algebra for general Gaussian processes \cite{booth_complete_2024,booth_graphical_2024}. Another potential proposal is to introduce diagrams representing non-unitary operations such as damping operator $\mathcal{N}(\gamma)=\exp(-\gamma \hat{a}^\dagger\hat{a})$. Using this 
operator, the correspondence between infinite- and finitely squeezed states is given by the equation below~\cite{walshe_continuous-variable_2020}:
\begin{equation}
    \mathcal{N}(\gamma) \ket{0}_q \propto \widehat{Sq}\left(-\frac{\log \tanh\gamma}{2}\right) \ket{0}
\end{equation}
and thus the damping operation transforms an unphysical state into physical state (up to some infinite scalar factor).
Since the damping operation could be interpreted as a type of decoherece, this process may also be formalized in an extensive framework of ZX calculus generalized for representation of mixed states, decoherence, and classical information \cite{carette_completeness_2021,coecke_categorical_2016}.

A different extension of the graphical calculus may be found in the direction of quantum error correction codes. As we mentioned in the introduction, CV information processing is highly compatible with error correction, and it is natural to expect the CV calculus model to be unifed with DV computing. One possible approach for such application is to introduce new diagrams representing encoded qubit states. For example, the code space of the Gottesman-Kitaev-Preskill qubit~\cite{gottesman_encoding_2001} is stabilized by operators
\begin{equation}
    \left\{\exp(-2i\sqrt{\pi}\hat{p}), \exp(2i\sqrt{\pi}\hat{x})\right\}
    .
\end{equation}
Thus, one could define the notion of a GKP diagram, denoted as a yellow diamond in the following equation, such that 
\begin{equation}
    \begin{ZX}
   \zxN{}\ar[r,marrow=<]&[\zxwCol,2mm] \zxGKP{}
\end{ZX} = \begin{ZX}
   \zxN{}\ar[r,marrow=<]&[\zxwCol,1mm]\zxZ{2\sqrt{\pi}}\ar[r,marrow=<]&[\zxwCol,1mm] \zxGKP{}
\end{ZX}= \begin{ZX}
   \zxN{}\ar[r,marrow=<]&[\zxwCol,1mm]\zxX{2\sqrt{\pi}}\ar[r,marrow=<]&[\zxwCol,1mm] \zxGKP{}
\end{ZX}
\end{equation}
holds. Moreover, the logical state $\ket{0_\mathrm{GKP}}$ is uniquely characterized by the following relation
\begin{equation}
   \ket{0_\mathrm{GKP}}=\widehat{Sq}(2)\hat{F}\ket{0_\mathrm{GKP}},
\end{equation}
thus it can be specified with an additional axiom as follows:
\begin{equation}
    \begin{ZX}
   \zxN{}\ar[r,marrow=<]&[\zxwCol,2mm] \zxGKP{}
\end{ZX} = \begin{ZX}
   \zxN{}\ar[r,marrow=<]&[\zxwCol,1mm]\zxBox{Sq(2)}\ar[r,marrow=<]&[\zxwCol,1mm]\zxF{}\ar[r,marrow=<]&[\zxwCol,1mm] \zxGKP{}
\end{ZX}
\end{equation}
These equations represent the principal properties of the GKP qubit and thus would be sufficient for graphical reasonings of the fundamental error correction schemes proposed in Ref.~\cite{gottesman_encoding_2001}. It is an interesting question to consider how these non-unitary diagrams and code states could interact with other elements through rewrite rules to be derived in future work.

\subsection{Incompleteness of the set of rewrite rules}

The set of rewrite rules we presented in Sec.~\ref{ssec:CV_basic_rules} are determined not only as an extension of the framework of the ZX~calculus but also as a handy rule set to conduct diagrammatic calculus operations for what we consider to be the 
principal processes in CV quantum information processing. Though we have investigated various properties of the CV system within our rule set, it is unknown what rule set can achieve a wider range of application for more general processes. In fact, in the last section, we discussed the completeness of the graphical calculus for 1-mode Gaussian processes in Sec.~\ref{ssec:gauss_univ}, but in the discussion we restricted attention to diagrams composed only of 1-to-1 quadratic spiders, and it cannot be directly extended to a general form of Gaussian diagrams including multi-mode spiders. For instance, direct calculation suggests
\begin{equation}
    \left\llbracket\begin{ZX}
&[\zxwCol,1mm]&[\zxwCol,2mm]\zxBox{Sq(\tau)}\ar[dr,-N,inarrow=<]    &[\zxwCol,2mm] &[\zxwCol,1mm]\\
        \zxN{}\ar[r,marrow=<]   &\zxX{}\ar[ur,N-,outarrow=<]\ar[dr,N-,outarrow=<]   &  & \zxZ{}\ar[r,marrow=<] &\zxN{}\\
&  &\zxBox{Sq(\tau')}\ar[ur,-N,inarrow=<]
    \end{ZX} \right\rrbracket =
    \left\llbracket \begin{ZX}
\zxN{}\ar[r,marrow=<]&[\zxwCol,1mm]\zxBox{Sq\left(\tau+\tau'\right)}\ar[r,marrow=<]&[\zxwCol,1mm]\zxN{}
    \end{ZX}  \right\rrbracket
\end{equation}
should hold. However, rewritability of the two diagrams under the rule set given in this paper is unclear so far.

Previous research has revealed that the qubit ZX~calculus is complete for Clifford processes and that all Clifford diagrams can be efficiently rewritten into a standard form~\cite{duncan_graph_2009,duncan_graph-theoretic_2020}, and this fact is used in various applications, including graphical characterization of Clifford process and circuit extraction. As properties of Gaussian processes in a CV system are analogous to that of Clifford operations for qubits in some ways~\cite{asavanant_optical_2022}, it is an interesting question whether completeness for Gaussian diagrams can be boiled down to similar graphical manipulations.

\section{Conclusion}

In this paper, we proposed a graphical calculus to model CV dynamics by convertibility of simple diagrams that represent each quantum process under a set of rewrite rules. Although the model contains subtle issues regarding infinities, the framework we propose is a starting point for further investigations.
As
with the original ZX~calculus, we expect our model to have
numerous 
applications in
CV quantum information processing.

One potential application is as the foundation of computer software for automated circuit optimization and interactive theorem proving. Recently, the ZX~calculus has been used in computing packages that treat DV computations, such as PyZX~\cite{kissinger_pyzx_2020}, Quantomatic~\cite{kissinger_quantomatic_2015}, PennyLane~\cite{bergholm_pennylane_2018} and others. Similar tools for CV quantum computing are far more sparse. As we have seen in this paper, our graphical calculus is powerful enough to interpret various CV protocols, and it may be utilized as a language for abstraction of CV quantum information processing.

Another interesting possibility might be found in the reasoning of MBQC and quanutm error-correcting codes. Since one-way quantum computing and topological codes are highly compatible with diagrammatic representation \cite{backens_there_2021,kissinger_universal_2019,horsman_quantum_2011}, a number of investigations reported advantageous in using the (qubit) ZX~calculus as a graphical language. Our CV version may be helpful in the analysis and design of CV measurement-based quantum computing architectures~\cite{asavanant_generation_2019,larsen_deterministic_2019} and applications.

\begin{acknowledgments}

We thank Peter van Loock, Takaya Matsuura and Blayney Walshe for meaningful discussions. This work was
supported by the Japan Science and Technology (JST) Agency (Moonshot R \& D) Grant No.JPMJMS2064 and JPMJMS2066, UTokyo Foundation, and donations from Nichia Corporation of Japan. H.N.\ and R.I.\ acknowledge financial support from The Forefront Physics and Mathematics Program to Drive Transformation (FoPM). H.N.\ and W.A.\ acknowledges funding from the Japan Society for the Promotion of Science KAKENHI (No.\ 23K13040, 24KJ0745). W.A.\ acknowledges funding from Research Foundation for OptoScience and Technology. This work was supported by the Australian Research Council through the Centre of Excellence for Quantum Computation and Communication Technology (Project No.\ CE170100012).
\end{acknowledgments}

\appendix

\section{Consistency of graphical calculus and CV operators}

\subsection{Quantum gates}\label{sec:consistency}

In this section, we will confirm that the graphical representations of CV quantum gates given in Sec.~\ref{ssec:resp_gates} is consistent with the definitions listed in Table~\ref{tab:table_gate}.

\subsubsection{Displacement gate}

\begin{align}
    &\left\llbracket
    \begin{ZX}
        \zxN{}&[\zxWCol] \zxX{\sqrt{2}\Re(\alpha)x} \lar[inarrow=>] \rar[inarrow=<] &[\zxWCol] \zxZ{\sqrt{2}\Im(\alpha)x}\rar[marrow=<]&[\zxWCol]\zxN{}
    \end{ZX}
    \right\rrbracket\\
    =&
    \scalebox{0.9}{$\displaystyle \left(\int_{\mathbb{R}}\dd{t} e^{-i\sqrt{2}\Re(\alpha)t} \subscripts{}{\ket{t}}{p}\subscripts{p}{\bra{t}}{}\right)\left(\int_{\mathbb{R}}\dd{s} e^{i\sqrt{2}\Im(\alpha)t} \subscripts{}{\ket{s}}{q}\subscripts{q}{\bra{s}}{}\right)$}\\
    =&
    e^{-i\sqrt{2}\Re(\alpha)\hat{p}}e^{i\sqrt{2}\Im(\alpha)\hat{q}}\\
    \sim& e^{i(-\sqrt{2}\Re(\alpha)\hat{p}+\sqrt{2}\Im(\alpha)\hat{q})}\\
    =&e^{i(\alpha^*\hat{a}-\alpha\hat{a}^\dagger)}
\end{align}
Here we used Baker-Campbell-Hausdorff formula under the canonical commutation relation $[\hat{q},\hat{p}]=i$.

\subsubsection{Phase rotation gate}

Instead of directly calculating operator multiplications in the Schr\"odinger picture, here we employ the Heisenberg picture and compute the quadrature transformation function defined in Eq.~\eqref{eq:trans}. For a quadratic Gaussian operator without displacement, one only needs to compute its representation matrix for a unique identification of the operator (up to global scalar factor). Let $M(\hat{A})$ denote the symplectic matrix representing the Heisenberg action of an operator $\hat{A}$ with respect to the vector of quadratures~$(\hat q, \hat p)^T$. Then,
\begin{align}
    &
    M\left(
        \left\llbracket
    \begin{ZX}
        \zxN{}&[\zxWCol] \zxX{\frac{\tan(\theta/2)}{2}x^2} \lar[inarrow=>] \rar[inarrow=<] &[\zxWCol] \zxZ-{\frac{\sin\theta}{2}x^2} \rar[marrow=<]&[\zxWCol] \zxX{\frac{\tan(\theta/2)}{2}x^2} \rar[marrow=<]&[\zxWCol]\zxN{}
    \end{ZX}
    \right\rrbracket
    \right)
    \\
    =&
    \mat{1 & \tan{\frac{\theta}{2}}\\ 0 & 1}
    \mat{1 & 0 \\ -\sin\theta & 1}
    \mat{1 & \tan{\frac{\theta}{2}}\\ 0 & 1}
    \\
    =&
    \mat{\cos\theta & \sin\theta\\ -\sin\theta & \cos\theta}
\end{align}
This is the rotation matrix for the quadrature operators as is defined in Table~\ref{tab:table_gate}.

\subsubsection{1-mode squeezing gate}

\begin{align}
    &
    M\left(
        \left\llbracket
    \begin{ZX}
        \zxN{}&[\zxWCol] \zxX{\frac{\tau(1-\tau)}{4}x^2} \lar[inarrow=>] \rar[inarrow=<] &[\zxWCol] \zxZ-{\frac{1}{\tau}x^2} \rar[marrow=<]&[\zxWCol] \zxX{\frac{(\tau-1)}{4}x^2} \rar[marrow=<]&[\zxWCol] \zxZ{x^2} \rar[marrow=<]&[\zxWCol]\zxN{}
    \end{ZX}
    \right\rrbracket
    \right)
    \\
    =&
    \mat{1 & \frac{\tau(1-\tau)}{2}\\ 0 & 1}
    \mat{1 & 0 \\ -\frac{2}{\tau} & 1}
    \mat{1 & \frac{\tau-1}{2}\\ 0 & 1}
    \mat{1 & 0 \\ 2 & 1}
    \\
    =&
    \mat{\tau & 0\\ 0 & \frac{1}{\tau}}
\end{align}
This matrix transformation achieves a squeezing transformation for the quadrature operators, as defined in Table~\ref{tab:table_gate} with $\tau = e^{-r}$.

\subsubsection{Controlled-sum gate}

For the unbiased ($g=1$) controlled-sum gate,
\begin{align}
    &
    \left\llbracket
    \begin{ZX}
    \zxN{}\ar[drrr,-N,outarrow=<]&[\zxwCol]&[\zxwCol]&[\zxWCol]&[\zxwCol]&[\zxwCol]\\
    &&&\zxZ{} \ar[rr,marrow=<]&&\zxN{}\\[\zxWRow]
    \zxN{}\ar[rr,marrow=<] &&\zxX{}\ar[ru,s.,marrow=<]\ar[drrr,N-,inarrow=<]\\
    &&&&&\zxN{}
    \end{ZX}
    \right\rrbracket
    \\
    =&\left(\int_{\mathbb{R}}\dd{t} \subscripts{}{\ket{t}}{p_2}\subscripts{p_{\mathrm{anc}},p_2}{\bra{t,t}}{}\right)\left(\int_{\mathbb{R}}\dd{s} \subscripts{}{\ket{s,s}}{q_1,q_{\mathrm{anc}}}\subscripts{q_1}{\bra{s}}{}\right)
    \\
    =&\int_{\mathbb{R}}\dd{s}\int_{\mathbb{R}}\dd{t} \subscripts{p_{\mathrm{anc}}}{\braket{t}{s}}{q_{\mathrm{anc}}} \subscripts{}{\ket{s,t}}{q_1,p_2}\subscripts{q_1,p_2}{\bra{s,t}}{}
    \\
    =&\int_{\mathbb{R}}\dd{s}\int_{\mathbb{R}}\dd{t} e^{-ist}\subscripts{}{\ket{s,t}}{q_1,p_2}\subscripts{q_1,p_2}{\bra{s,t}}{}
    \\
    =&e^{-i\hat{q}_1\hat{p}_2}
\end{align}
realizes the unitary transformation defined in Table~\ref{tab:table_gate}. For biased gates, sandwiching the unbiased CSUM gate with squeezing gates yields
\begin{align}
    &
    \left\llbracket
\begin{ZX}
    \zxN{}\ar[r,marrow=<]&[\zxwCol,1mm]\zxBox{Sq(g^{-1})}\ar[drr,-N,inarrow=<]&[\zxwCol]&[\zxWCol]&[\zxwCol,1mm]&[\zxwCol,1mm]\\[\zxZeroRow,-4mm]
    &&&\zxZ{} \ar[r,marrow=<]&\zxBox{Sq(g)}\ar[r,marrow=<]&\zxN{}\\[\zxWRow]
    \zxN{}\ar[rr,marrow=<] &&\zxX{}\ar[ru,s.,marrow=<]\ar[drrr,N-,inarrow=<]\\
    &&&&&\zxN{}
    \end{ZX}
    \right\rrbracket
    \\
    =&\left(\int_{\mathbb{R}}\dd{s} \subscripts{}{\ket{s}}{q}\subscripts{q}{\bra{gs}}{}\right)e^{-i\hat{q}_1\hat{p}_2}\left(\int_{\mathbb{R}}\dd{s'} \subscripts{}{\ket{gs'}}{q}\subscripts{q}{\bra{s'}}{}\right)
    \\
    =&e^{-ig\hat{q}_1\hat{p}_2}.
\end{align}

\subsubsection{Controlled-Z gate}
The unbiased CZ gates can be expanded as
\begin{align}
    &\left\llbracket
    \begin{ZX}
    &[\zxwCol]&[\zxwCol]&[\zxWCol]&[\zxWCol]&[\zxwCol]&[\zxwCol]\\
    \zxN{}\ar[rr,marrow=<] &&\zxZ{}\ar[rd,s.,marrow=<]\ar[urrrr,N-,inarrow=<]\\[\zxWRow]
    &&&\zxF{} \ar[rd,s.,marrow=<]&&\zxN{}\\[\zxWRow]
    &&&&\zxZ{} \ar[rr,marrow=<]&&\zxN{}\\
    \zxN{}\ar[urrrr,-N,outarrow=<]
    \end{ZX}
    \right\rrbracket
    \\
    =&\left(\int_{\mathbb{R}}\dd{s} \subscripts{}{\ket{s}}{q_1}\subscripts{q_1,q_{\mathrm{anc}}}{\bra{s,s}}{}\right)\hat{F}_{\mathrm{anc}}\left(\int_{\mathbb{R}}\dd{s'} \subscripts{}{\ket{s}}{q_{\mathrm{anc}}}\subscripts{q_2}{\bra{s'}}{}\right)
    \\
    =&\int_{\mathbb{R}}\dd{s}\int_{\mathbb{R}}\dd{s'} \subscripts{q_{\mathrm{anc}}}{\mel{s}{\hat{F}}{s'}}{q_{\mathrm{anc}}} \subscripts{}{\ket{s,s'}}{q_1,q_2}\subscripts{q_1,q_2}{\bra{s,s'}}{}
    \\
    =&\int_{\mathbb{R}}\dd{s}\int_{\mathbb{R}}\dd{s'} \subscripts{q_{\mathrm{anc}}}{\braket{s}{s'}}{p_{\mathrm{anc}}}\subscripts{}{\ket{s,s'}}{q_1,q_2}\subscripts{q_1,q_2}{\bra{s,s'}}{}
    \\
    =&\int_{\mathbb{R}}\dd{s}\int_{\mathbb{R}}\dd{s'} e^{iss'}\subscripts{}{\ket{s,s'}}{q_1,q_2}\subscripts{q_1,q_2}{\bra{s,s'}}{}
    \\
    =&e^{i\hat{q}_1\hat{q}_2}
\end{align}
and sandwiching this by squeezing gates works for biased ones in the same way as CSUM gates.

\begin{widetext}

\subsubsection{Beamsplitter gate}\label{ssec:bs_decom}

\begin{align}
    &
    M\left(\left\llbracket
    \begin{ZX}
        \zxN{}\ar[r,marrow=<]&[\zxwCol,2mm]\zxBox{Sq\left(\frac{1}{\tan\theta}\right)}\ar[drr,-N,inarrow=<]&[\zxWCol,1mm]&[\zxwCol]&[\zxwCol,1mm]&[\zxwCol,1mm]&[\zxwCol]&[\zxwCol,1mm]&[\zxwCol,1mm]\\[\zxZeroRow,-2mm]
        &&&\zxZ{} \ar[r,marrow=<]&\zxBox{Sq\left(\frac{\sin^2\theta}{\cos\theta}\right)}\ar[drr,-N,inarrow=<]&\zxN{}&\\[\zxZeroRow,-2mm]
        &&&&&&\zxX{} \ar[r,marrow=<]&\zxBox{Sq\left(\frac{1}{\tan\theta}\right)}\ar[r,marrow=<]&\zxN{}\\
        \zxN{}\ar[rr,marrow=<]&&\zxX{}\ar[ruu,s.,marrow=<]\ar[drr,N-,outarrow=<]\\[\zxZeroRow]
        &&&&\zxBox{Sq\left(\frac{1}{\cos\theta}\right)}\ar[r,marrow=<]&\zxZ{}\ar[ruu,s.,marrow=<]\ar[drrr,N-,inarrow=<]\\
        &&&&&&&&\zxN{}
        \end{ZX}
    \right\rrbracket\right)\\
    =&\mat{\frac{1}{\tan\theta}&0&0&0\\0&\tan\theta&0&0\\0&0&1&0\\0&0&0&1}\mat{1&0&0&0\\0&1&0&-1\\1&0&1&0\\0&0&0&1}\mat{\frac{\sin^2\theta}{\cos\theta}&0&0&0\\0&\frac{\cos\theta}{\sin^2\theta}&0&0\\0&0&\frac{1}{\cos\theta}&0\\0&0&0&\cos\theta}\mat{1&0&-1&0\\0&1&0&0\\0&0&1&0\\0&1&0&1}\mat{\frac{1}{\tan\theta}&0&0&0\\0&\tan\theta&0&0\\0&0&1&0\\0&0&0&1}\\
    =&\mat{\cos\theta&0&-\sin\theta&0\\0&\cos\theta&0&-\sin\theta\\\sin\theta&0&\cos\theta&0\\0&\sin\theta&0&\cos\theta}
\end{align}
\end{widetext}
\subsubsection{Cubic phase gate}

\begin{equation}
    \left\llbracket
    \begin{ZX}
        \zxN{}&[\zxwCol,1mm] \zxZ{\gamma x^3} \lar[inarrow=>] \rar[inarrow=<] &[\zxwCol,1mm]\zxN{}
    \end{ZX}\right\rrbracket
    =\int_{\mathbb{R}}\dd{s} e^{i\gamma x^3}\subscripts{q}{\ket{s}\bra{s}}{q}=e^{i \gamma \hat{q}^3}
\end{equation}

\subsection{Soundness of basic rewriting rules}\label{sec:rewrite_sound}
As is discussed in Sec~\ref{ssec:ZX_qubit}, to prove soundness of a graphical calculus one just needs to verify that the individual rewrite rules are sound. In this section, we will show soundness for each basic rewriting rules given in \ref{ssec:CV_basic_rules} by direct calculaion.

\subsubsection{Identity rule}
\begin{align}
    \left\llbracket\begin{ZX}
       \zxN{}\ar[r,marrow=<]&[\zxwCol,1mm]\zxZ{a}\ar[r,marrow=<]&[\zxwCol,1mm]\zxN{}
    \end{ZX}\right\rrbracket=&\int_{\mathbb{R}}\dd{s} e^{ia} \subscripts{}{\ket{s}}{q}\subscripts{q}{\bra{s}}{}\sim\mathrm{id}_{\mathcal{H}}\\
     \left\llbracket
    \begin{ZX}
        \zxN{}\ar[r,marrow=<]&[\zxwCol,1mm]\zxX{b}\ar[r,marrow=<]&[\zxwCol,1mm]\zxN{}
        \end{ZX}\right\rrbracket=&\int_{\mathbb{R}}\dd{s} e^{ib} \subscripts{}{\ket{t}}{p}\subscripts{p}{\bra{t}}{}\sim\mathrm{id}_{\mathcal{H}}
\end{align}
where $\mathrm{id}_{\mathcal{H}}$ denotes the identity operation. 

\subsubsection{Fusion rule}
\begin{align}
    &
    \left\llbracket
    \begin{ZX}
   \leftArrowedManyDots{n}\zxZ{f(x)}\ar[dd,C=1,start anchor=south,end anchor=north]\ar[dd,C-=1,start anchor=south,end anchor=north]\rightArrowedManyDots{m}\\
   &\ldots\\
   \leftArrowedManyDots{n'}\zxZ{g(x)}\rightArrowedManyDots{m'}
\end{ZX}
\right\rrbracket\\
=&
    \int_{\mathbb{R}} \dd{s}\int_{\mathbb{R}}\dd{s'}e^{if(s)}e^{ig(s')}\bigl(\subscripts{q}{\braket{s}{s'}}{q}\bigr)^k\bigl(\subscripts{q}{\braket{s'}{s}}{q}\bigr)^{k'}
    \nonumber \\
    &
    \qquad \times
    \bigl(\subscripts{}{\ket{s}}{q}\bigr)^{\otimes n}\bigl(\subscripts{q}{\bra{s}}{}\bigr)^{\otimes m}\otimes\bigl(\subscripts{}{\ket{s'}}{q}\bigr)^{\otimes n'}\bigl(\subscripts{q}{\bra{s'}}{}\bigr)^{\otimes m'}
\\
=&
    \int_{\mathbb{R}} \dd{s}\int_{\mathbb{R}} \dd{s'}e^{if(s)}e^{ig(s')}\delta(s-s')^{k+k'}
    \nonumber \\
    &
    \qquad \times
    \bigl(\subscripts{}{\ket{s}}{q}\bigr)^{\otimes n}\bigl(\subscripts{q}{\bra{s}}{}\bigr)^{\otimes m}\otimes\bigl(\subscripts{}{ \ket{s'}}{q}\bigr)^{\otimes n'}\bigl(\subscripts{q}{ \bra{s'}}{}\bigr)^{\otimes m'}
    \\
=&\int_{\mathbb{R}} \dd{s}e^{if(s)}e^{ig(s)}\delta(0)^{k+k'-1}\bigl(\subscripts{}{\ket{s}}{q}\bigr)^{\otimes n+n'}\bigl(\subscripts{q}{\bra{s}}{}\bigr)^{\otimes m+m'}\\
\sim&\left\llbracket
\begin{ZX}
   \leftArrowedManyDots{n+n'}\zxZ{(f+g)(x)} \rightArrowedManyDots{m+m'}
\end{ZX}
\right\rrbracket
\end{align}
where $k$ and $k'$ denote the number of upward and downward wires between the two spiders, respectively. The colour-flipped counterpart can be shown to be sound in the same way. Note that here we have neglected the infinite scalar term~$\delta(0)^{k+k'-1}$.

\begin{widetext}

\subsubsection{Bialgebra rule}
\begin{align}
&
\left\llbracket
 \begin{ZX}
    &[\zxwCol,2mm]&[\zxwCol,1mm]\zxN{}&[\zxwCol,1mm]&[\zxwCol,2mm]\\
    \zxN{}\ar[r,marrow=<]&\zxX{}\ar[drr,s.,numarrow={<}{0.25},numarrow={<}{0.75}]\ar[rr,o'={-=.2,L=.25},marrow=<]&&\zxZ{}\ar[r,marrow=<]&\zxN{}\\[\zxWRow,.7mm]
    \zxN{}\ar[r,marrow=<]&\zxX{}\ar[rr,o.={-=.2,L=.25},marrow=<]\ar[urr,s.,numarrow={<}{0.25},numarrow={<}{0.75}]&&\zxZ{}\ar[r,marrow=<]&\zxN{}
 \end{ZX}
 \right\rrbracket
\\
=&\int_{\mathbb{R}}\dd{s_1}\int_{\mathbb{R}}\dd{s_2}\int_{\mathbb{R}}\dd{t_1}\int_{\mathbb{R}}\dd{t_2}
\subscripts{}{\ket{t_1,t_2}}{p_1p_2}\subscripts{pppp}{\braket{t_1,t_1,t_2,t_2}{s_1,s_2,s_1,s_2}}{qqqq}\subscripts{q_1q_2}{\bra{s_1,s_2}}{}\\
\sim&\int_{\mathbb{R}}\dd{s_1}\int_{\mathbb{R}}\dd{s_2}\int_{\mathbb{R}}\dd{t_1}\int_{\mathbb{R}}\dd{t_2}
e^{-i(s_1t_1+s_1t_2+s_2t_1+s_2t_2)}\subscripts{}{\ket{t_1,t_2}}{p_1p_2}\subscripts{q_1q_2}{\bra{s_1,s_2}}{}\\
=&\int_{\mathbb{R}}\dd{s_1}\int_{\mathbb{R}}\dd{s_2}\int_{\mathbb{R}}\dd{t_1}\int_{\mathbb{R}}\dd{t_2}
e^{-i(s_1+s_2)(t_1+t_2)}\subscripts{}{\ket{t_1,t_2}}{p_1p_2}\subscripts{q_1q_2}{\bra{s_1,s_2}}{}\\
=&\int_{\mathbb{R}}\dd{s_1}\int_{\mathbb{R}}\dd{s_2}\left(\int_{\mathbb{R}}\dd{t_1}e^{-i(s_1+s_2)t_1}\subscripts{}{\ket{t_1}}{p_1}\right)\left(\int_{\mathbb{R}}\dd{t_2}e^{-i(s_1+s_2)t_2}\subscripts{}{\ket{t_2}}{p_2}\right)
\subscripts{q_1q_2}{\bra{s_1,s_2}}{}\\
\sim&\int_{\mathbb{R}}\dd{s_1}\int_{\mathbb{R}}\dd{s_2}\subscripts{}{\ket{s_1+s_2}}{q_1}\subscripts{}{\ket{s_1+s_2}}{q_2}
\subscripts{q_1q_2}{\bra{s_1,s_2}}{}\\
=&
\int_{\mathbb{R}}\dd{s} \int_{\mathbb{R}}\dd{s_1}\int_{\mathbb{R}}\dd{s_2}
\delta (s - s_1 - s_2)
\subscripts{}{\ket{s,s}}{q_1q_2} \subscripts{q_1q_2}{\bra{s_1,s_2}}{}\\
=&\left(\int_{\mathbb{R}}\dd{s}\subscripts{}{\ket{s,s}}{q_1q_2}\subscripts{q}{\bra{s}}{}\right)\left(\int_{\mathbb{R}}\dd{s_1}\int_{\mathbb{R}}\dd{s_2}\subscripts{}{\ket{s_1+s_2}}{q}\subscripts{q_1q_2}{\bra{s_1,s_2}}{}\right)\\
=&
\left\llbracket
\begin{ZX}
    \zxN{}\ar[dr,-N, outarrow=<]&[\zxwCol,1mm]&[\zxwCol,2mm]&[\zxwCol,1mm]\zxN{}\\[\zxWRow]
    &  \zxZ{}\ar[r,marrow=<]&\zxX{}\ar[dr,N-, inarrow=<]\ar[ur,N-, inarrow=<]\\[\zxWRow]
    \zxN{}\ar[ur,-N, outarrow=<]&&&\zxN{}
 \end{ZX}
 \right\rrbracket
 \end{align}

\end{widetext}

\subsubsection{Fourier rule}
By definition, the Fourier transform operator~$\hat{F}$ can be expanded as
\begin{equation}
    \hat{F}=\int_{\mathbb{R}}\dd{u}\subscripts{}{\ket{u}}{p}\subscripts{q}{\bra{u}}{}=\int_{\mathbb{R}}\dd{u}\subscripts{}{\ket{-u}}{q}\subscripts{p}{\bra{u}}{}.
\end{equation}
This operator has the Heisenberg action on the
quadrature operators $\hat{q}$ and $\hat{p}$ is as follows:
\begin{align}
    \hat{q}\mapsto\hat{F}^\dagger\hat{q}\hat{F} =& \left(\int_{\mathbb{R}}\dd{u}\subscripts{}{\ket{u}}{p}\subscripts{q}{\bra{-u}}{}\right)\hat{q}\left(\int_{\mathbb{R}}\dd{u'}\subscripts{}{\ket{-u'}}{q}\subscripts{p}{\bra{u'}}{}\right)\\
    =&\int_{\mathbb{R}}\dd{u}\int_{\mathbb{R}}\dd{u'}(-u)\delta(u-u')\subscripts{}{\ket{u}}{p}\subscripts{p}{\bra{u'}}{}\\
    =& -\int_{\mathbb{R}}\dd{u} u\subscripts{}{\ket{u}}{p}\subscripts{p}{\bra{u}}{}=-\hat{p},\\
    \hat{p}\mapsto\hat{F}^\dagger\hat{p}\hat{F} =& \left(\int_{\mathbb{R}}\dd{u}\subscripts{}{\ket{u}}{q}\subscripts{p}{\bra{u}}{}\right)\hat{q}\left(\int_{\mathbb{R}}\dd{u'}\subscripts{}{\ket{u'}}{p}\subscripts{q}{\bra{u'}}{}\right)\\
    =&\int_{\mathbb{R}}\dd{u}\int_{\mathbb{R}}\dd{u'}(u)\delta(u-u')\subscripts{}{\ket{u}}{q}\subscripts{q}{\bra{u}}{}\\
    =& \int_{\mathbb{R}}\dd{u} u\subscripts{}{\ket{u}}{q} \subscripts{q}{\bra{u}}{} =\hat{q}.
\end{align}
Thus, one obtains
\begin{align}
    M\left(\left\llbracket
        \begin{ZX}\\[\zxHRow]
            \zxN{}\rar[inarrow=<]&[\zxWCol]\zxF{}\rar[inarrow=<]&[\zxWCol]\zxN{}\\[\zxHRow]
            \end{ZX}
    \right\rrbracket\right)
    =&\mat{0&-1\\1&0}
    \\=& \left.M\left(\left\llbracket
        \begin{ZX}
            \zxN{}&[\zxWCol] \zxBox{R(\theta)} \lar[inarrow=>] \rar[inarrow=<] & [\zxWCol] \zxN{}
        \end{ZX}
        \right\rrbracket\right)\right|_{\theta=-\frac{\pi}{2}}
\end{align}
and the same works for $\hat{F}^\dagger$ and $\hat{F}^2$.

\subsubsection{Copy rule}

\begin{align}
    \left\llbracket\begin{ZX}
   \zxN{}\ar[dr,-N, outarrow=<]&[\zxwCol,1mm]&[\zxwCol,1mm]\\
   \makebox[0pt][l]{\scalebox{0.8}{$\cvdotsCenterMathline$}}& \zxZ{f(x)}\ar[r,marrow=<]&\zxX{}\\
   \zxN{}\ar[ur,-N, outarrow=<]
\end{ZX}\right\rrbracket
\sim&\int_{\mathbb{R}}\dd{s} e^{if(s)} \subscripts{}{\ket{s\ldots s}}{q_1\ldots q_n}\subscripts{q}{\braket{s}{0}}{q}\\
=&e^{if(s)}\subscripts{}{\ket{0\ldots 0}}{q_1\ldots q_n}\\
\sim&{\subscripts{}{\ket{0}}{q}}^{\otimes n}\\
=&\left\llbracket\begin{ZX}
    \zxN{}\ar[r,marrow=<]&[\zxwCol,1mm]\zxX{}\\
    &\makebox[0pt][l]{\scalebox{0.8}{$\cvdotsCenterMathline$}}&\\
    \zxN{}\ar[r,marrow=<]&\zxX{}
 \end{ZX}\right\rrbracket
\end{align}
The same works for its colour-flipped counterpart. 

\subsubsection{Displacement rule}

\begin{align}
    &
    \left\llbracket\begin{ZX}
    \zxN{}\ar[dr,-N, outarrow=<]&[\zxwCol,1mm]&[\zxwCol,1mm]\zxX{ax}\ar[r,marrow=<]&[\zxwCol,1mm]\zxN{}\\
    \makebox[0pt][l]{\scalebox{0.8}{$\cvdotsCenterMathline$}}&  \zxZ{f(x)}\ar[ur,N-,inarrow=<]\ar[dr,N-,inarrow=<]&\makebox[0pt][r]{\scalebox{0.8}{$\cvdotsCenterMathline$}}\\
    \zxN{}\ar[ur,-N, outarrow=<]&&\zxX{ax}\ar[r,marrow=<]&\zxN{}
 \end{ZX}\right\rrbracket
 \\
=&\int_{\mathbb{R}}\dd{s}e^{if(s)}\left(\subscripts{}{\ket{s}}{q}\right)^{\otimes n}\left(\subscripts{q}{\bra{s}e^{-ia\hat{p}}}{}\right)^{\otimes m}\\
=&\int_{\mathbb{R}}\dd{s}e^{if(s)}\left(\subscripts{}{\ket{s}}{q}\right)^{\otimes n}\left(\subscripts{q}{\bra{s-a}}{}\right)^{\otimes m}\\
=&\int_{\mathbb{R}}\dd{s}e^{if(s+a)}\left(\subscripts{}{\ket{s+a}}{q}\right)^{\otimes n}\left(\subscripts{q}{\bra{s}}{}\right)^{\otimes m}\\
=&\int_{\mathbb{R}}\dd{s}e^{if(s+a)}\left(\subscripts{}{e^{-ia\hat{p}}\ket{s}}{q}\right)^{\otimes n}\left(\subscripts{q}{\bra{s}}{}\right)^{\otimes m}\\
=&
\left\llbracket
\begin{ZX}
    \zxN{}\ar[r,marrow=<]&[\zxwCol,1mm]\zxX{ax}\ar[dr,-N, marrow=<]&[\zxwCol,1mm]&[\zxwCol,2mm]\zxN{}\\
    &\makebox[0pt][l]{\scalebox{0.8}{$\cvdotsCenterMathline$}}&  \zxZ{f(x+a)}\ar[ru,N-,marrow=<]\ar[rd,N-,marrow=<]&\makebox[0pt][r]{\scalebox{0.8}{$\cvdotsCenterMathline$}}\\
    \zxN{}\ar[r,marrow=<]&\zxX{ax}\ar[ur,-N, marrow=<]&&\zxN{}
\end{ZX}
    \right\rrbracket
\end{align}
The colour-flipped counterpart holds as well.

\subsubsection{Antipode rule}

\begin{align}
    \left\llbracket
    \begin{ZX}
        \zxN{}&[\zxwCol]&[\zxwCol,2mm]&[\zxwCol]\\
        && \zxZ{} \ar[ull,N-, inarrow=>] \ar[dl,s.={L=.4}, marrow=>]\\
        &\zxX{}\ar[drr,N-, inarrow=<]&\\
        &&&\zxN{}
     \end{ZX}
     \right\rrbracket
   & = \int_\mathbb{R} \dd{s}  \int_\mathbb{R} \dd{t} \subscripts{}{\ket{s}}{q_{out}} \subscripts{p}{\braket{t}{s}}{q}  \subscripts{p_{in}}{\bra{t}}{}\\
   & = \int_\mathbb{R} \dd{s}  \int_\mathbb{R} \dd{t} \subscripts{}{\ket{-s}}{p_{out}} \subscripts{p}{\braket{t}{s}}{p} \subscripts{p_{in}}{\bra{t}}{}\\
   & = \int_\mathbb{R} \dd{s} \subscripts{}{\ket{-s}}{p_{out}}\subscripts{p_{in}}{\bra{s}}{}\\
   =&
   \left\llbracket
   \begin{ZX}
    \zxN{}&[\zxWCol] \zxFsq{} \lar[inarrow=>] \rar[inarrow=<] & [\zxWCol] \zxN{}
\end{ZX}
    \right\rrbracket
\end{align}

\subsubsection{Squeezing rule}

\begin{align}
    &
    \left\llbracket
    \begin{ZX}
   \zxN{}\ar[dr,-N, outarrow=<]&[\zxwCol,1mm]&[\zxwCol,1mm]\zxBox{Sq\left(\tau\right)}\ar[r,marrow=<]&[\zxwCol,1mm]\zxN{}\\
   \makebox[0pt][l]{\scalebox{0.8}{$\cvdotsCenterMathline$}}&  \zxZ{f(x)}\ar[ur,N-,numarrow={<}{0.4}]\ar[dr,N-,numarrow={<}{0.4}]&\makebox[0pt][r]{\scalebox{0.8}{$\cvdotsCenterMathline$}}\\
   \zxN{}\ar[ur,-N, outarrow=<]&&\zxBox{Sq\left(\tau\right)}\ar[r,marrow=<]&\zxN{}
\end{ZX}
\right\rrbracket
\\
=&\int_{\mathbb{R}}\dd{s}e^{if(s)}\left(\subscripts{}{\ket{s}}{q}\right)^{\otimes n}\left(\subscripts{q}{\bra{s}\widehat{Sq}(\tau)}{}\right)^{\otimes m}\\
\sim&\int_{\mathbb{R}}\dd{s}e^{if(s)}\left(\subscripts{}{\ket{s}}{q}\right)^{\otimes n}\left(\subscripts{q}{\bra{\frac{s}{\tau}}}{}\right)^{\otimes m}\\
\sim&\int_{\mathbb{R}}\dd{s}e^{if\left(\tau s\right)}\left(\subscripts{}{\ket{\tau s}}{q}\right)^{\otimes n}\left(\subscripts{q}{\bra{s}}{}\right)^{\otimes m}\\
\sim&\int_{\mathbb{R}}\dd{s}e^{if\left(\tau s\right)}\left(\subscripts{}{\widehat{Sq}(\tau)\ket{s}}{q}\right)^{\otimes n}\left(\subscripts{q}{\bra{s}}{}\right)^{\otimes m}\\*
=&
    \left\llbracket
    \begin{ZX}
   \zxN{}\ar[r,marrow=<]&[\zxwCol,1mm]\zxBox{Sq(\tau)}\ar[dr,-N, numarrow={<}{0.6}]&[\zxwCol,1mm]&[\zxwCol,2mm]\zxN{}\\
   &\makebox[0pt][l]{\scalebox{0.8}{$\cvdotsCenterMathline$}}&  \zxZ{f(x\tau)}\ar[ru,N-,marrow=<]\ar[rd,N-,marrow=<]&\makebox[0pt][r]{\scalebox{0.8}{$\cvdotsCenterMathline$}}\\
   \zxN{}\ar[r,marrow=<]&\zxBox{Sq(\tau)}\ar[ur,-N, numarrow={<}{0.6}]&&\zxN{}
\end{ZX}
\right\rrbracket
\end{align}
Again, the colour-flipped counterpart holds in the same way. 

\subsubsection{Quadratic rule}

\begin{align}
    &
    M\left(
    \left\llbracket\begin{ZX}
        \zxN{}&[\zxWCol] \zxZ{ax^2} \lar[inarrow=>] \rar[inarrow=<] &[\zxWCol] \zxX{bx^2} \rar[marrow=<]&[\zxWCol] \zxZ{cx^2} \rar[marrow=<]&[\zxWCol]\zxN{}
    \end{ZX}\right\rrbracket
    \right)
    \\=&\mat{1&0\\2a&1}\mat{1&2b\\0&1}\mat{1&0\\2c&1}
    \\=&\mat{4bc+1& 2b \\ 8abc+2a+2c& 4ab+1}
\end{align}
On the other hand,
\begin{align}
    &M\left(\left\llbracket
    \begin{ZX}
        \zxN{}&[\zxWCol] \zxX{\frac{bc}{4abc+a+c}x^2} \lar[inarrow=>] \rar[inarrow=<] &[\zxWCol] \zxZ{(4abc+a+c)x^2} \rar[marrow=<]&[\zxWCol] \zxX{\frac{ab}{4abc+a+c}x^2} \rar[marrow=<]&[\zxWCol]\zxN{}
    \end{ZX}\right\rrbracket
    \right)
    \\=&\mat{1&\frac{2bc}{4abc+a+c}\\0&1}\mat{1&0\\2(4abc+a+c)&1}\mat{1&\frac{2bc}{4abc+a+c}\\0&1}
    \\=&\mat{4bc+1& 2b \\ 8abc+2a+2c& 4ab+1}
\end{align}
holds. Thus, these two unitary operators are equivalent up to a scalar factor.

\subsubsection{Inversion rule}
\begin{align}
    &
    \left\llbracket
    \begin{ZX}
   \zxN{}\ar[r,marrow=<]&[\zxwCol,1mm]\zxF{}\ar[dr,-N, marrow=<]&[\zxwCol,1mm]&[\zxwCol,1mm]\zxFdag{}\ar[r,marrow=<]&[\zxwCol,1mm]\zxN{}\\
   &\makebox[0pt][l]{\scalebox{0.8}{$\cvdotsCenterMathline$}}&  \zxZ{f(x)}\ar[ur,N-,marrow=<]\ar[dr,N-,,marrow=<]&\makebox[0pt][r]{\scalebox{0.8}{$\cvdotsCenterMathline$}}\\
  \zxN{}\ar[r,marrow=<]&\zxF{}\ar[ur,-N, marrow=<]&&\zxFdag{}\ar[r,marrow=<]&\zxN{}
\end{ZX}
\right\rrbracket
\\
=&\int_{\mathbb{R}}\dd{u}e^{if(u)}\left(\subscripts{}{\hat{F}\ket{u}}{q}\right)^{\otimes n}\left(\subscripts{q}{\bra{u}\hat{F}^\dagger}{}\right)^{\otimes m}\\
=&\int_{\mathbb{R}}\dd{u}e^{if(u)}\left(\subscripts{}{\ket{u}}{p}\right)^{\otimes n}\left(\subscripts{p}{\bra{u}}{}\right)^{\otimes m}\\
=&
    \left\llbracket
    \begin{ZX}
        \zxN{}\ar[dr,-N, marrow=<]&[\zxwCol,1mm]&[\zxwCol,2mm]\zxN{}\\[\zxWRow]
        \makebox[0pt][l]{\scalebox{0.8}{$\cvdotsCenterMathline$}}&  \zxX{f(x)}\ar[ru,N-,marrow=<]\ar[rd,N-,marrow=<]&\makebox[0pt][r]{\scalebox{0.8}{$\cvdotsCenterMathline$}}\\[\zxWRow]
        \zxN{}\ar[ur,-N, marrow=<]&&\zxN{}
\end{ZX}
\right\rrbracket
\end{align}

\begin{widetext}

\section{Graphical calculus of 1-mode Gaussian gates}

\subsection{Associativity of squeezing gates}\label{ssec:sq_assoc}
\begin{thm}\label{thm:sq_assoc}
    For arbitrary nonzero real $\tau$ and $\kappa$,
    \begin{equation}
    \begin{ZX}
        \zxN{}&[\zxWCol] \zxBox{Sq(\tau)} \lar[inarrow=>] \rar[marrow=<] & [\zxWCol] \zxBox{Sq(\kappa)} \rar[inarrow=<] & [\zxWCol] \zxN{}
    \end{ZX}=\begin{ZX}
        \zxN{}&[\zxWCol] \zxBox{Sq(\tau\kappa)} \lar[inarrow=>] \rar[inarrow=<] & [\zxWCol] \zxN{}
    \end{ZX}\label{eq:sq_assoc}
\end{equation}
holds.
\end{thm}

\begin{prf*}
    Using Theorem.~\ref{thm:sq_eq}, one obtains
\begin{equation}
    \begin{ZX}
        \zxN{}&[\zxWCol] \zxBox{Sq(\tau)} \lar[inarrow=>] \rar[inarrow=<] & [\zxWCol] \zxN{}
    \end{ZX}
    =
    \begin{ZX}
        \zxN{}&[\zxWCol] \zxX{\frac{\tau(1-\tau)}{4}x^2} \lar[inarrow=>] \rar[inarrow=<] &[\zxWCol] \zxZ-{\frac{1}{\tau}x^2} \rar[marrow=<]&[\zxWCol] \zxX{\frac{\tau-1}{4}x^2} \rar[marrow=<]&[\zxWCol] \zxZ{x^2} \rar[marrow=<]&[\zxWCol]\zxN{}
    \end{ZX}
\end{equation}
and
\begin{equation}
    \begin{ZX}
        \zxN{}&[\zxWCol] \zxBox{Sq(\kappa)} \lar[inarrow=>] \rar[inarrow=<] & [\zxWCol] \zxN{}
    \end{ZX}
    =\begin{ZX}
        \zxN{}&[\zxWCol] \zxZ{-x^2} \lar[inarrow=>] \rar[inarrow=<] & \zxX{\frac{\kappa-1}{4\kappa}x^2} \rar[marrow=<]&[\zxWCol] \zxZ{\kappa x^2} \rar[marrow=<]&[\zxWCol] \zxX{-\frac{\kappa-1}{4\kappa^2}x^2} \rar[marrow=<]&[\zxWCol]\zxN{}
    \end{ZX}
\end{equation}
Therefore,
\begin{align}
    &\begin{ZX}
        \zxN{}&[\zxWCol] \zxBox{Sq(\tau)} \lar[inarrow=>] \rar[marrow=<] & [\zxWCol] \zxBox{Sq(\kappa)} \rar[inarrow=<] & [\zxWCol] \zxN{}
    \end{ZX}
    \\\overset{(\hyperlink{li:fusion}{f})}{=}&
\begin{ZX}
        \zxN{}&[\zxWCol] \zxX{\frac{\tau(1-\tau)}{4}x^2} \lar[inarrow=>] \rar[inarrow=<] &[\zxWCol] \zxZ-{\frac{1}{\tau}x^2} \rar[marrow=<]&[\zxWCol] \zxX{\frac{\tau-1}{4}x^2} \rar[marrow=<]&[\zxWCol] \zxX{\frac{\kappa-1}{4\kappa}x^2} \rar[marrow=<]&[\zxWCol] \zxZ{\kappa x^2} \rar[marrow=<]&[\zxWCol] \zxX{-\frac{\kappa-1}{4\kappa^2}x^2} \rar[marrow=<]&[\zxWCol]\zxN{}
    \end{ZX}
    \\\overset{(\hyperlink{li:fusion}{f})}{=}&
\begin{ZX}
        \zxN{}&[\zxWCol] \zxX{\frac{\tau(1-\tau)}{4}x^2} \lar[inarrow=>] \rar[inarrow=<] &[\zxWCol] \zxZ-{\frac{1}{\tau}x^2} \rar[marrow=<]&[\zxWCol] \zxX{\frac{\kappa\tau-1}{4\kappa}x^2} \rar[marrow=<]&[\zxWCol] \zxZ{\kappa x^2} \rar[marrow=<]&[\zxWCol] \zxX{-\frac{\kappa-1}{4\kappa^2}x^2} \rar[marrow=<]&[\zxWCol]\zxN{}
    \end{ZX}
    \\
\overset{(\hyperlink{li:quad}{q})}{=}&
\begin{ZX}
        \zxN{}&[\zxWCol] \zxX{\frac{\tau(1-\tau)}{4}x^2} \lar[inarrow=>] \rar[inarrow=<] &[\zxWCol] \zxZ-{\frac{1}{\tau}x^2} \rar[marrow=<]&[\zxWCol] \zxZ{\frac{1-\kappa}{\tau-1}x^2} \rar[marrow=<]&[\zxWCol] \zxX{\frac{\tau-1}{4\kappa} x^2} \rar[marrow=<]&[\zxWCol] \zxZ{\frac{\kappa(\kappa\tau-1)}{\tau-1}x^2} \rar[marrow=<]&[\zxWCol]\zxN{}
    \end{ZX}\\
\overset{(\hyperlink{li:fusion}{f})}{=}&\begin{ZX}
        \zxN{}&[\zxWCol] \zxX{\frac{\tau(1-\tau)}{4}x^2} \lar[inarrow=>] \rar[inarrow=<] &[\zxWCol] \zxZ{\frac{1-\kappa\tau}{\tau(\tau-1)}x^2} \rar[marrow=<]&[\zxWCol] \zxX{\frac{\tau-1}{4\kappa} x^2} \rar[marrow=<]&[\zxWCol] \zxZ{\frac{\kappa(\kappa\tau-1)}{\tau-1}x^2} \rar[marrow=<]&[\zxWCol]\zxN{}
    \end{ZX}
\end{align}
which is equivalent to
\begin{align}
    \begin{ZX}
        \zxN{}&[\zxWCol] \zxBox{Sq(\tau\kappa)} \lar[inarrow=>] \rar[inarrow=<] & [\zxWCol] \zxN{}
    \end{ZX}
    =\begin{ZX}
        \zxN{}&[\zxWCol] \zxX{\frac{\tau\kappa(1-\tau\kappa)}{4k}x^2} \lar[inarrow=>] \rar[inarrow=<] &[\zxWCol] \zxZ-{\frac{k}{\tau\kappa}x^2} \rar[marrow=<]&[\zxWCol] \zxX{\frac{\tau\kappa-1}{4k}x^2} \rar[marrow=<]&[\zxWCol] \zxZ{kx^2} \rar[marrow=<]&[\zxWCol]\zxN{}
    \end{ZX}
\end{align}
by substituting $k=\frac{\kappa(\kappa\tau-1)}{\tau-1}$ in Theorem.~\ref{thm:sq_eq}.
\hfill \qed
\end{prf*}


\subsection{Associativity of rotation gates}\label{ssec:rot_assoc}

\begin{thm}
    For arbitrary real $\theta$ and $\phi$,
    \begin{equation}\label{eq:rot_assoc}
        \begin{ZX}
            \zxN{} \rar[marrow=<]&[\zxWCol] \zxBox{R(\theta)} \rar[marrow=<] & [\zxWCol] \zxBox{R(\phi)} \rar[marrow=<] & [\zxWCol] \zxN{}
        \end{ZX}
        =
        \begin{ZX}
            \zxN{}\rar[marrow=<]&[\zxWCol] \zxBox{R(\theta+\phi)}  \rar[marrow=<] &[\zxWCol] \zxN{}
        \end{ZX}
    \end{equation}
    can be shown under the identity rule~(\hyperlink{li:id}{id}), the fusion rule~(\hyperlink{li:fusion}{f}), and the quadratic rule~(\hyperlink{li:quad}{q}).
\end{thm}

\begin{prf*}
    If either $\theta= 2n\pi$ or $\phi = 2n\pi$ holds, then the rotation is equivalent to identity operation, and thus \eqref{eq:rot_assoc} is obvious. Likewise, if $\theta = (2n+1)\pi$ for some $n\in \mathbb{Z}$, then
    \begin{align}
        &
        \begin{ZX}
            \zxN{} \rar[marrow=<]&[\zxWCol] \zxBox{R((2n+1)\pi)} \rar[marrow=<] & [\zxWCol] \zxBox{R(\phi)} \rar[inarrow=<] & [\zxWCol] \zxN{}
        \end{ZX}
        \\\overset{\mathclap{\eqref{def:rot_sp}}}{=}&
\begin{ZX}
            \zxN{} \rar[marrow=<]&[\zxWCol] \zxBox{Sq(-1)} \rar[marrow=<] & [\zxWCol] \zxX{\frac{\tan(\theta/2)}{2}x^2} \rar[marrow=<] &[\zxWCol] \zxZ-{\frac{\sin\theta}{2}x^2} \rar[marrow=<]&[\zxWCol] \zxX{\frac{\tan(\theta/2)}{2}x^2} \rar[marrow=<]&[\zxWCol]\zxN{}
        \end{ZX}
        \\\overset{\mathclap{\eqref{eq:sq_k_Z}}}{=}&\begin{ZX}
        \zxN{}&[\zxWCol] \zxZ{\frac{1}{\tan(\phi/2)}x^2} \lar[marrow=>] \rar[marrow=<] &[\zxWCol] \zxX-{\frac{\tan(\phi/2)}{2}x^2} \rar[marrow=<]&[\zxWCol] \zxZ{\frac{1}{\tan(\phi/2)}x^2}\rar[marrow=<] & [\zxWCol] \zxX-{\frac{\tan(\theta/2)}{2}x^2}\rar[marrow=<] & [\zxWCol] \zxX{\frac{\tan(\theta/2)}{2}x^2} \rar[marrow=<] &[\zxWCol] \zxZ-{\frac{\sin\phi}{2}x^2} \rar[marrow=<]&[\zxWCol] \zxX{\frac{\tan(\phi/2)}{2}x^2} \rar[marrow=<]&[\zxWCol]\zxN{}
    \end{ZX}\\
    \overset{(\hyperlink{li:fusion}{f})}{=}&
        \begin{ZX}
        \zxN{}&[\zxWCol] \zxZ{\frac{1}{\tan(\phi/2)}x^2} \lar[marrow=>] \rar[marrow=<] &[\zxWCol] \zxX-{\frac{\tan(\phi/2)}{2}x^2} \rar[marrow=<]&[\zxWCol] \zxZ{\left(\frac{1}{\tan(\phi/2)}-\frac{\sin\phi}{2}\right)x^2} \rar[marrow=<]&[\zxWCol] \zxX{\frac{\tan(\phi/2)}{2}x^2} \rar[marrow=<]&[\zxWCol]\zxN{}
    \end{ZX}\\
    \overset{(\hyperlink{li:quad}{q})}{=}&
\begin{ZX}
        \zxN{}&[\zxWCol] \zxZ{\frac{1}{\tan(\phi/2)}x^2} \lar[marrow=>] \rar[marrow=<] &[\zxWCol] \zxZ-{\frac{1}{2\tan(\phi/2)}x^2} \rar[marrow=<]&[\zxWCol] \zxX-{\frac{\sin\phi}{2}x^2} \rar[marrow=<]&[\zxWCol] \zxZ{\frac{1}{2\tan(\phi/2)}x^2} \rar[marrow=<]&[\zxWCol]\zxN{}
    \end{ZX}\\
    \overset{(\hyperlink{li:fusion}{f})}{=}&
\begin{ZX}
        \zxN{}\rar[marrow=<] &[\zxWCol] \zxZ{\frac{1}{2\tan(\phi/2)}x^2} \rar[marrow=<]&[\zxWCol] \zxX-{\frac{\sin\phi}{2}x^2} \rar[marrow=<]&[\zxWCol] \zxZ{\frac{1}{2\tan(\phi/2)}x^2} \rar[marrow=<]&[\zxWCol]\zxN{}
    \end{ZX}\\
    \overset{(\hyperlink{li:quad}{q})}{=}&\begin{ZX}
        \zxN{}&[\zxWCol] \zxX-{\frac{1}{2\tan(\theta/2)}x^2} \lar[inarrow=>] \rar[inarrow=<] &[\zxWCol] \zxZ{\frac{\sin\theta}{2}x^2} \rar[marrow=<]&[\zxWCol] \zxX-{\frac{1}{2\tan(\theta/2)}x^2} \rar[marrow=<]&[\zxWCol]\zxN{}
    \end{ZX}\\
    \overset{\mathclap{\eqref{def:rot_sp}}}{=}&
\begin{ZX}
        \zxN{}\rar[marrow=<] &[\zxWCol] \zxBox{R(\phi+\pi)} \rar[inarrow=<] & [\zxWCol] \zxN{}
    \end{ZX}
    \end{align}    
    for all $\phi$ satisfying $\sin\phi \neq 0$, and
    \begin{align}
        &\begin{ZX}
            \zxN{} \rar[marrow=<]&[\zxWCol] \zxBox{R((2n+1)\pi)} \rar[marrow=<] & [\zxWCol] \zxBox{R((2n+1)\pi)} \rar[inarrow=<] & [\zxWCol] \zxN{}
        \end{ZX}
        \\=&\begin{ZX}
            \zxN{} \rar[marrow=<]&[\zxWCol] \zxBox{Sq(-1)} \rar[marrow=<] &[\zxWCol] \zxBox{Sq(-1)} \rar[marrow=<] &[\zxWCol]\zxN{}
        \end{ZX}
        \\\overset{\mathclap{\eqref{eq:sq_assoc}}}{=}&\begin{ZX}
            \zxN{} \rar[marrow=<]&[\zxWCol] \zxBox{Sq(1)} \rar[marrow] &[\zxWCol]\zxN{}
        \end{ZX}
        \\=&
\begin{ZX}
            \zxN{}&[\zxWCol,5mm] \zxN{} \lar[marrow=>]
        \end{ZX}\label{eq:anti_self}
    \end{align}
    holds (by Theorem.~\ref{thm:sq_assoc}). The same argument applies when $\phi = (2n+1)\pi$. Therefore, all we need to confirm is when $\sin\theta \neq 0$ and $\sin \phi \neq 0$.

\begin{enumerate}[(i)]
  \item If $\sin(\theta + \phi) \neq 0$, then
\begin{align}
    &\begin{ZX}
        \zxN{}&[\zxWCol] \zxBox{R(\theta)} \lar[inarrow=>] \rar[marrow=<] & [\zxWCol] \zxBox{R(\phi)} \rar[inarrow=<] & [\zxWCol] \zxN{}
    \end{ZX}
        \\\overset{(\hyperlink{li:fusion}{f})}{=}&\begin{ZX}
        \zxN{}&[\zxWCol] \zxX{\frac{\tan(\theta/2)}{2}x^2} \lar[inarrow=>] \rar[inarrow=<] &[\zxWCol] \zxZ-{\frac{\sin\theta}{2}x^2} \rar[marrow=<]&[\zxWCol] \zxX{\frac{\tan(\theta/2)+\tan(\phi/2)}{2}x^2} \rar[marrow=<]&[\zxWCol] \zxZ-{\frac{\sin\phi}{2}x^2} \rar[marrow=<]&[\zxWCol] \zxX{\frac{\tan(\phi/2)}{2}x^2} \rar[marrow=<]&[\zxWCol]\zxN{}
    \end{ZX}
    \\
    \overset{(\hyperlink{li:quad}{q})}{=}&
\begin{ZX}
        \zxN{}&[\zxWCol] \zxX{\frac{\tan(\theta/2)}{2}x^2} \lar[inarrow=>] \rar[inarrow=<] &[\zxWCol] \zxX{\left(\frac{\tan((\theta+\phi)/2)-\tan(\theta/2)}{2}\right)x^2} \rar[marrow=<]&[\zxWCol] \zxZ-{\frac{\sin(\theta+\phi)}{2}x^2} \rar[marrow=<]&[\zxWCol] \zxX{\left(\frac{\tan((\theta+\phi)/2)-\tan(\phi/2)}{2}\right)x^2} \rar[marrow=<]&[\zxWCol] \zxX{\frac{\tan(\phi/2)}{2}x^2} \rar[marrow=<]&[\zxWCol]\zxN{}
    \end{ZX}
    \\\overset{(\hyperlink{li:fusion}{f})}{=}&\begin{ZX}
        \zxN{}&[\zxWCol]\zxX{\frac{\tan((\theta+\phi)/2)}{2}x^2}\lar[inarrow=>]  \rar[marrow=<]&[\zxWCol] \zxZ-{\frac{\sin(\theta+\phi)}{2}x^2} \rar[marrow=<]&[\zxWCol] \zxX{\frac{\tan((\theta+\phi)/2)}{2}x^2}\rar[marrow=<]&[\zxWCol]\zxN{}
    \end{ZX}
        \\\overset{\mathclap{\eqref{def:rot_sp}}}{=}&\begin{ZX}
        \zxN{}&[\zxWCol] \zxBox{R(\theta+\phi)} \lar[inarrow=>] \rar[inarrow=<] & [\zxWCol] \zxN{}
    \end{ZX}
\end{align}
holds. Here we applied the quadratic rule~(\hyperlink{li:quad}{q}) under the assumption $\sin(\theta + \phi) \neq 0$.

\item If $\theta + \phi = 2n\pi$, then $\sin\theta=-\sin\phi$ and $\tan(\phi/2)=\tan(n\pi -\theta/2)= -\tan(\theta/2)$,
and thus
\begin{align}
    &
    \begin{ZX}
        \zxN{}&[\zxWCol] \zxBox{R(\theta)} \lar[inarrow=>] \rar[marrow=<] & [\zxWCol] \zxBox{R(\phi)} \rar[inarrow=<] & [\zxWCol] \zxN{}
    \end{ZX}
    \\\overset{(\hyperlink{li:fusion}{f})}{=}&\begin{ZX}
        \zxN{}&[\zxWCol] \zxX{\frac{\tan(\theta/2)}{2}x^2} \lar[inarrow=>] \rar[inarrow=<] &[\zxWCol] \zxZ-{\frac{\sin\theta}{2}x^2} \rar[marrow=<]&[\zxWCol] \zxX{\frac{\tan(\theta/2)+\tan(\phi/2)}{2}x^2} \rar[marrow=<]&[\zxWCol] \zxZ-{\frac{\sin\phi}{2}x^2} \rar[marrow=<]&[\zxWCol] \zxX{\frac{\tan(\phi/2)}{2}x^2} \rar[marrow=<]&[\zxWCol]\zxN{}
    \end{ZX}
    \\=&\begin{ZX}
        \zxN{}&[\zxWCol] \zxX{\frac{\tan(\theta/2)}{2}x^2} \lar[inarrow=>] \rar[inarrow=<] &[\zxWCol] \zxZ-{\frac{\sin\theta}{2}x^2} \rar[marrow=<]&[\zxWCol] \zxX{0} \rar[marrow=<]&[\zxWCol] \zxZ{\frac{\sin\theta}{2}x^2} \rar[marrow=<]&[\zxWCol] \zxX-{\frac{\tan(\theta/2)}{2}x^2} \rar[marrow=<]&[\zxWCol]\zxN{}
    \end{ZX}
    \\\overset{(\hyperlink{li:id}{id})}{=}&\begin{ZX}
        \zxN{}&[\zxWCol] \zxX{\frac{\tan(\theta/2)}{2}x^2} \lar[inarrow=>] \rar[inarrow=<] &[\zxWCol] \zxZ-{\frac{\sin\theta}{2}x^2} \rar[marrow=<]&[\zxWCol] \zxZ{\frac{\sin\theta}{2}x^2} \rar[marrow=<]&[\zxWCol] \zxX-{\frac{\tan(\theta/2)}{2}x^2} \rar[marrow=<]&[\zxWCol]\zxN{}
    \end{ZX}
    \\\overset{(\hyperlink{li:fusion}{f})}{=}&\begin{ZX}
        \zxN{}&[\zxWCol] \zxX{\frac{\tan(\theta/2)}{2}x^2} \lar[inarrow=>] \rar[inarrow=<]&[\zxWCol] \zxX-{\frac{\tan(\theta/2)}{2}x^2} \rar[marrow=<]&[\zxWCol]\zxN{}
    \end{ZX}
    \\
    \overset{(\hyperlink{li:fusion}{f})}{=}&\begin{ZX}
        \zxN{}&[\zxWCol,5mm] \zxN{} \lar[marrow=>]
    \end{ZX}
    \\
    =&\begin{ZX}
        \zxN{}&[\zxWCol] \zxBox{R(2n\pi)} \lar[inarrow=>] \rar[marrow=<] & [\zxWCol] \zxN{}
    \end{ZX}
\end{align}
holds.

\item If $\theta + \phi = (2n+1)\pi$, then $\sin\phi=\sin\theta$ and $\tan(\phi/2)= \frac{1}{\tan(\theta/2)}$,
and thus
\begin{align}
    &\begin{ZX}
        \zxN{}&[\zxWCol] \zxBox{R(\theta)} \lar[inarrow=>] \rar[marrow=<] & [\zxWCol] \zxBox{R(\phi)} \rar[inarrow=<] & [\zxWCol] \zxN{}
    \end{ZX}
    \\\overset{(\hyperlink{li:fusion}{f})}{=}&\begin{ZX}
        \zxN{}&[\zxWCol] \zxX{\frac{\tan(\theta/2)}{2}x^2} \lar[inarrow=>] \rar[inarrow=<] &[\zxWCol] \zxZ-{\frac{\sin\theta}{2}x^2} \rar[marrow=<]&[\zxWCol] \zxX{\frac{\tan(\theta/2)+\tan(\phi/2)}{2}x^2} \rar[marrow=<]&[\zxWCol] \zxZ-{\frac{\sin\phi}{2}x^2} \rar[marrow=<]&[\zxWCol] \zxX{\frac{\tan(\phi/2)}{2}x^2} \rar[marrow=<]&[\zxWCol]\zxN{}
    \end{ZX}
    \\=&\begin{ZX}
        \zxN{}&[\zxWCol] \zxX{\frac{\tan(\theta/2)}{2}x^2} \lar[inarrow=>] \rar[inarrow=<] &[\zxWCol] \zxZ-{\frac{\sin\theta}{2}x^2} \rar[marrow=<]&[\zxWCol] \zxX{\left(\frac{\tan(\theta/2)}{2}+\frac{1}{2\tan(\theta/2)}\right)x^2} \rar[marrow=<]&[\zxWCol] \zxZ-{\frac{\sin\theta}{2}x^2} \rar[marrow=<]&[\zxWCol] \zxX{\frac{1}{2\tan(\theta/2)}x^2} \rar[marrow=<]&[\zxWCol]\zxN{}
    \end{ZX}
    \\\overset{(\hyperlink{li:quad}{q})}{=}&
\begin{ZX}
        \zxN{}&[\zxWCol] \zxX{\frac{\tan(\theta/2)}{2}x^2} \lar[inarrow=>] \rar[inarrow=<] &[\zxWCol] \zxZ-{\frac{\sin\theta}{2}x^2} \rar[marrow=<]&[\zxWCol] \zxZ-{\frac{\sin\theta}{2\tan^2(\theta/2)}x^2} \rar[marrow=<]&[\zxWCol] \zxX{\frac{\tan(\theta/2)}{2}x^2} \rar[marrow=<]&[\zxWCol] \zxZ-{\frac{1}{\tan(\theta/2)}x^2} \rar[marrow=<]&[\zxWCol]\zxN{}
    \end{ZX}
    \\\overset{(\hyperlink{li:fusion}{f})}{=}&\begin{ZX}
        \zxN{}&[\zxWCol] \zxX{\frac{\tan(\theta/2)}{2}x^2} \lar[inarrow=>] \rar[inarrow=<] &[\zxWCol] \zxZ-{\frac{1}{\tan(\theta/2)}x^2} \rar[marrow=<]&[\zxWCol] \zxX{\frac{\tan(\theta/2)}{2}x^2} \rar[marrow=<]&[\zxWCol] \zxZ-{\frac{1}{\tan(\theta/2)}x^2} \rar[marrow=<]&[\zxWCol]\zxN{}
    \end{ZX}
    \\\overset{\mathclap{\eqref{eq:sq_k_X}}}{=}&\begin{ZX}
        \zxN{}& [\zxWCol] \zxBox{Sq(-1)} \lar[inarrow=>]\rar[inarrow=<] & [\zxWCol] \zxN{}
    \end{ZX}
    \\\overset{(\hyperlink{fozurier}{F})}{=}&
    \begin{ZX}
        \zxN{}&[\zxWCol] \zxBox{R(\pi)} \lar[inarrow=>]\rar[inarrow=<] & [\zxWCol] \zxN{}
    \end{ZX}\\=&
    \begin{ZX}
        \zxN{}&[\zxWCol] \zxBox{R((2n+1)\pi)} \lar[inarrow=>]\rar[inarrow=<] & [\zxWCol] \zxN{}
    \end{ZX}
\end{align}
holds.
\end{enumerate}
\end{prf*}

\end{widetext}




\bibliography{zx_bibs}
\end{document}